%% file: main.tex
\documentclass{article}

\usepackage{PRIMEarxiv}

\usepackage[utf8]{inputenc} 
\usepackage[T1]{fontenc}    
\usepackage{hyperref}       
\usepackage{url}            
\usepackage{booktabs}       
\usepackage{amsfonts}       
\usepackage{nicefrac}       
\usepackage{microtype}      
\usepackage{lipsum}
\usepackage{fancyhdr}       
\usepackage{graphicx}       
\graphicspath{{media/}}  
\usepackage{lipsum}
\usepackage{multicol}
\usepackage{crop}%
\usepackage{amsmath,amssymb,amsfonts,upref,endnotes}
\usepackage{amsthm}
\usepackage{marginnote}
\usepackage{soul}
\RequirePackage{graphicx}
\usepackage[figuresright]{rotating}
\usepackage{color}
\usepackage{amsmath,amsfonts,amssymb,amsthm}
\usepackage{lipsum}
\usepackage{graphicx,color}
\usepackage[noend]{algorithmic}
\usepackage{algorithm} 
\usepackage{color,soul}
\usepackage{wrapfig}
\usepackage{subcaption}
\title{Social Stratification in Networks: Insights from Co-Authorship Networks
}

\author{
  Zeinab S. Jalali, Josh Introne, and Sucheta Soundarajan \\
  Syracuse University \\
  \texttt{\{zsaghati, jeintron, susounda\}syr@edu} \\
}

\begin{document}
\maketitle

\begin{abstract}
It has been observed that real-world social networks often exhibit stratification along economic or other lines, with consequences for class mobility and access to opportunities. 
With the rise in human interaction data and extensive use of online social networks, the structure of social networks (representing connections between individuals) can be used for measuring stratification. 
However, although stratification has been studied extensively in the social sciences, there is no single, generally applicable metric for measuring the level of stratification in a network.  

In this work, we first propose the novel \texttt{Stratification Assortativity (StA)} metric, which measures the extent to which a network is stratified into different tiers.  
Then, we use the \texttt{StA} metric to perform an in-depth analysis of the stratification of five co-authorship networks.
We examine the evolution of these networks over 50 years and show that these fields demonstrate an increasing level of stratification over time, and, correspondingly, the trajectory of a researcher's career is increasingly correlated with her entry point into the network.
\end{abstract}


\section{Introduction}
Human social networks play a critical role in the trajectories of people's lives.  A highly-desirable property of dynamic societal networks is that they should allow for individuals to rise and fall on the basis of their own merits, rather than their inherited positional inequalities~\cite{van2009social}.  This basic property has been described many times in the sociological and philosophical literature~\cite{savage1997social}.

However, in real societies, individuals in human networks are commonly divided into a hierarchical arrangement based on different attributes such as importance, wealth, knowledge and power.  This phenomenon is known as \textit{social stratification}~\cite{gupte2011finding}, and stratification along economic or class-based lines has been one of the most important topics of study in the modern social sciences~\cite{Kerbo2017Social}.  Social stratification, as well as its counterpart \textit{social mobility}, governs the trajectories of people's lives, including the extent of prejudice that they face~\cite{hodler2020measuring}, their careers and occupations~\cite{warren2002occupational}, and the likelihood that they will experience violence~\cite{lamertz2004social}.  

Historically, social science researchers studying stratification have been armed with domain knowledge about the nature of stratification in the system: in particular, knowledge of the classes of interest, such as upper, middle, and lower socioeconomic classes). These classes are often defined with respect to occupation, living conditions, socioeconomic status, etc~\cite{savage2013new, leo2016socioeconomic}.
Existing empirical analysis of stratification tends to either study social mobility as a proxy for stratification or, if network connections are known,
examine inter-class connections between predefined classes~\cite{van2010historical}.  However, these methods tend to be ad hoc, relying on \textit{a priori} knowledge about the classes. 
 
Although stratification has been extensively documented (especially in non-network settings), to our knowledge, there is no single metric to quantify the level of stratification in a network and there is a need for an interpretable, quantitative metric to summarize network stratification in a single number. 
There are several reasons for this.  (1) In many cases, one may not know the classes ahead of time.  While economic classes are well-established, this may not be the case for other domains.  
(2) When comparing different ecosystems (e.g., Facebook vs. Twitter), it is useful to have a single number in order to perform a quantitative comparison.  In the long run, this may lead to discovery of universal laws about stratification.

In the first part of our work, we introduce the \texttt{Stratification Assortativity (StA)} metric, which measures a network's stratification with respect to an attribute of interest.  \texttt{StA} differs from existing assortativity metrics in that it is based on scalar characteristics that give rise to a set of ordered classes,
whereas other assortativity metrics are either based on categorical characteristics that divide a network into non-ordered groups
or scalar characteristics that do not consider group memberships~\cite{newman2003mixing}.

In the second part of our work, we apply our proposed metric to understand stratification in scientific co-authorship networks.
A fair amount of work has examined collaboration networks with respect to stratification and equality. Most of these studies use centrality metrics to show the correlation between an individual's position within the collaboration network and their success; and the ramifications of such correlation to stratification and inequality~\cite{abbasi2012egocentric, liu2022scientific}. 
none of them have studied the evolution of co-authorship networks with respect to long-term stratification and the success of researchers. 

We perform an in-depth analysis of five co-authorship networks with respect to author $h$-index\footnote{The maximum number $h$ such that the author has at least $h$ papers with at least $h$ citations each.}, examining the evolution of these networks over 50 years, and demonstrate that networks evolve into a highly-stratified state.  Using our proposed \texttt{StA} metric, we show that these fields demonstrate high levels of stratification. Interestingly, we also find that while other types of assortativity decrease over time, stratification increases: in other words, while individuals collaborate with a more diverse set of researchers (with respect to $h$-index) over time, stratification actually increases.  Moreover, as stratification increases, the trajectory of a researcher's career becomes increasingly correlated with her entry point into the network.

The major contributions of this work are as follows: (a) We propose the novel \texttt{StA} metric and introduce an algorithm for identifying classes in a stratified network,
(b) We perform an extensive analysis of stratification in scientific co-authorship networks
 and show how networks evolve into a highly stratified state as they age, using both topological properties and success score of researchers.

\section{Background}
\label{sec: net_stra}
In this section, we provide background on social stratification in societies and social networks and describe existing research on stratification in collaboration networks.

\subsection{Social Stratification Overview}
{The study of social stratification has been one of the most important topics in modern sociology and economics}~\cite{Kerbo2017Social}.
{{While, historically, there has been disagreement on what exactly constitutes a stratified system, there is some consensus that, at the least, a stratified system requires a ranking or hierarchy of people and groups; acknowledgement, acceptance, and legitimation of that ranking; and a correlation between one's position in the ranking and access to power, prestige, or resources}}~\cite{kerbo2006social}.  {{Note that social stratification is not synonymous with social inequality: while social inequality can be a cause of social stratification}}~\cite{leo2016socioeconomic}, {there are societies in which such inequalities have not created the separate classes present in a stratified society}~\cite{pakulski2005foundations}.

{{In the Marxian perspective, stratification or class divisions occur due to the division of individuals based on "control and ownership of  the means of production and labor power"}}~\cite{marx1975marx, kerbo2006social}.  {{Weber extended this one-dimensional view of social stratification to multi dimensions and considered other types of ownership such as skills, status, and organizational power}}~\cite{gerth1946politics, kerbo2006social}.  {{These views are in contrast to the functional view of stratification espoused by Durkheim, who distinguished external inequalities (those imposed by society) from internal inequalities (based on personal merits, such as talent), and believed that the latter type are necessary for the functioning of society}}~\cite{kerbo2006social}.  

Alternative notions of class also exist: for example, Bourdieu suggested that a class is a set of people with similar nature and living conditions~\cite{savage2013new}.  
{In Bourdieu's perspective, class structure in a society is a multidimensional space which is shaped by the distributions of different forms of social, economic, and cultural capital.  For Bourdieu, social classes are constructed as social groups, through `articulation, representation, and mobilization, by relevant parties or unions'
}~\cite{flemmen2013putting, bourdieu1987makes}.

{Intrinsically connected to the notion of social stratification is that of social class.  When analyzing the stratification of a society, it is useful to identify specific social classes.}  Defined social classes can directly be used to measure social stratification or be used as a basis of other analysis. For instance, some studies have examined social mobility between classes~\cite{van2010historical}.  Social class refers to hierarchical social categories arising from different relationships in the society~\cite{krieger1997measuring}. These social classes either divide individuals into categories where boundaries are clearly identified ~\cite{lambert2018social} or divide individuals into points along a one-dimensional scale~\cite{lambert2018social}.

The study of social stratification is closely related to that of social mobility~\cite{grusky2019social}, and social mobility (or the lack thereof) is a driving process behind social stratification~\cite{featherman1981social}. 
It is known that social stratification can influence prejudice~\cite{hodler2020measuring}, social capital~\cite{kim2009networks}, probability of victimization~\cite{lamertz2004social}, 
occupation~\cite{warren2002occupational}, 
and other crucial factors in the lives of individuals.

{{Many existing measures of social networks can provide evidence of stratification, but we are not aware of any measures that confirm its presence. For instance, aggregate-level social metrics like the Gini coefficient measure inequality}~\cite{dorfman1979formula}.

{Another well-studied type of inequality is the Matthew effect (the rich-get-richer or preferential attachment phenomenon)}~\cite{merton1988matthew}, {and numerous works have explored quantifying the strength of such a phenomenon}~\cite{perc2014matthew}. }

With the rise in human interaction data and extensive use of online social networks, the structure of social networks can be used to study social classes and social stratification~\cite{leo2016socioeconomic}. For instance, the capability of individuals to be upwardly mobile can be estimated by examining their connections to higher status individuals in networks~\cite{son2012network}. 
In other words, networks provide an opportunity to study the emergence of social stratification~\cite{gupte2011finding}, which can help to understand how decisions of individuals can lead to a socially stratified network.

Some empirical analyses on networks have examined social mobility as a proxy for stratification or, if network connections are known, individually examine inter-class connections between predefined classes~\cite{van2010historical}. 
These methods are closer to the phenomenon of stratification then measures of inequality, but they are nonetheless indirect (inferred from its consequences).  Moreoever, measuring social mobility requires long-term temporal data, and looking at inter-class connections between predefined classes is not generalizable. 

{{For instance, Holder \textit{et al.} use an index which measures the probability that for any random pair of individuals, the poorer individual is deprived of opportunities associated with ethnic class boundaries. This metric uses ethnicity as a proxy for social distance and examines connections between individuals in the context of wealth
}}
\cite{hodler2020measuring}.
{{However, the phenomenon targeted by Holder's metric could arise for other reasons (e.g., racism), and it does not explicitly examine the division of the network into social classes. }}

{{Various measures of homophily are also related to stratification, but do not distinguish between ordered strata. For example, the Scalar Assortativity Coefficient (SAC)}~\cite{newman2003mixing} {is like the Matthew effect in that it can reveal a process of preferential attachment, but is better understood as a measure of} {{overall inequality without considering the division of the network into any specific classes. Modularity}}~\cite{newman2006modularity} {and the Discrete Assortativity Coefficient (DAC)}~\cite{newman2003mixing} {consider the division of network into classes, but these classes are not necessarily tiered and don't satisfy the social class definitions.}}
{Details of these metrics (Modularity, Discrete/Scalar Assortativity) are provided in the appendix section.
None of the existing metrics }
measure the extent to which the network is divided into \textit{ordered classes}.
Our first major contribution in this paper is to propose a metric that satisfies this requirement.

\subsection{Stratification in Collaboration Networks}\label{sec: Strat_in_collab}
 
The structures of collaboration networks are important for the diffusion and production of scientific knowledge, and can have an impact on the productivity and work done by the researchers~\cite{hoekman2013proximity}. Social connections in these networks affect the career trajectories of researchers~\cite{yin2006connection, dusdal2021higher}. Accordingly, the study of stratification and inequality in collaboration networks has attracted recent attention. 

Most of these studies use centrality metrics to analyze networks. For instance Yin \textit{et al.} studied COLLNET, a small collaboration network with respect to stratification. This work demonstrated that certain nodes in favored locations that are densely connected
cause inequality or stratification. Abbasi \textit{et al.} and Liu \textit{et al.} study co-authorship networks and show that the research performance of researchers in terms of their \textit{h-}index~\cite{abbasi2012egocentric, liu2022scientific}.
McCarty \textit{et al.} shows the importance of the 
positions of researchers in a co-authorship network by predicting the \textit{h-}index of authors in the future from their current position~\cite{mccarty2013predicting}. Most of the works in this area focus on analysing one snapshot of the network. However, there are some works on analyzing the evolution of collaboration networks. For instance, Servia \textit{et al.} study the academic success of researchers using co-authorship networks over time and showed the correlation between centrality and citations~\cite{servia2015evolution}.

Other works have studied stratification and inequality in collaboration networks from different perspectives. Some of these works examine the inequality of citations among researchers. Dong \textit{et al.} used the Gini index and quantified inequality in citation impact in different stages of academic life, and showed that the majority of citations come from a small percentage of researchers~\cite{dong2021inequality}. 

\section{Proposed Metric: Stratification Assortativity (StA)}

In this section we first introduce the problem of measuring social stratification in networks. 
Then, we propose \texttt{Stratification Assortativity (StA)}, an assortativity metric for measuring network social stratification. 

\subsection{Goal}
Assume that we are given an undirected, unweighted network/graph $G(V,E)$ with vertices $V$ and edges $E$ and binary adjacency matrix $\mathbf{A}$.  Nodes in $G$ have a numeric \textit{characteristic attribute} of interest.  This characteristic attribute represents a node's status (e.g., wealth, professional success, etc.), with higher values indicating a higher status. This attribute is used to identify the social class hierarchy in the networks that causes social stratification.  (Note that while the attribute is required, specific class boundaries are not.)  
 
 \begin{table*}
\footnotesize
\centering
 \caption{Summary of desired properties and whether different metrics satisfy them. }
\label{table:metrics_comparision}
\begin{tabular}{l|l|l|l|l } 
\textbf{Metrics}
 &\textbf{Scalar } 
  &\textbf{Ordered }  
  &\textbf{Class } 
 &\textbf{Mutual}
 \\
 &\textbf{ Values} 
  &\textbf{ Tiers} 
 &\textbf{ Membership}  
 & \textbf{Segregation}
 \\
\hline

\textbf{Modularity}
 & No    & No & \textbf{Yes} & No 
\\
\hline
\textbf{Discrete}
 & No   &No  & \textbf{Yes}& No 
\\
\textbf{Assortativity}
&   &  &   
\\
\hline
\textbf{Scalar}
 &  \textbf{Yes}  & No & No & No
\\
\textbf{Assortativity}
&   &  &  
\\
\hline
\textbf{Stratification}
  & \textbf{Yes}   & \textbf{Yes} & \textbf{Yes}  & \textbf{Yes} 
\\
\textbf{Assortativity}
&   &  &  
\\

\end{tabular}
\vspace{0.5cm}
\end{table*}
 
\textit{\textbf{Our goal is to propose an assortativity metric for measuring {network social stratification}, the division of a network into ordered tiers (social classes) based on the attribute of interest.}}

At a high-level, a network is socially stratified if nodes can be partitioned into social classes corresponding to contiguous intervals of the attribute value, such that (a) those classes are separated in the graph topology and (b) individuals tend to connect to others with similar attribute values.  This latter requirement suggests that to the extent that inter-class connections exist, they should be between nodes in similar classes.   
For instance, later in this study we will consider co-authorship networks, where nodes represent authors and edges represent collaborations.  In such networks, a reasonable attribute of interest is the $h$-index of authors. 
Such a network may be considered stratified if nodes separate into, e.g., classes (groups) of high, medium, and low $h$-indices, and both inter- and intra-class connections tend to be with nodes of similar scores.

\subsubsection{Desired Properties} \label{sec: prop}
The proposed metric should have the following properties: 

\textbf{Property 1:} The characteristic attribute should be a \textbf{scalar characteristic} and the \textbf{actual values} of the characteristic attribute should be taken into account. The metric should more greatly reward connections between nodes with very similar attribute scores, and more greatly penalize connections with very different attribute scores.  This property ensures that inter-class connections between nodes in very different tiers are penalized more greatly than those in adjacent tiers.

\textbf{Property 2:} The scalar characteristic should be
tied to a status feature that allows for grouping into multiple, \textbf{ordered tiers} (social classes).  Categorical properties are not appropriate, unless they can be associated with a numerical status-related property.  Separation with respect to such properties is better measured using other assortativity metrics.

\textbf{Property 3:}  The metric should use \textbf{class membership}; these classes can be known/unknown ahead of time.
Like other assortativity metrics, it should be based on connections within and between classes, where more connections within the same class and fewer connections between different classes tend to increase the metric value.
Note that this property is different from Property 2, as Property 2 takes orders into account, while Property 3 does not (e.g., modularity/DAC are based on class-membership, and so satisfy Property 3, but not Property 2).

\textbf{Property 4:} 
\textbf{Mutual} \textbf{segregation} is needed for stratification of two classes. If, due to differences in class sizes, the fraction of connections inside a class A is much higher than the fraction of connection between A and B,
but the fraction of connections within B is not higher than those from B to A, then A and B are not considered highly segregated. For instance, suppose that A is large and B is small, and that there are $n_1$ intra-class connections inside A, $n_2$ intra-class connections inside B and $n_3$ inter-class connections. If $n_1 >> n_3$, the fraction of intra-class connections is very high. However, this high fraction does not necessarily correspond to a high level of stratification, because while $ n_3>= n_2$, members of B are connected to members A as much as they are connected to each other.  Such a network should not be considered highly stratified.

A real world example of this can be seen in socioeconomic classes where the size of the upper class is much smaller than the middle class and lower class. Whether the upper class is segregated or not has a huge impact on the social stratification of the society.

\subsection{Measuring Stratification Assortativity}
\label{sec:metric}

The \texttt{StA} metric is inspired by \textit{DAC} and \textit{SAC}~\cite{newman2003mixing} which measures the tendency of nodes in the networks to connect with like minded others and is reformulated in a way that captures the desired properties as explained in the previous section. 
 Table~\ref{table:metrics_comparision} shows a summary of desired properties and whether these metrics satisfy them.  As we see, the \textit{DAC} and \textit{SAC} metrics do not satisfy all the desired properties.

\texttt{StA} measures stratification in the given network by first defining the network \textit{ stratification score} as the average stratification over all classes. It considers a class to be maximally stratified if nodes of the class only have connections to nodes inside the same class, and minimally stratified if nodes of the class only have connections to nodes in the most distant classes. Then, the \textit{stratification score} is compared to the expected \textit{stratification score} of a random graph with similar properties in order to determine whether the observed network structure is statistically surprising. This is based on the idea that if the number of inter-class connections is significantly more or significantly less than what we expect in a random network, then something interesting is happening in the network~\cite{newman2006modularity}. Finding the statistical significance of a property by comparing it with a random structure is common in different network science studies~\cite{wilson2013measuring, palowitch2019computing}.
Finally, the final score is normalized to have a value between -1 and 1.

\subsubsection{Input} 
\begin{enumerate}
    \item  An undirected, unweighted graph $G=(V,E)$, where each node $u$ has a characteristic attribute value, denoted by $s(u)$. (Definitions can be easily modified for a weighted graph.)
    \item \textit{Social similarity function} $w(s_1,s_2)$ which measures the distance between characteristic attribute scores $s_1$ and $s_2$. { In this work, we define $w(s(u),s(v))= 1- \frac{|s(u)- s(v)|}{max(S) -min(S)}$,  where $S$ is the score distribution for all nodes in the network $G$ and $u,v \in {V}$.} Depending on the domain, other functions may be more appropriate.  These weights are intended to capture the effect of the actual values of node scores (\textbf{Property 1}). 
    \item If classes are known: $k$ classes $C=\{c_1, c_2,..., c_k\}$, where each $c_i$ represents the nodes with scores within a contiguous portion of the range of possible $score$ values {(\textbf{Property 2})}. Let $c(u)$ represent the class membership of $u$ (i.e., if $s(u)$ is in the range represented by $c_i$, then $c(u) = i$) (\textbf{Property 3}) (In Section~\ref{sec:unknown}, 
    we explain how to find classes if they are not known ahead of time).  
\end{enumerate}

\subsubsection{Computation}

First, we define a weighted version of graph $G$, where the weight of each edge $(u,v) \in E$ is computed by the \textit{social similarity function}.  We then define the \texttt{StA} of the weighted network $G$ as:
$${\scriptstyle  StA(G) = \frac{S_{strat}(G) - E(S_{strat}(G'))}{Max(S_{strat}(G)) - E(S_{strat}(G'))}.}$$  
Here, $S_{strat}(G)$ is the \textit{stratification score} of the weighted network $G$, $E(S_{strat}(G'))$ is the expected stratification score of a random network $G'$ with the same weighted degree distribution as network $G$ and $Max(S_{strat}(G))$ is the maximum stratification score among all networks with the same density and set of nodes with the same scalar attribute and class membership as nodes in $G$. 
These values are computed as:\\

${\scriptstyle S_{strat}(G) = \sum_{c_i \in C} \sum_{(u,v) \in E} \frac{sim(u,v, c_i)}{sim(u,v,c_i) + dissim(u,v,c_i)}.}$\\

${\scriptstyle ES_{strat}(G) = \sum_{c_i \in C} \sum_{(u,v) \in E} \frac{E(sim(u,v,c_i))}{E(sim(u,v,c_i)) + E(dissim(u,v,c_i))}.}\\$

 ${\scriptstyle Max(S_{strat}(G))=k.}\\$

\noindent where $sim(u,v)$ and  $dissim(u,v)$ are the similarity and dissimilarity  of nodes u and v respectively, and $E(sim(u,v))$ and $E(dissim(u,v)$ are the expected similarity and dissimilarity of nodes $u$ and $v$ in networks with the same weighted degree distribution as $G$.  Details on how these values are defined are provided in the Appendix.

\begin{figure*}
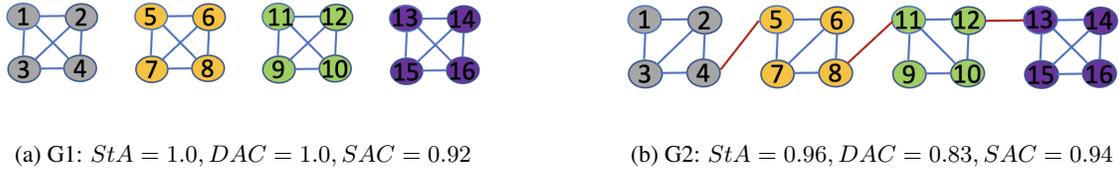

  \centering
  \vspace{-0.05cm}
  \begin{subfigure}[b]{0.4\textwidth}  \includegraphics[width=\textwidth]{figs/1.pdf}
  \caption{G1: $StA = 1.0, DAC = 1.0, SAC= 0.92$}
  \label{fig:g1}
  \end{subfigure}
  ~~~~~~~~~~~~~~~~~~
    \begin{subfigure}[b]{0.4\textwidth}  \includegraphics[width=\textwidth]{figs/2.pdf}  \caption{G2: $StA = 0.96, DAC = 0.83, SAC= 0.94$}
    \label{fig:g2}
  \end{subfigure}
  \caption{Stratification comparison between $G_1$ and $G_2$. {Node scores are written inside the nodes. There is no inter-class connections in  $G_1$, thus $G_1$ is more stratified than $G_2$ but SAC of $G_1$ is lower than $G_2$.}} 
  \label{fig:examples_1}
\end{figure*}

\begin{figure*}
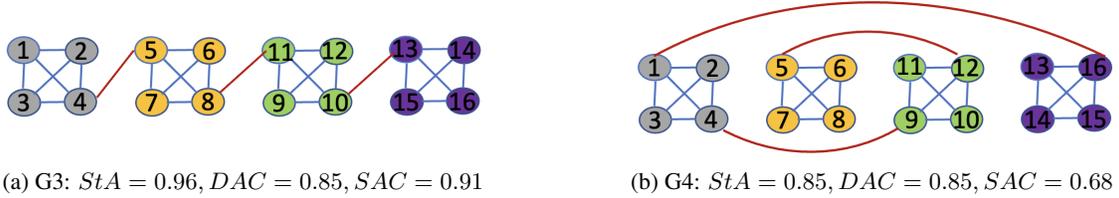

  \centering
  \vspace{-0.1cm}
    \begin{subfigure}[b]{0.4\textwidth}  \includegraphics[width=\textwidth]{figs/3.pdf}
  \caption{G3:  $StA = 0.96, DAC = 0.85, SAC= 0.91$}
   \label{fig:g3}
  \end{subfigure}
  ~~~~~~~~~~~~~~~~~~
      \begin{subfigure}[b]{0.4\textwidth}  \includegraphics[width=\textwidth]{figs/4.pdf}
  \caption{G4: $StA = 0.85, DAC = 0.85, SAC= 0.68$}
   \label{fig:g4}
  \end{subfigure}
  \caption{Stratification comparison between $G_3$ and $G_4$. {Node scores are written inside the nodes. Distance between low score nodes and high score nodes in $G_3$ is higher than $G_4$, thus $G_3$ is more stratified than $G_2$ but DAC of $G_3$ is equal to $G_4$.}}
  \label{fig:examples_2}
\end{figure*}

\begin{figure*}
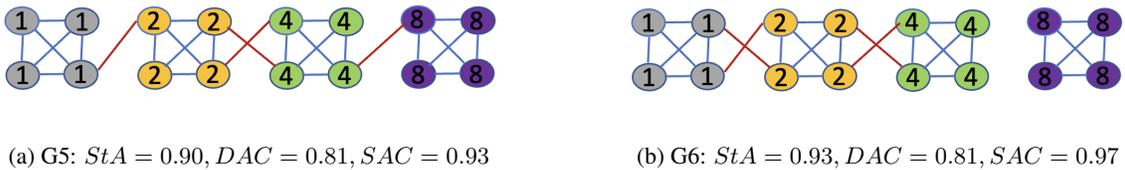

  \centering
  \vspace{-0.05cm}
        \begin{subfigure}[b]{0.4\textwidth}  \includegraphics[width=\textwidth]{figs/5.pdf}
  \caption{G5: $StA = 0.90, DAC = 0.81, SAC= 0.93$}
   \label{fig:g5}
  \end{subfigure}
  ~~~~~~~~~~~~~~~~~~
        \begin{subfigure}[b]{0.4\textwidth}  \includegraphics[width=\textwidth]{figs/6.pdf}
  \caption{G6: $StA = 0.93, DAC = 0.81, SAC= 0.97$}
   \label{fig:g6}
  \end{subfigure}
  \caption{Stratification comparison between $G_5$ and $G_6$. {Node scores are written inside the nodes. High score nodes are segregated from the rest of the network in $G_6$, thus $G_6$ is more stratified than $G_5$ but DAC of $G_6$ is equal to $G_5$. }}
    \label{fig:examples_3}
\end{figure*}

\subsection{Properties of StA Metric} \label{sec: properties}

\texttt{StA} is a real number between {-1} and 1, with 1 representing a network that is fully stratified and {-1} corresponding to a dis-stratified network and 0 corresponding to a network with balanced inter- vs intra- class connections.

In a fully stratified network with $k$ classes, all connections are between nodes of one class (intra-class connections) and there are no connections between nodes of different classes (inter-class connections). In a fully unstratified network, all connections are inter-class connections. \texttt{StA} is greater than 0 if there are  more normalized weighted connections between similar nodes from one class than dissimilar nodes from different classes and \texttt{StA} is lower than 0 if there are  more normalized weighted connections between dissimilar nodes from different classes than similar nodes from one class.

Next, we explain why \texttt{StA} properly measures stratification, while other assortativity metrics (\textit{DAC} and \textit{SAC}) cannot measure it.  

For purposes of this discussion, suppose that we have a set of networks with 16 nodes where each node has a merit score (here, the node ID is equal to its score).
Suppose nodes are divided into 4 classes: low score nodes (nodes 1 to 4), medium-low score nodes (nodes 5 to 8), medium-high score nodes (nodes 9 to 12) and high score nodes (nodes 13 to 16). 

Figure~\ref{fig:examples_1} shows two networks with these nodes and different topologies.
In graph $G_1$, there are no inter-class connections. As $S_{strat}(G) = Max(S_{strat}(G))$ when there are no inter-class connections, the \texttt{StA} of $G_1$ is equal to 1 as expected. 
In graph $G_2$, we see three inter-class connection and the \texttt{StA} of $G_2$ is equal to $0.96$ which is lower than $G_1$ as expected. However, if we compare the \textit{SAC} o of graphs $G_1$ and $G_2$, $SAC(G_1)<SAC(G_2)$. This happens because \textit{SAC} does not consider specific classes (property 3) and ordered tiers (property 2). 
In graph $G_2$, there is a path from nodes in any class to nodes from other classes, whereas in graph $G_1$, there is no path from nodes of one class to the other classes. Thus, network $G_1$ is more stratified, as described by \texttt{StA}.

Next, consider graphs $G_3$ and $G_4$ as showed in Figure~\ref{fig:examples_2}. The node properties of these networks are similar to $G_1$ and $G_2$. If we compare graphs $G_3$ to $G_4$, we see that the number of intra-class connections and inter-class connections are the same. However, inter-class connections in $G_3$ are edges between similar nodes whereas intra-class connections in $G_4$ are  edges between dis-similar nodes. In other words, the distance between low score nodes and high score nodes is higher in $G_3$ compared to $G_4$. Thus, we expect \texttt{StA} of $G_3$ to be higher than  $G_4$. We see that $StA(G_3)>StA(G_4)$ as expected. However, if we compare the \textit{DAC} of graphs $G_3$ and $G_4$, $DAC(G_3)=DAC(G_4)$. This happens because \textit{DAC} does not consider scalar values (property 1) and ordered tiers (property 2).

Finally, consider graphs $G_5$ and $G_6$ as shown in Figure~\ref{fig:examples_3}. The node properties of these networks are similar to $G_1$ to $G_4$ except that the node number is not equal to its score, Nodes 1 to 4 have score equal to 1, nodes 5 to 8 have score equal to 2, nodes 9-12 have score equal to 4 and nodes 13-16 have score equal to 8. If we compare graphs $G_5$ to $G_6$, we see that although the number of inter vs intra- class connections in $G_5$ and $G_6$ are the same, high-score class nodes are segregated from the rest of the network in $G_6$ (there is no connection between high score nodes and the rest of the network), while in $G_5$, high score nodes have access to each other. Thus, we expect the stratification of $G_6$ to be higher than $G_5$.
 We see that $StA(G_5)<StA(G_6)$ as expected. However, if we compare the \textit{DAC} of graphs $G_5$ and $G_6$, $DAC(G_5)=DAC(G_6)$. This happens because \textit{DAC} does not take mutual segregation into account
 and is based on the overall inter- vs intra- class connections (property 4). Note that we set scores in these networks in a way that inter- and intra- class connections in both networks have the same weight because we wanted to consider class impact and didn't want other factors to interfere in the comparison process.

 \begin{table*}
\footnotesize
\centering \caption{Dataset Statistics. }
\label{table:stat}
\begin{tabular}{l|l|l|l } 
\textbf{Networks}
 &\textbf{ \# authors} 
 &\textbf{\# connections}  
 & \textbf{\# connections}
 \\
 \textbf{}
 &\textbf{} 
 &\textbf{(distinct)}  
 & \textbf{(all)}
 \\
\hline
\textbf{Computational
}&11k&22k&23k\\
\textbf{Linguistics
}& &&
\\
\hline
\textbf{NLP
}&137k &398k& 476k
\\

\hline

\textbf{Computational
}& 174k&1.6M&1.7M\\
\textbf{Biology}& &&
\\
\hline

\textbf{Biomedical
}&410k &1.3M&1.4M\\
\textbf{Engineering}&&
\\
\end{tabular}
\vspace{0.5cm}
\end{table*}

 \subsubsection{{Identifying Class Boundaries}\label{sec:unknown}}

Depending on the application under study, the social class hierarchy may or may not be available to the researcher.  In many real applications, these classes are known ahead of time. For example, it is conventional in economic analysis of Western societies to define a lower, middle, and upper class~\cite{cannadine1998beyond}.  However, in other networks, such as interactions in meetings or conferences~\cite{gupte2011finding}, it is possible that neither the class boundaries nor even the number of classes is known ahead of time.  

If classes are not given, one must first detect their boundaries. Here, we are interested in social classes that result in the greatest \texttt{StA}.  For instance, in a society that is actually divided into lower, middle and upper class, there might be many ways of partitioning the network into non-stratified classes, but the existence of a stratified partition is of great importance.

 To find these classes, we use the \texttt{MaxStrat} algorithm. \texttt{MaxStrat} is a dynamic programming-based heuristic that is based on maximizing the non-normalized version of the $StA$  ($StA'= {S_{strat}(G) - E(S_{strat}(G'))}$ ).  We explain the full details of \texttt{MaxStrat} in the Appendix.

\section{Social Stratification in Collaboration Networks}

Here, we study the social stratification over time in four co-authorship networks from the fields of Computational Linguistics, Natural Language Processing, Computational Biology, and Biomedical Engineering.  These fields have been active for approximately 50 years and were chosen because they are young enough that full data is available (in contrast to very old fields like general physics, where much of the early co-authorship information may not be accessible) but old enough for meaningful evolution to have occurred.  This 50-year period covers the bulk of the active period for these fields:  before 1966, there were very few papers in these fields, and the dataset is incomplete for papers published after 2015.

\subsection{Datasets}
\noindent \textbf{Graph Data:} To generate these networks, we extracted papers published in these four fields from the Microsoft Academic Graph (MAG)~\cite{sinha2015overview} over the time period from 1966 to 2015.\footnote{We used MAG because it outperforms Google Scholar in terms of structure, functionality, and richness of data~\cite{hug2017citation}.} Each node represents a researcher (author on a paper), and a link between two nodes indicates that they have published a paper together at least once.  We treat these networks as undirected and unweighted. \footnote{{Code and dataset can be accessed at: https://github.com/SaraJalali/StratificationAssortativity}}

\begin{figure}

\centering
\includegraphics[scale = 0.4]
{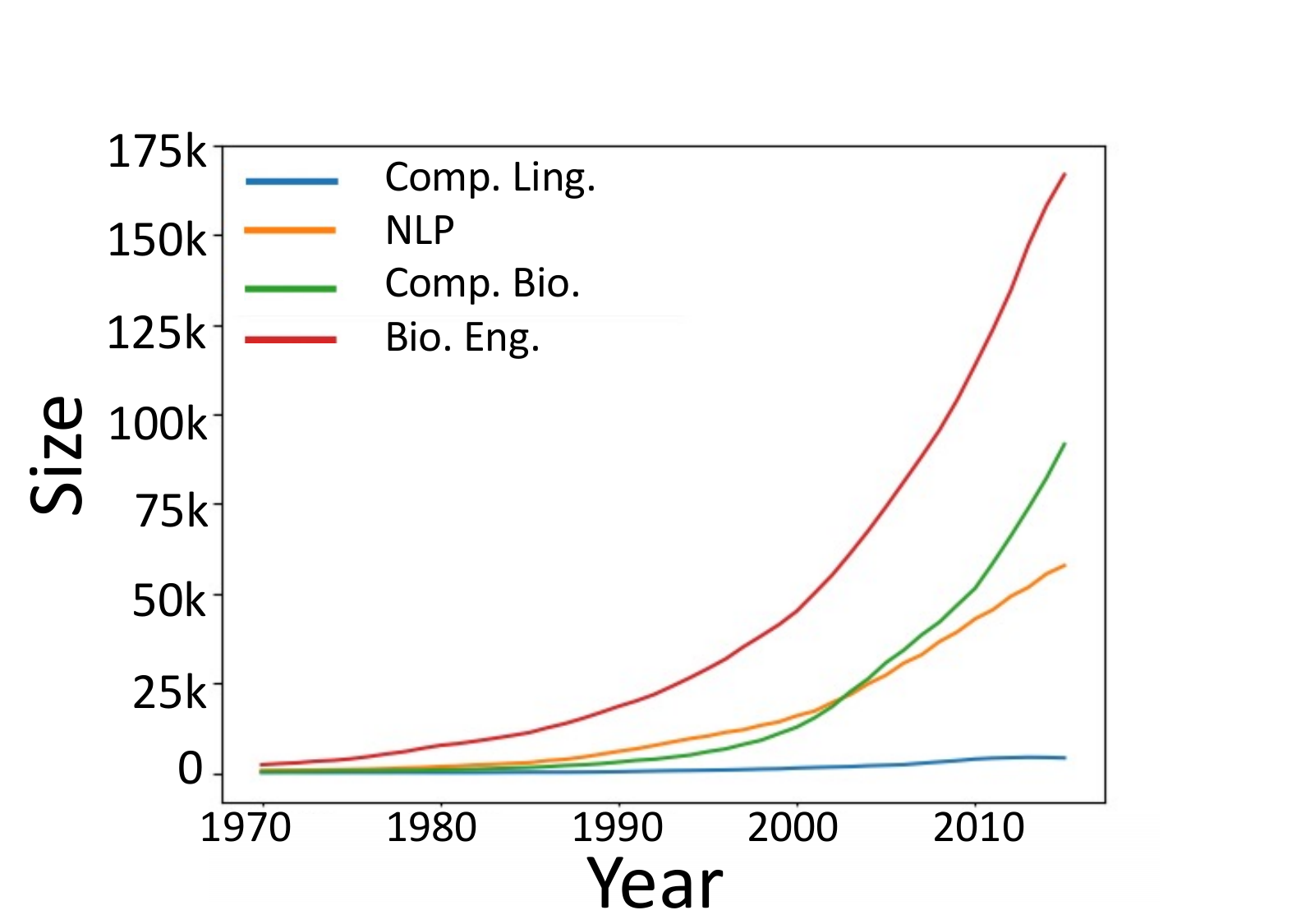}
\caption{Size of Networks (\# of Nodes).}
\label{fig:size1}
\end{figure}
\noindent \textbf{Snapshots:} For each field, we generated 5-year rolling network snapshots spanning 1966-2015 (each snapshot begins one year later than the previous). Each field contains at least 11k authors. Table~\ref{table:stat} shows dataset statistics and the size of the snapshots is provided in Figure~\ref{fig:size1}.

\noindent \textbf{Node Scores:}  Node scores are defined by author $h$-indices, computed using citation data within the field up to that year.

\subsection{Results} \label{sec: analysis}
In this section, we study stratification in these different fields and explore how stratification changes over time. To perform this analysis, we use the \texttt{StA} metric, which we earlier demonstrated on toy examples. 
\begin{figure*}
  \centering
   \centering
   \begin{subfigure}[b]{0.6\textwidth}  \includegraphics[width=\textwidth]{figs/leg1.pdf}
  \end{subfigure}
  
  \vspace{-0.05cm}
  \begin{subfigure}[b]{0.23\textwidth}  \includegraphics[width=\textwidth]{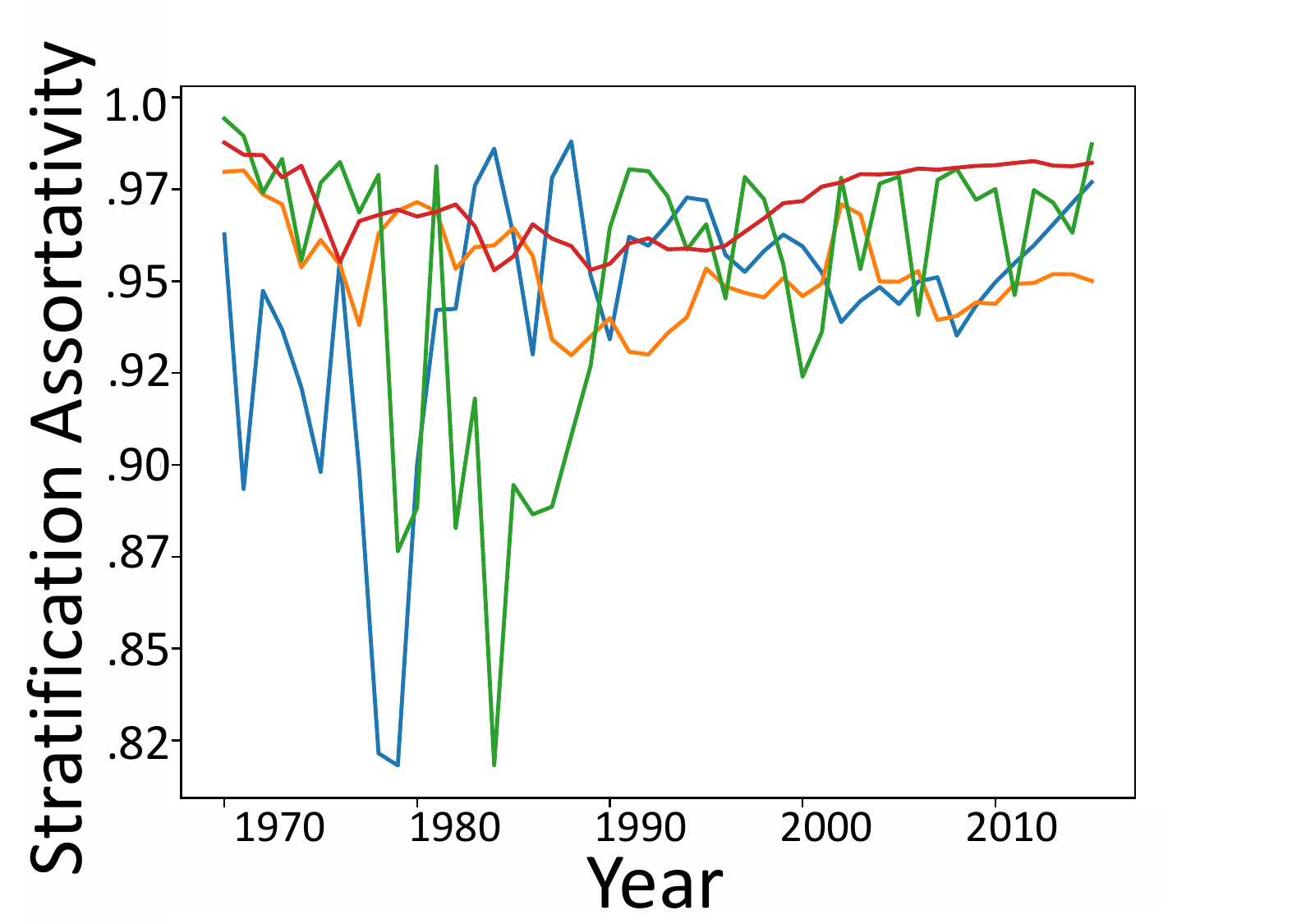}
  \caption{2 classes}
  \end{subfigure}
  \begin{subfigure}[b]{0.23\textwidth}  \includegraphics[width=\textwidth]{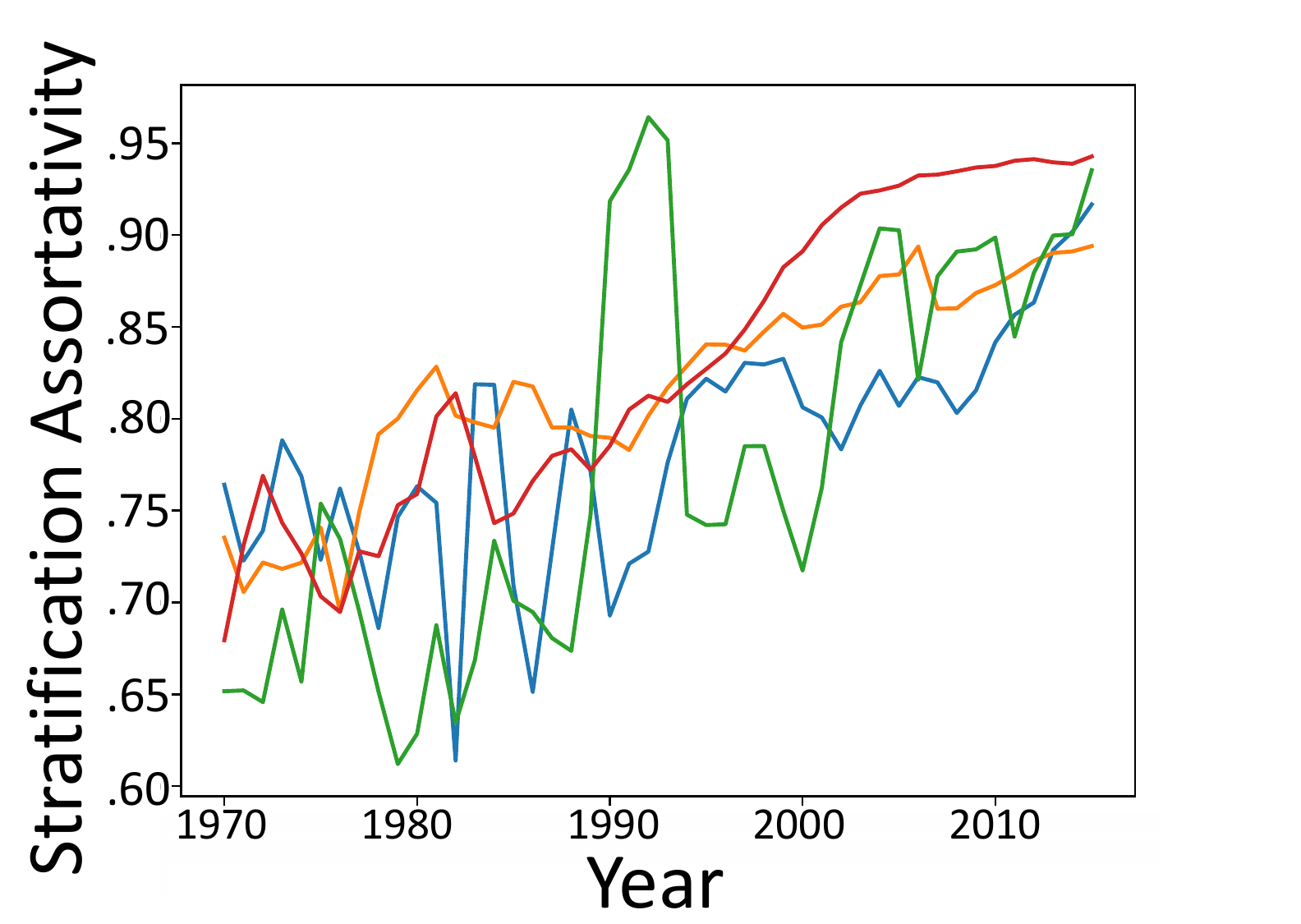}
  \caption{ 3 classes}
  \end{subfigure}
  \begin{subfigure}[b]{0.23\textwidth}  \includegraphics[width=\textwidth]{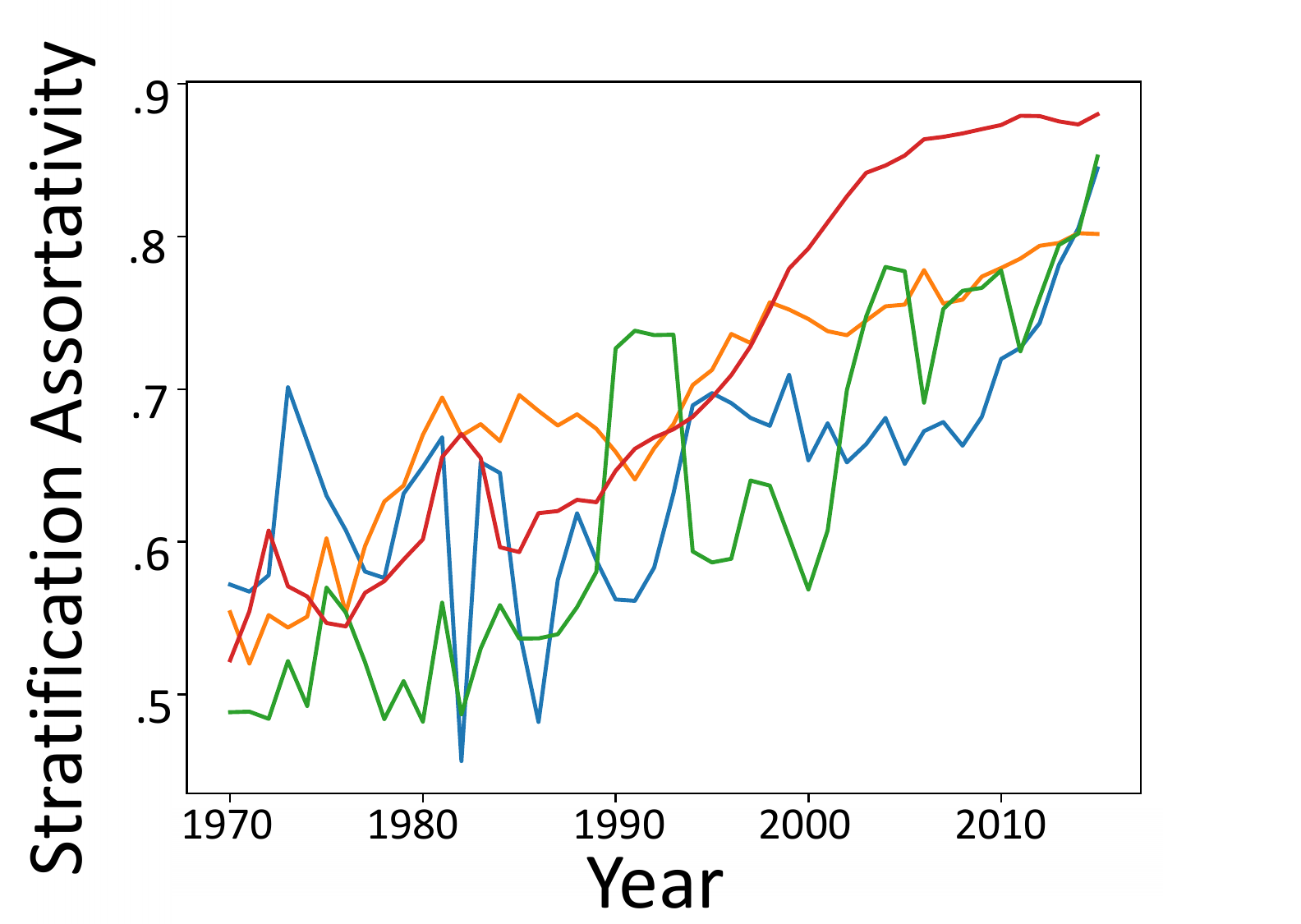}
  \caption{ 4 classes}
   \label{fig:SA_class_numbers_4}
  \end{subfigure}
  \begin{subfigure}[b]{0.23\textwidth}  \includegraphics[width=\textwidth]{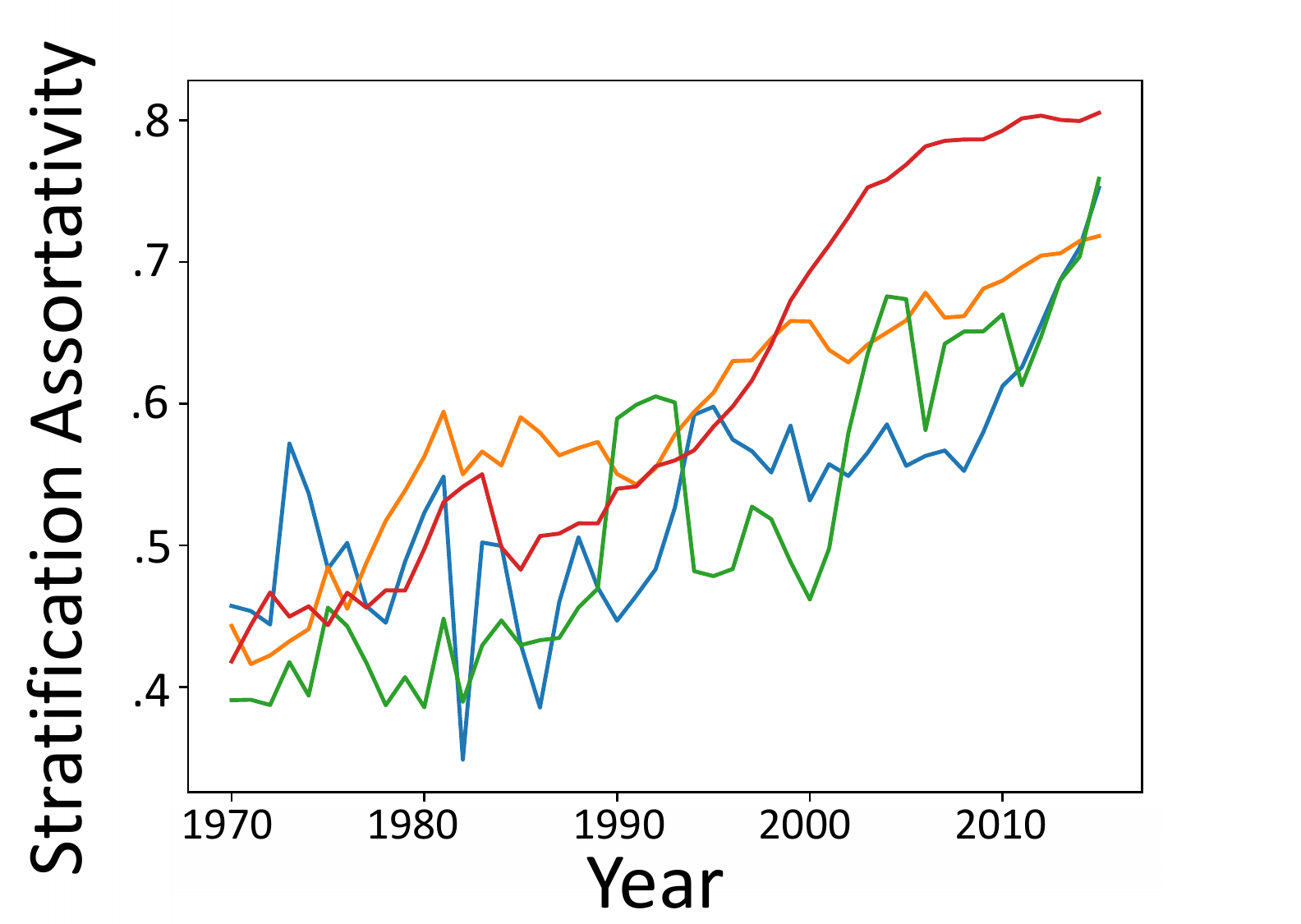}
  \caption{ 5 classes}
  \end{subfigure}
  \caption{Maximum \texttt{StA} for varying number of classes.}
  \label{fig:SA_class_numbers}
  \vspace{0.5cm}
\end{figure*}

\begin{figure*}
  \centering
  \begin{subfigure}[b]{0.23\textwidth}  \includegraphics[width=\textwidth]{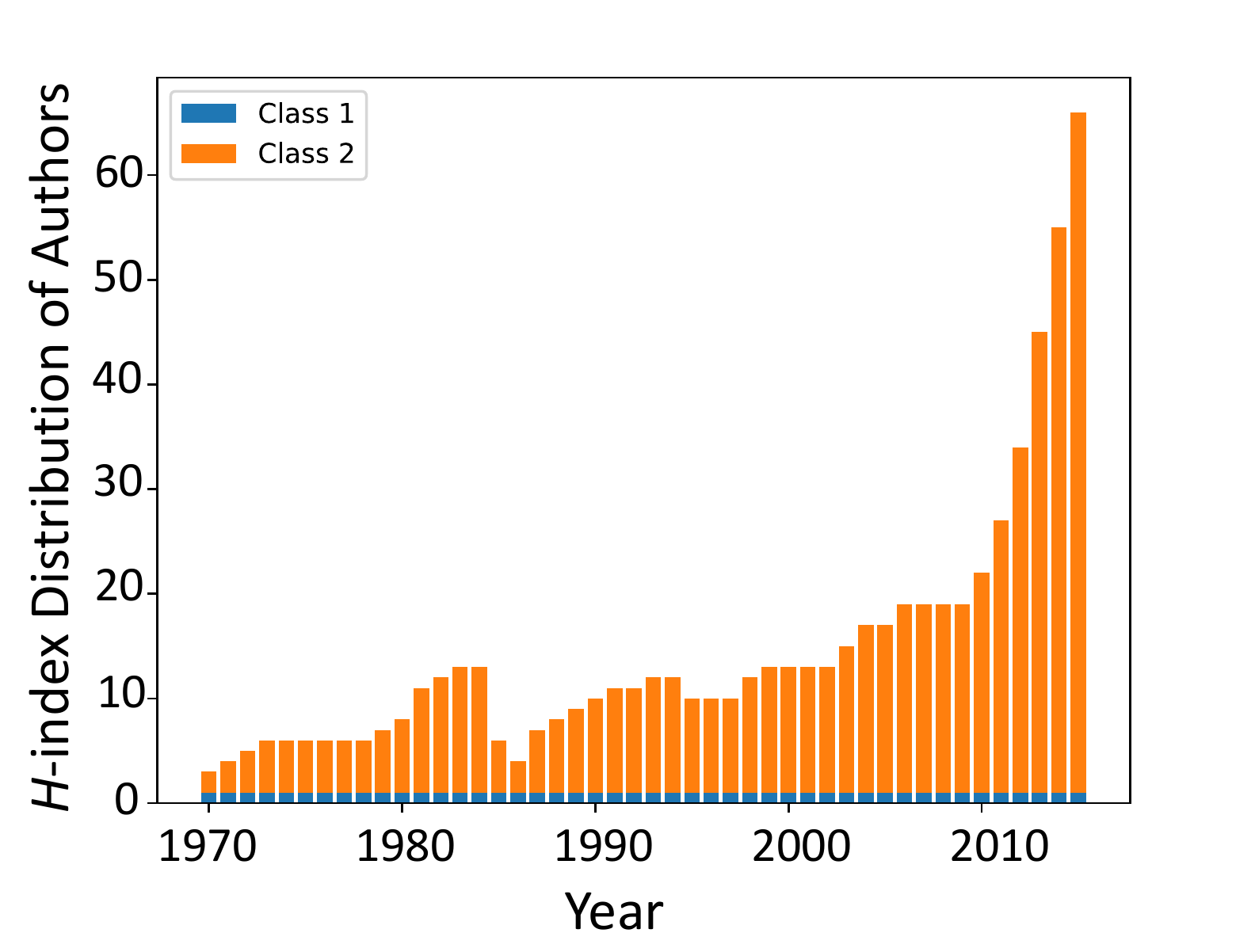}
  \caption{ 2 class boundaries .}
  \end{subfigure}
  \begin{subfigure}[b]{0.23\textwidth}  \includegraphics[width=\textwidth]{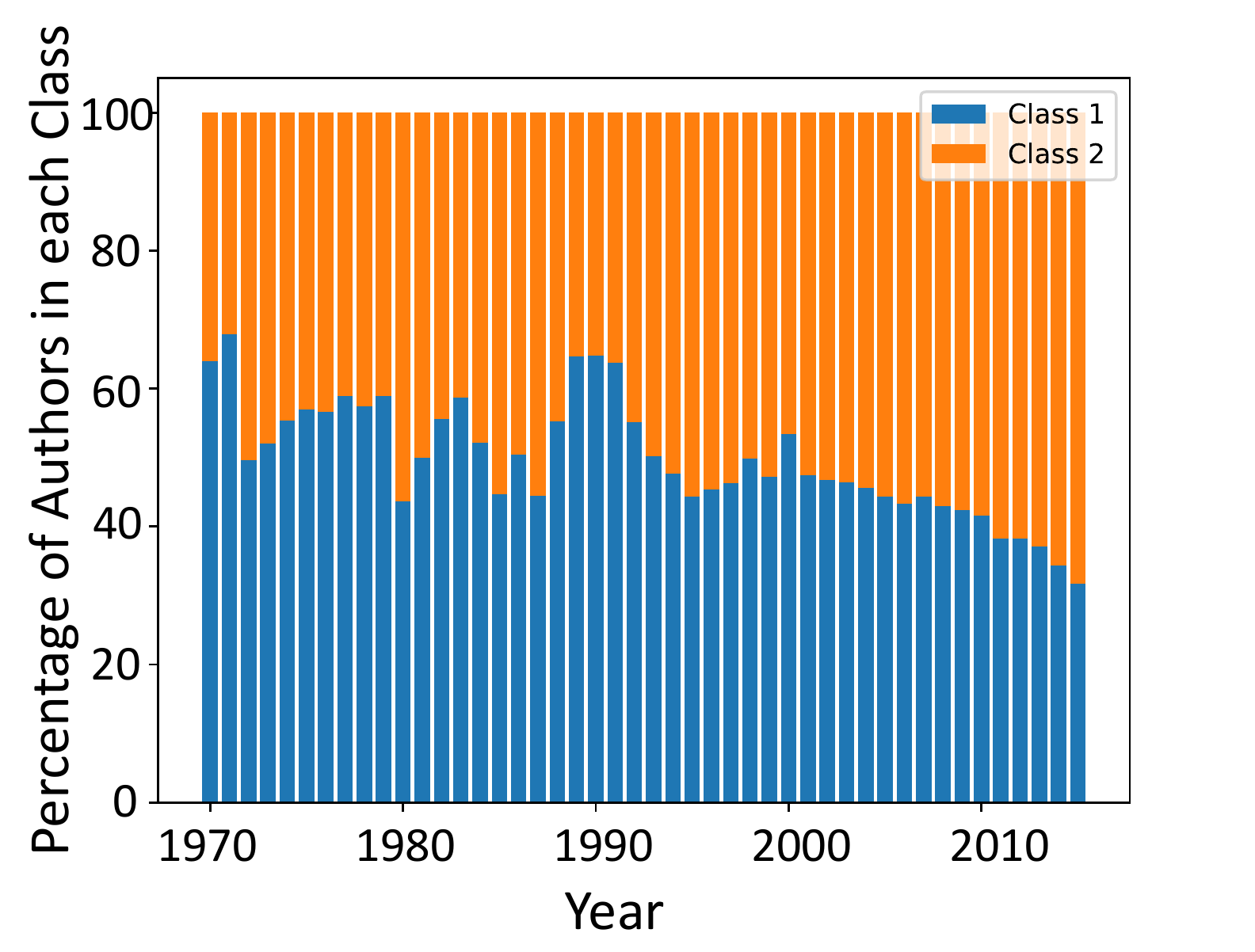} 
  \caption{Fraction of nodes.}
  \end{subfigure}
    \begin{subfigure}[b]{0.23\textwidth}  \includegraphics[width=\textwidth]{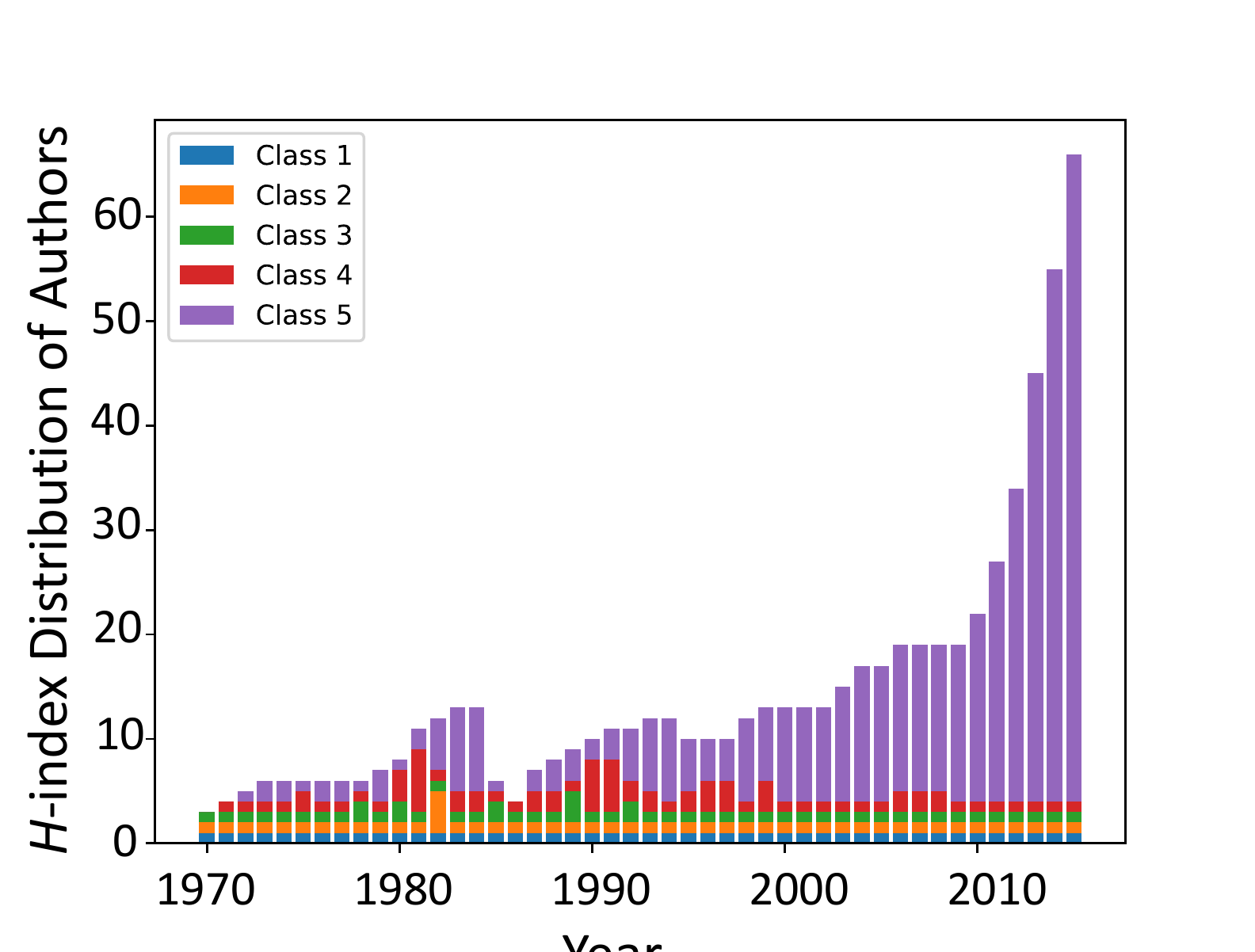}
  \caption{ 5 class boundaries.}
  \end{subfigure}
  \begin{subfigure}[b]{0.23\textwidth}  \includegraphics[width=\textwidth]{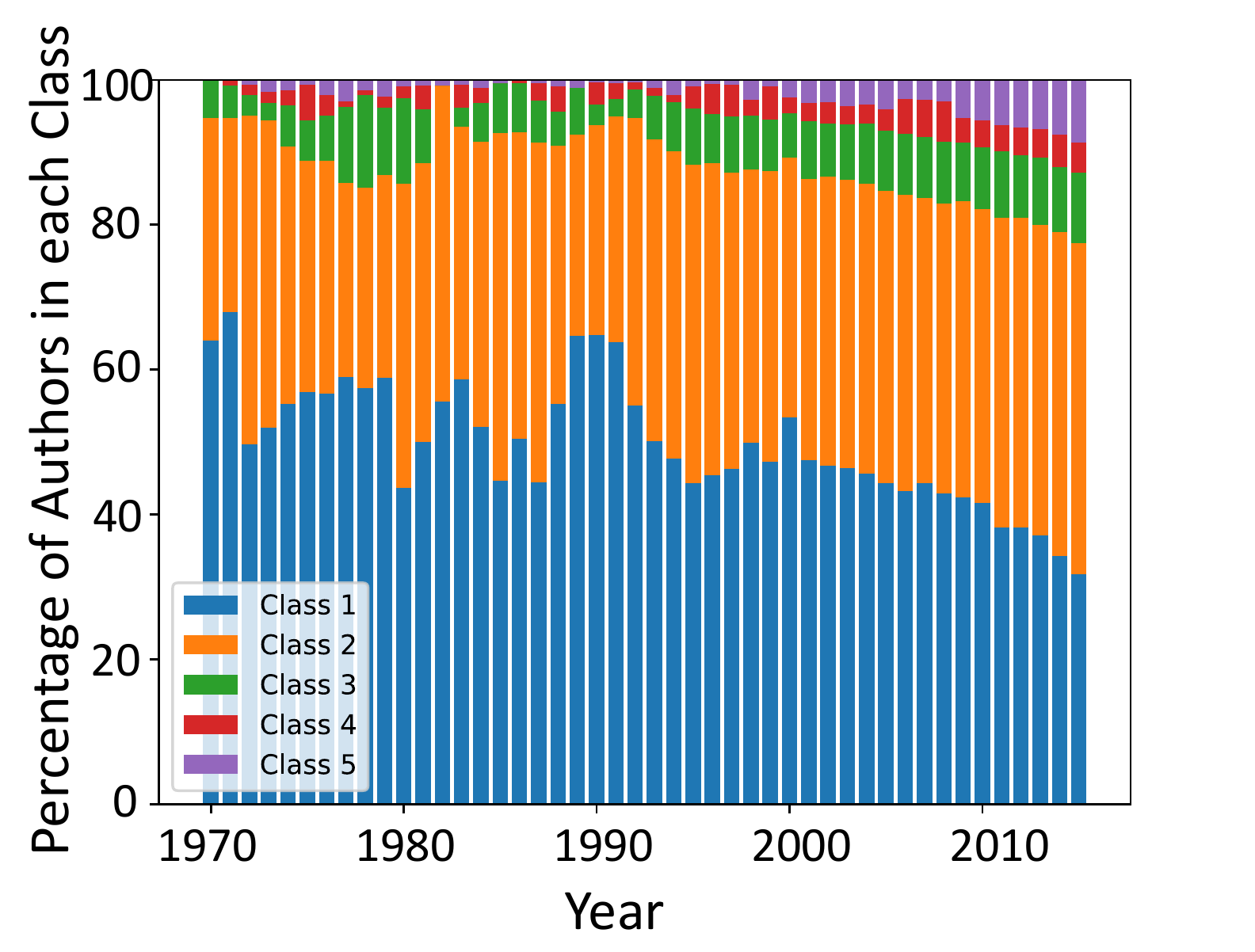}
  \caption{Fraction of nodes.}
  \end{subfigure}
  \caption{Optimal class boundaries and sizes for 2 and 5 classes on Computational Linguistics networks.}\label{fig:NLP_boundaries2}
  \vspace{0.5cm}
\end{figure*}

\subsubsection{StA of Different Fields}
Here, we examine the \texttt{StA} of different fields.  Our results show that as the network ages, people display a higher tendency to collaborate with members of their same class or nearby classes.  In particular, as the field evolves, high score nodes have a very strong tendency to collaborate with other high score nodes.  

First, we use the \texttt{MaxStrat} algorithm (defined in the Appendix) to identify class boundaries for 2 to 5 classes. Figure~\ref{fig:SA_class_numbers}  shows the \texttt{StA} of the 5-year snapshots from all four fields.  The results demonstrate that all networks have a fairly high level of stratification (above 0.45); and with the exception of the case when we divide into only two classes, \texttt{StA} \textit{increases over time.}

The highest values of \texttt{StA} are obtained with only two classes.  To understand why this is so, we examined the $h$-indices that fall into each class.  We observed that in almost all snapshots, over half of the nodes have an $h$-index of 0, and only a tiny minority have an $h$-index greater than two (0 to 13\% for Comp. Ling, 1 to 13\% for NLP, 0 to 7 \% for Comp. Bio and 1 to 8\% for Bio. Eng); however, because the $h$-indices are integer valued, finer granularity is impossible.  The most natural class division, thus, is to put every node with a low $h$-index (0 or 1) into one class, and every other node into a second class.  Further stratification within that upper class is outweighted by the stratification between the upper class and lower class.

\begin{figure*}
\vspace{-0.5cm}
  \centering
  \begin{subfigure}[b]{0.19\textwidth}  \includegraphics[width=\textwidth]{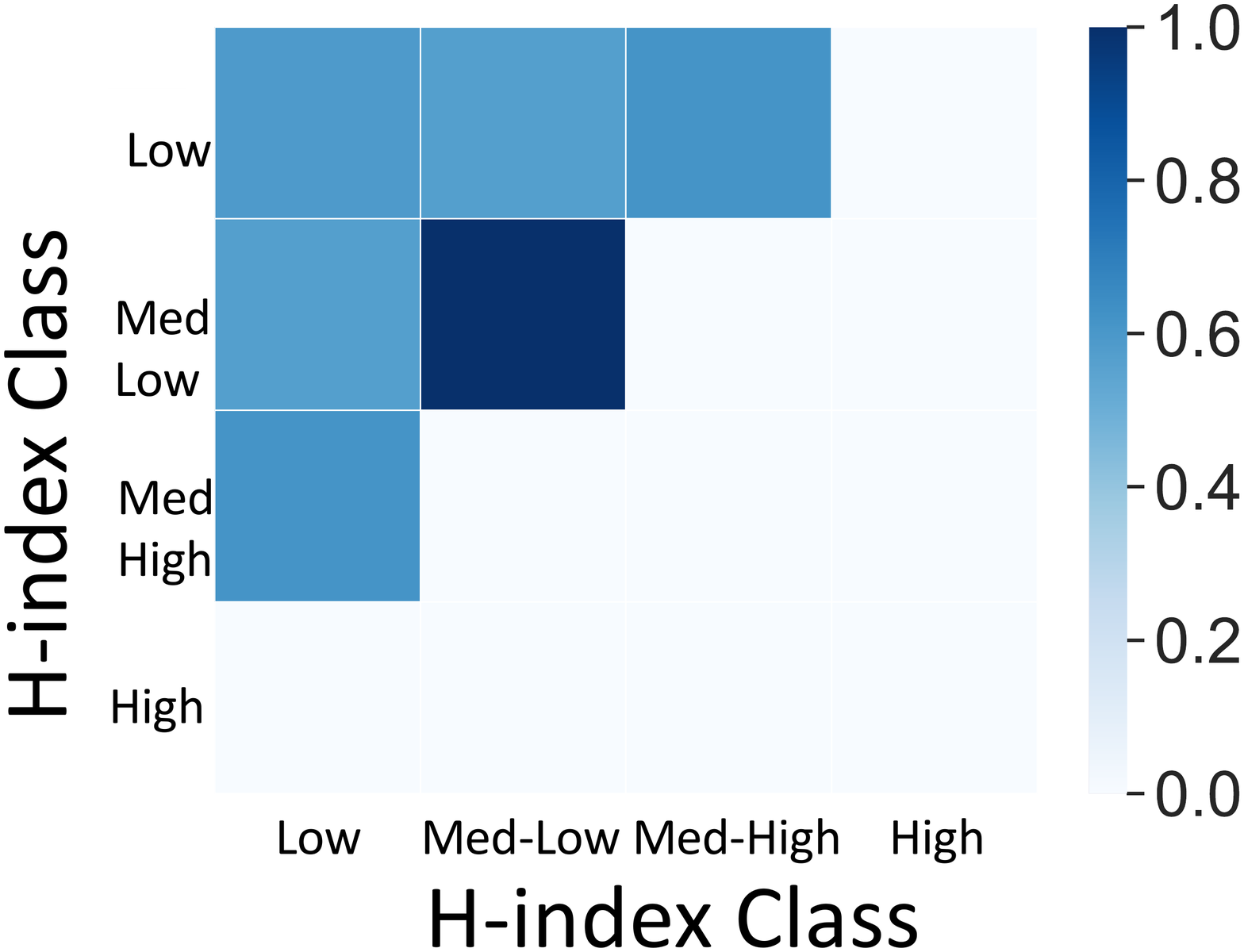}
  \caption{1966-1975}
  \end{subfigure}
  \begin{subfigure}[b]{0.19\textwidth}  \includegraphics[width=\textwidth]{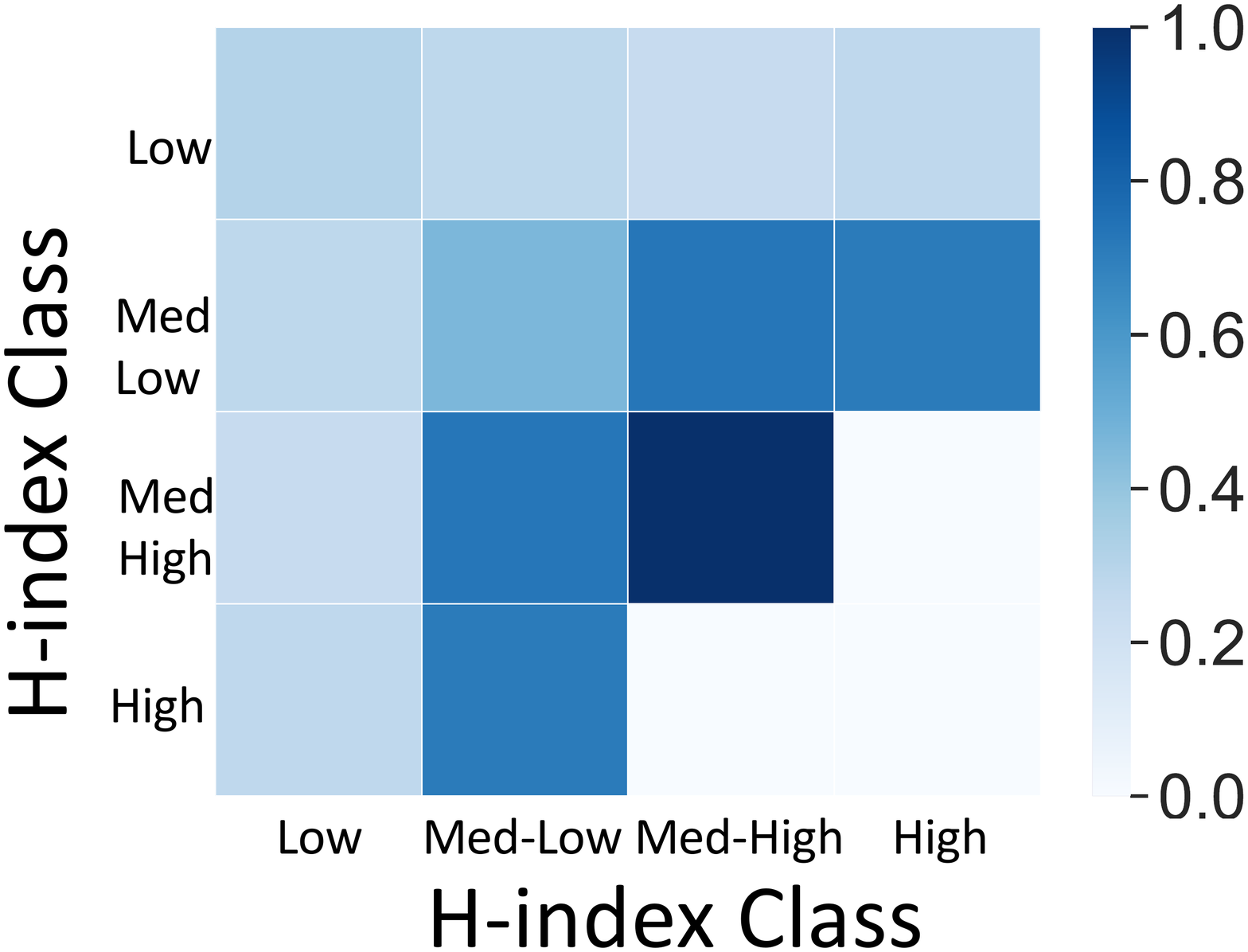}
  \caption{1976-1985}
  \end{subfigure}
    \begin{subfigure}[b]{0.19\textwidth}  \includegraphics[width=\textwidth]{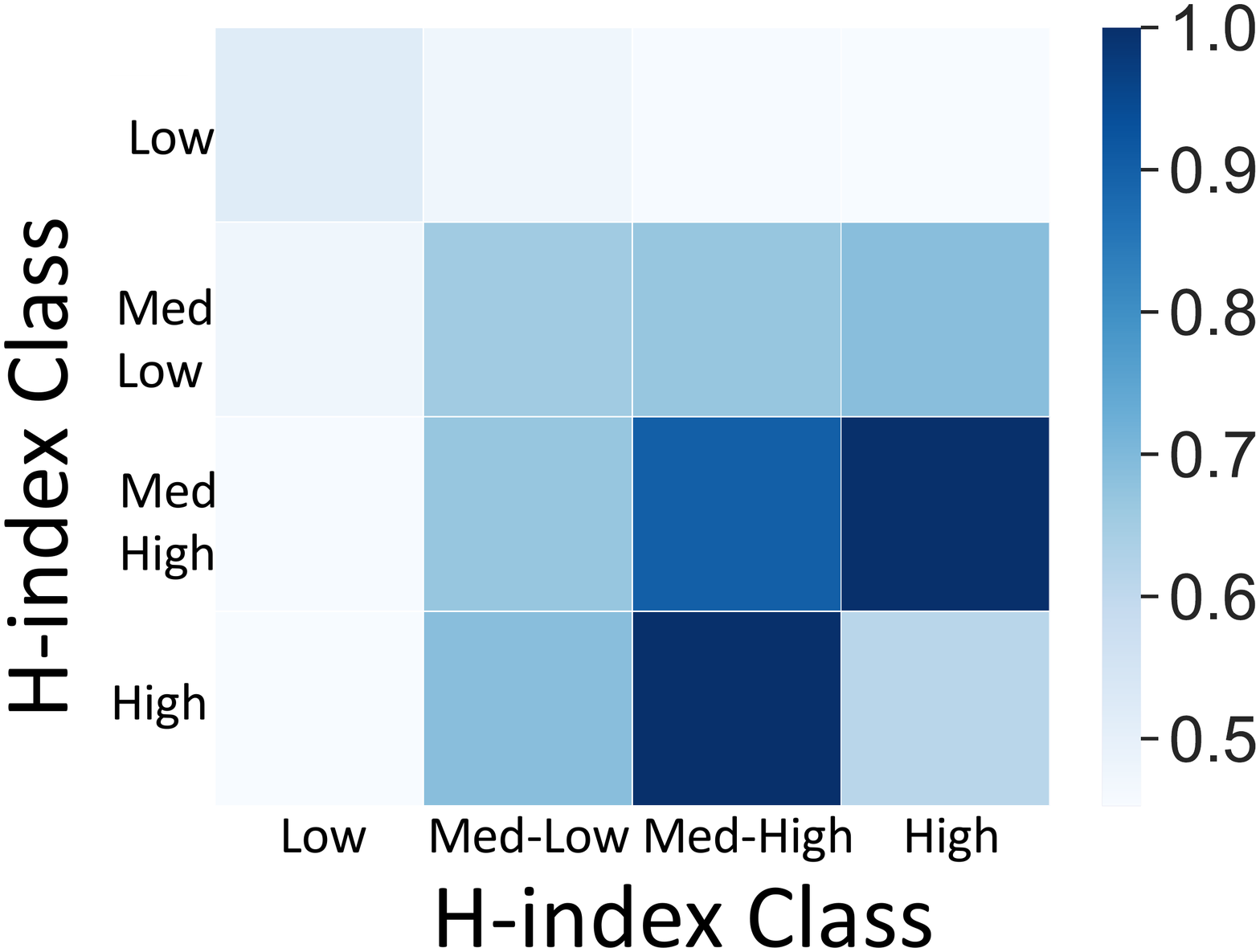}
  \caption{1986-1995}
  \end{subfigure}
    \begin{subfigure}[b]{0.19\textwidth}  \includegraphics[width=\textwidth]{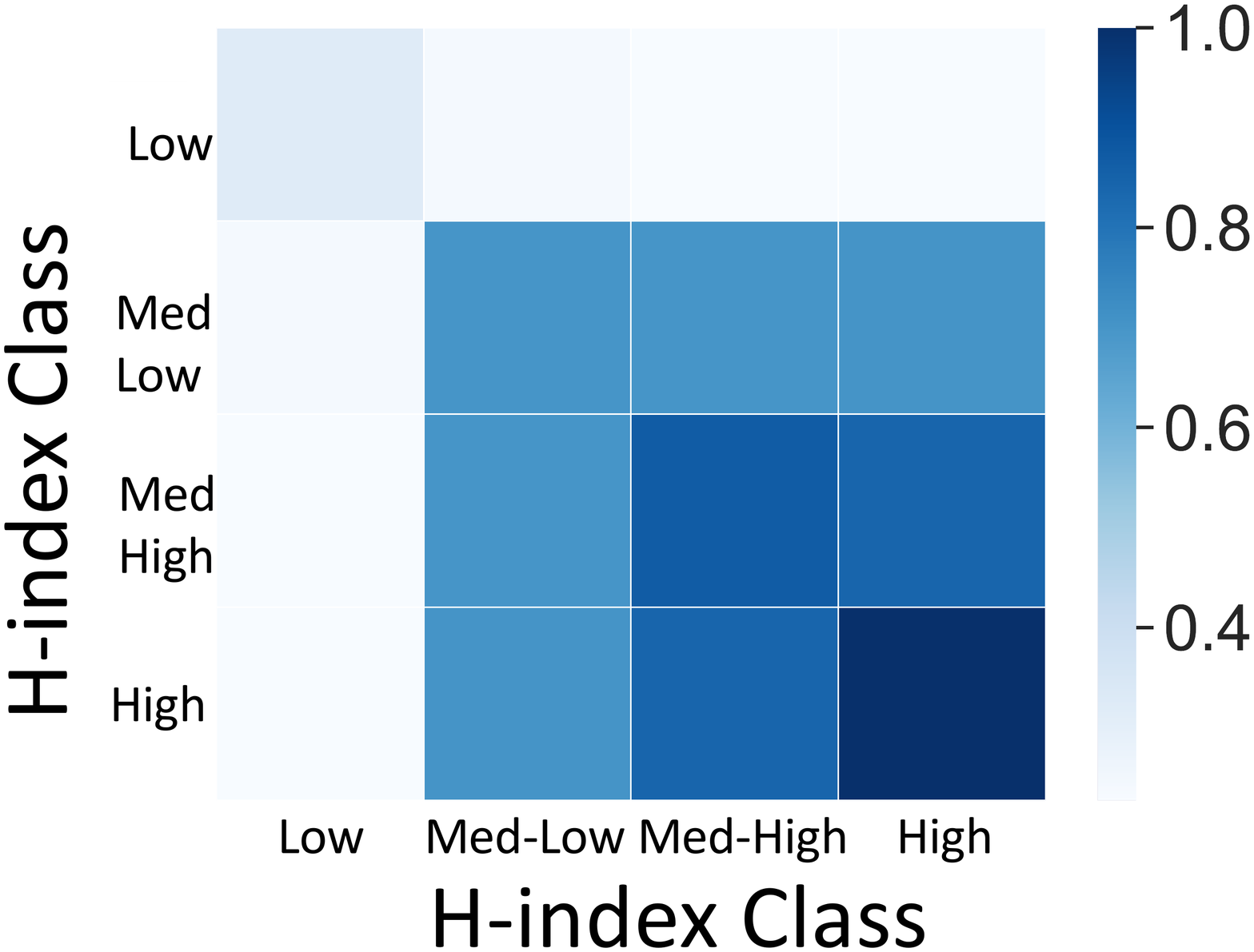}
  \caption{1996-2005}
  \end{subfigure}
    \begin{subfigure}[b]{0.19\textwidth}  \includegraphics[width=\textwidth]{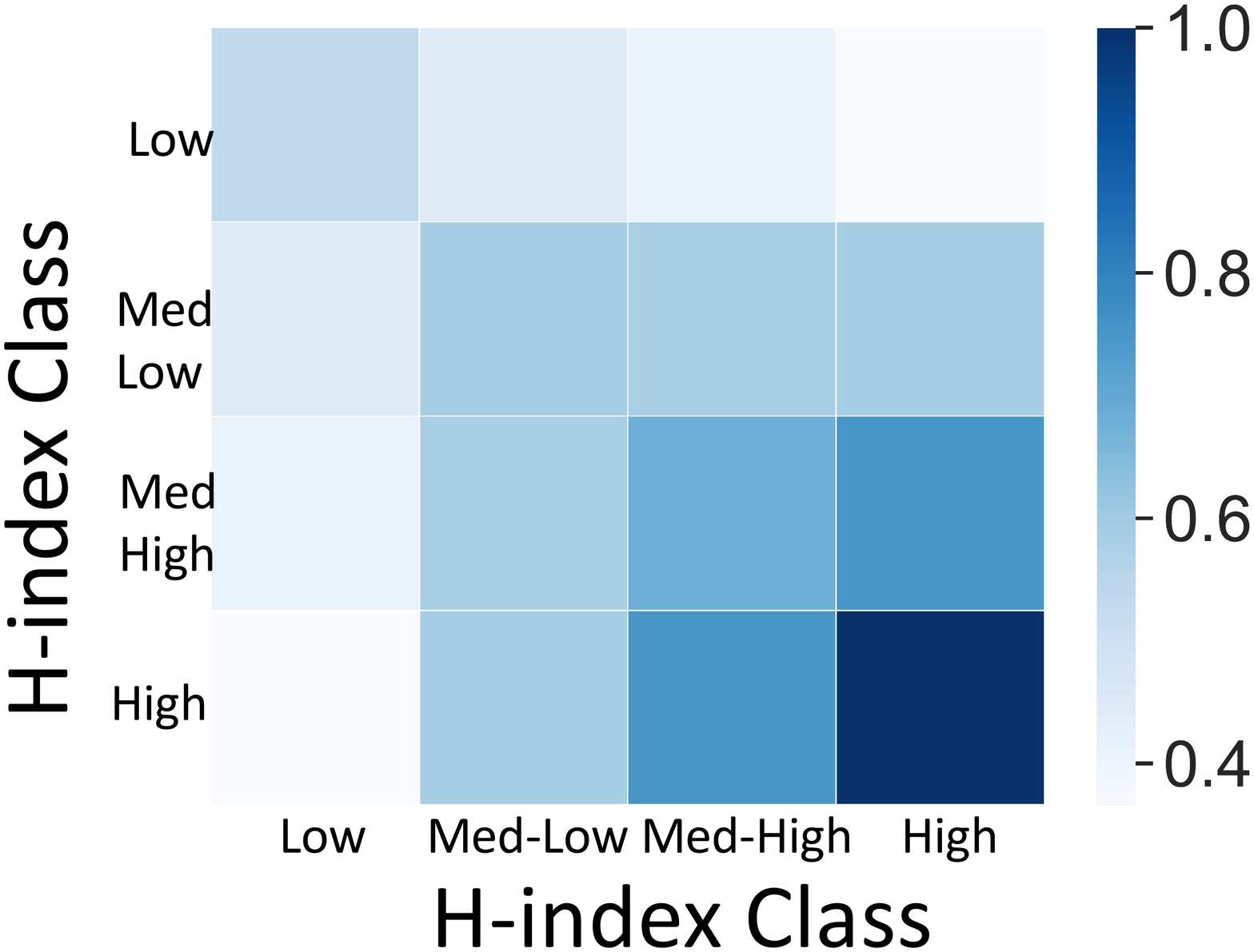}
  \caption{2006-2015}
  \end{subfigure}
  \caption{Collaborations from different h-index classes for NLP, normalized by degree (see description in text).}
  \label{fig:cor_nlp}
  \vspace{0.5cm}
\end{figure*}

\begin{figure*}
  \centering
   \begin{subfigure}[b]{0.6\textwidth}  \includegraphics[width=\textwidth]{figs/leg1.pdf}
  \end{subfigure}
  
  \vspace{-0.05cm}
  \begin{subfigure}[b]{0.3\textwidth}  \includegraphics[width=\textwidth]{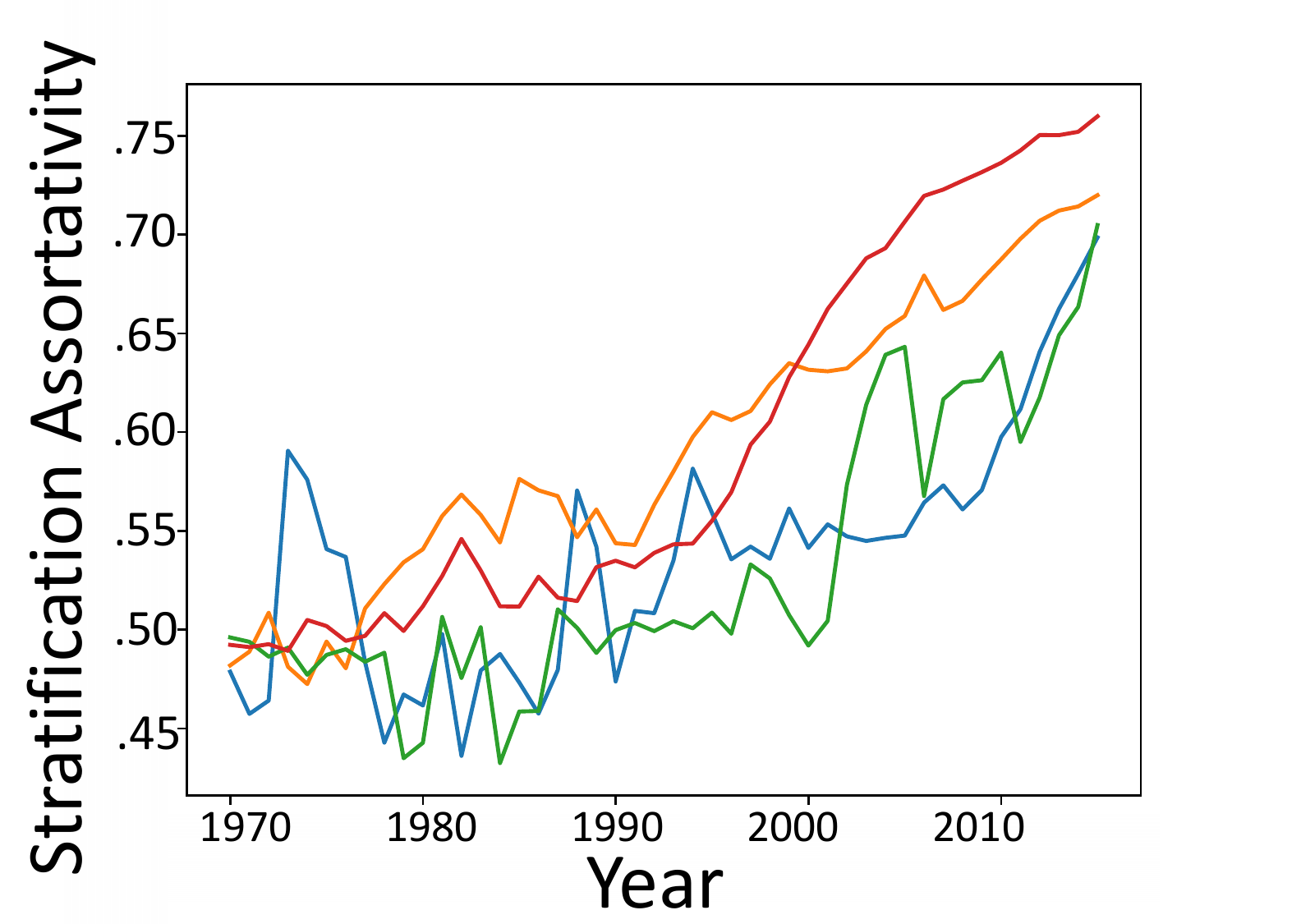}
  \caption{StA}
  \label{fig:st_mod}
  \end{subfigure}
    \begin{subfigure}[b]{0.3\textwidth}  \includegraphics[width=\textwidth]{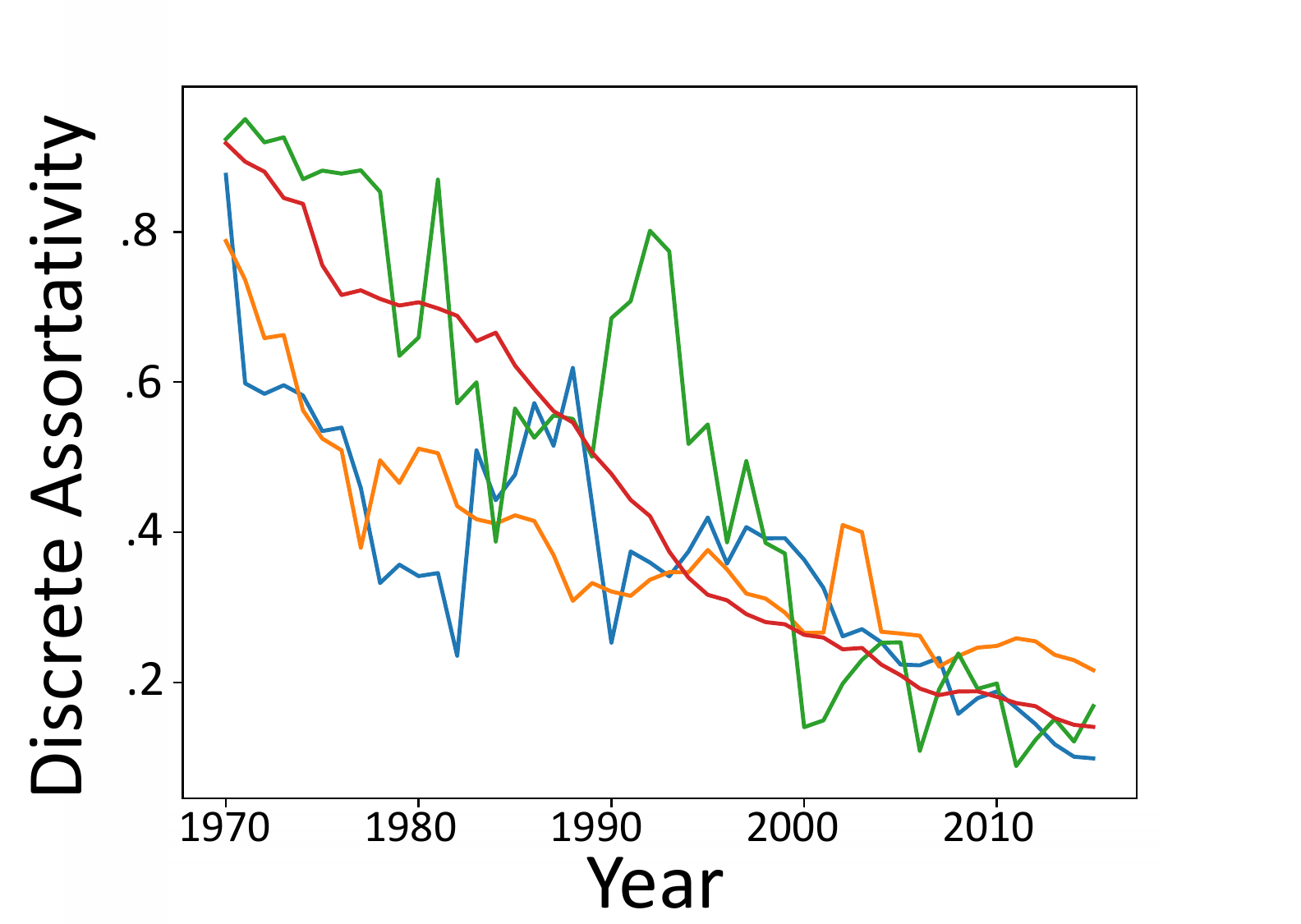}
  \caption{DAC }
  \end{subfigure}
    \begin{subfigure}[b]{0.3\textwidth}  \includegraphics[width=\textwidth]{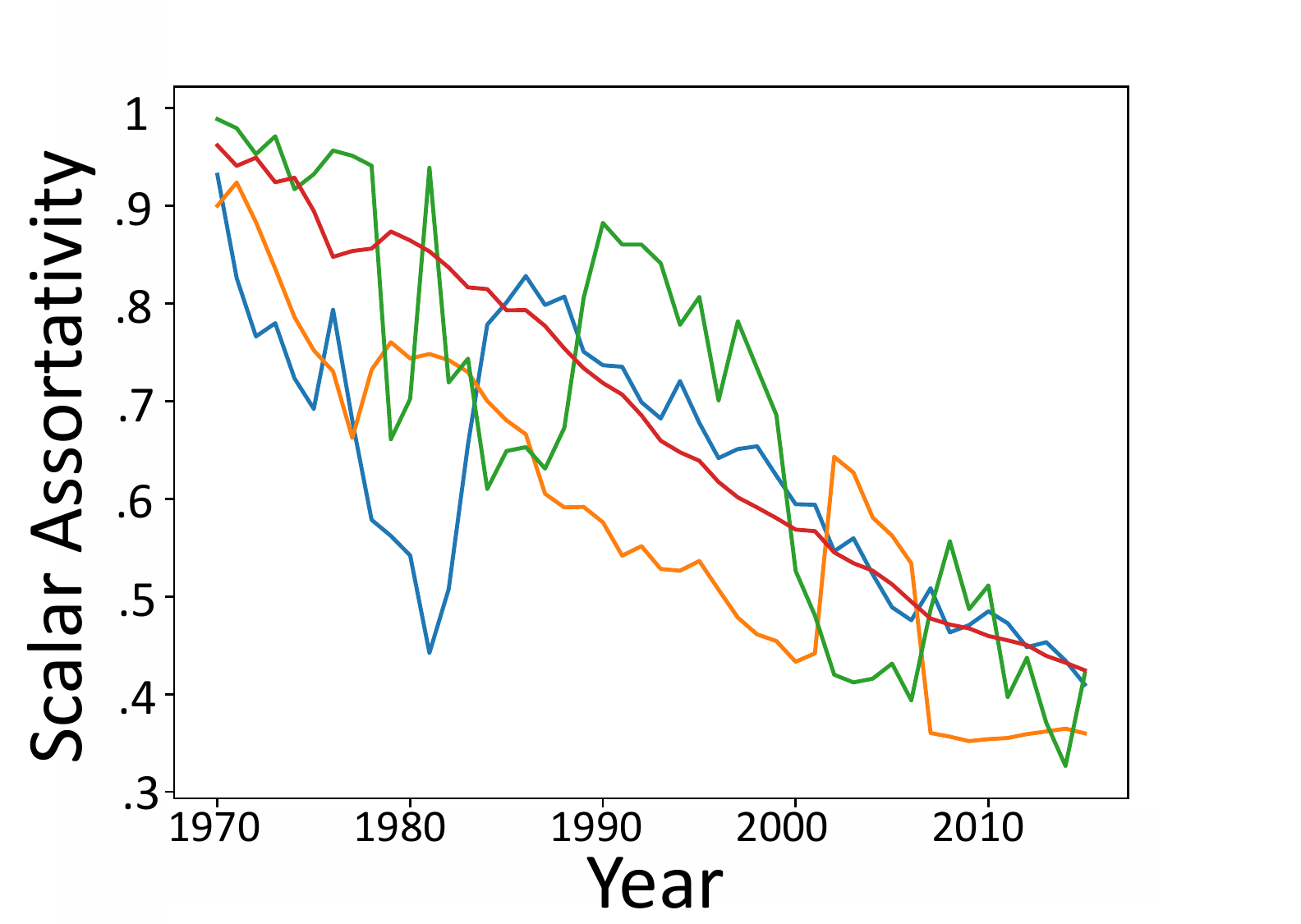}
  \caption{SAC}
  \end{subfigure}
    
  \caption{Metric Comparison}
  \label{fig:metrics}
  \vspace{0.5cm}
\end{figure*}

Figure~\ref{fig:NLP_boundaries2} shows optimal boundaries and class sizes for 2 and 5 classes for the NLP dataset (we observed similar results in other datasets).  In both cases, \texttt{MaxStrat} identifies nodes with score equal to zero as the lowest class for all datasets throughout all years, and the size of this class shrinks over time. Further exploration reveals that nodes with \textit{h}-index 0 are largely segregated from the rest of the networks and primarily collaborate with one another.

This explains why the highest values of \texttt{StA} are obtained with only two classes: because the lowest class ($h$-index=0) is so strongly segregated from the rest of the network, the clearest division is between that class and everything else. 
With a larger number of classes, stratification also appears among nodes with $h$-index greater than 0 and increases over time.  When the field is young, high $h$-index nodes collaborate with both medium and low $h$-index nodes, but such collaboration diversity wanes as the network gets older. One possible reason for this is that there are few high $h$-index researchers when the network is young, so high $h$-index must collaborate with lower $h$-index researchers. As more high $h$-index researchers become available, they prefer to work together and network gets more stratified.

To better understand the stratification of these fields over time, we break down collaborations across pairs of classes.  For simplicity, we use fixed class boundaries: in general, we find that it works well to partition nodes into four fixed classes of low score nodes ($h$-index=0), medium-low score nodes ($h$-index $\in \{1,2\}$), medium high score nodes (h-index $\in \{3,4,5,6\}$) and high score nodes (h-index$>6$): this gives high stratification across time periods, while allowing for more granularity than only having two classes (analysis with three classes is very similar). The results for fixed classes are very close to the results of \texttt{MaxStrat}.  We used fixed classes rather than the variable optimal classes for consistency of analysis across years. (See Figures~\ref{fig:SA_class_numbers_4} and~\ref{fig:st_mod}).

Figure~\ref{fig:cor_nlp} shows a heatmap describing results in the NLP field (results for other fields are provided in the Appendix).  This plot shows the frequency of collaborations between classes of authors. Here, each cell $(c_i, c_j)$ is the number of connections from class $c_i$ to class $c_j$, ($|(c_i,c_j)|$), normalized by the number of connections of the two classes ($cell (c_i, c_j) = \frac{|(c_i, c_j)|}{|c_i|\cdot|c_j|}$, where $|c_i|$ is the number of connections where at least one side is in class $c_i$). The $x$-axis and $y$-axis show the class of h-indices (like before, we divided the h-index scores into 4 classes).
The results show that as network gets older, high score nodes tend to collaborate with high score nodes more.

\begin{figure*}
  \centering
  \begin{subfigure}[b]{0.19\textwidth}  \includegraphics[width=\textwidth]{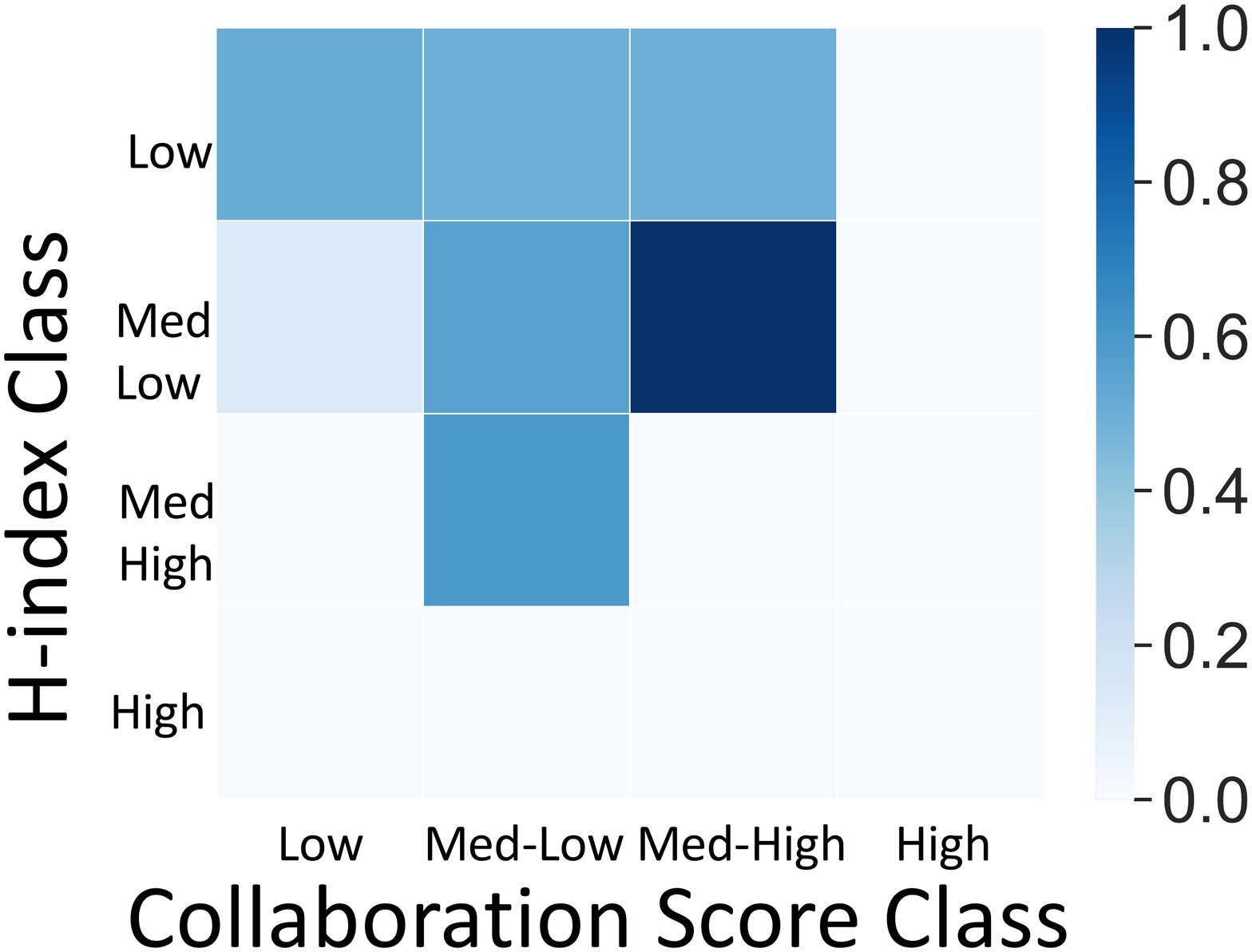}
  \caption{1966-1975}
  \end{subfigure}
  \begin{subfigure}[b]{0.19\textwidth}  \includegraphics[width=\textwidth]
  {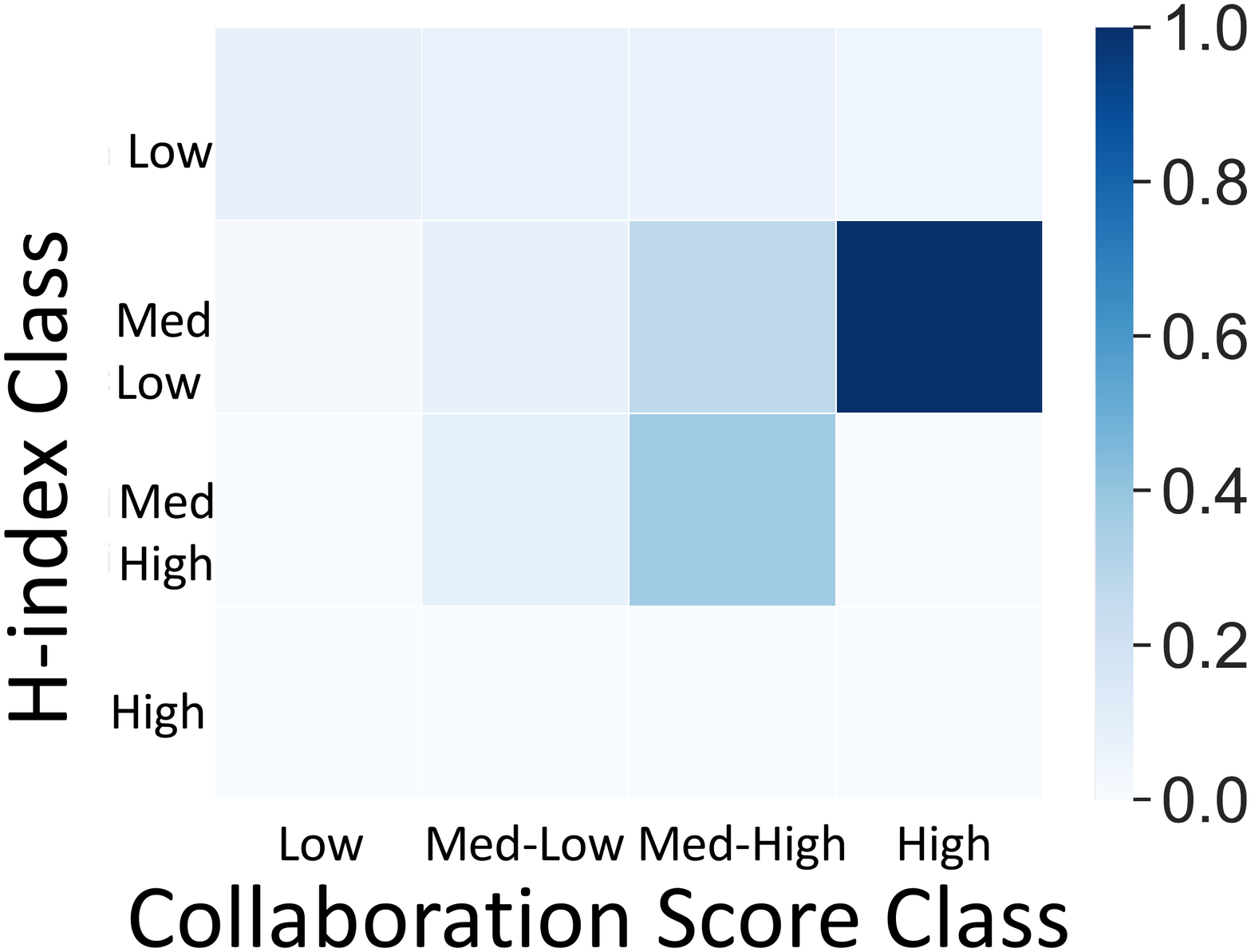}
  \caption{1976-1985}
  \end{subfigure}
    \begin{subfigure}[b]{0.19\textwidth}  \includegraphics[width=\textwidth]{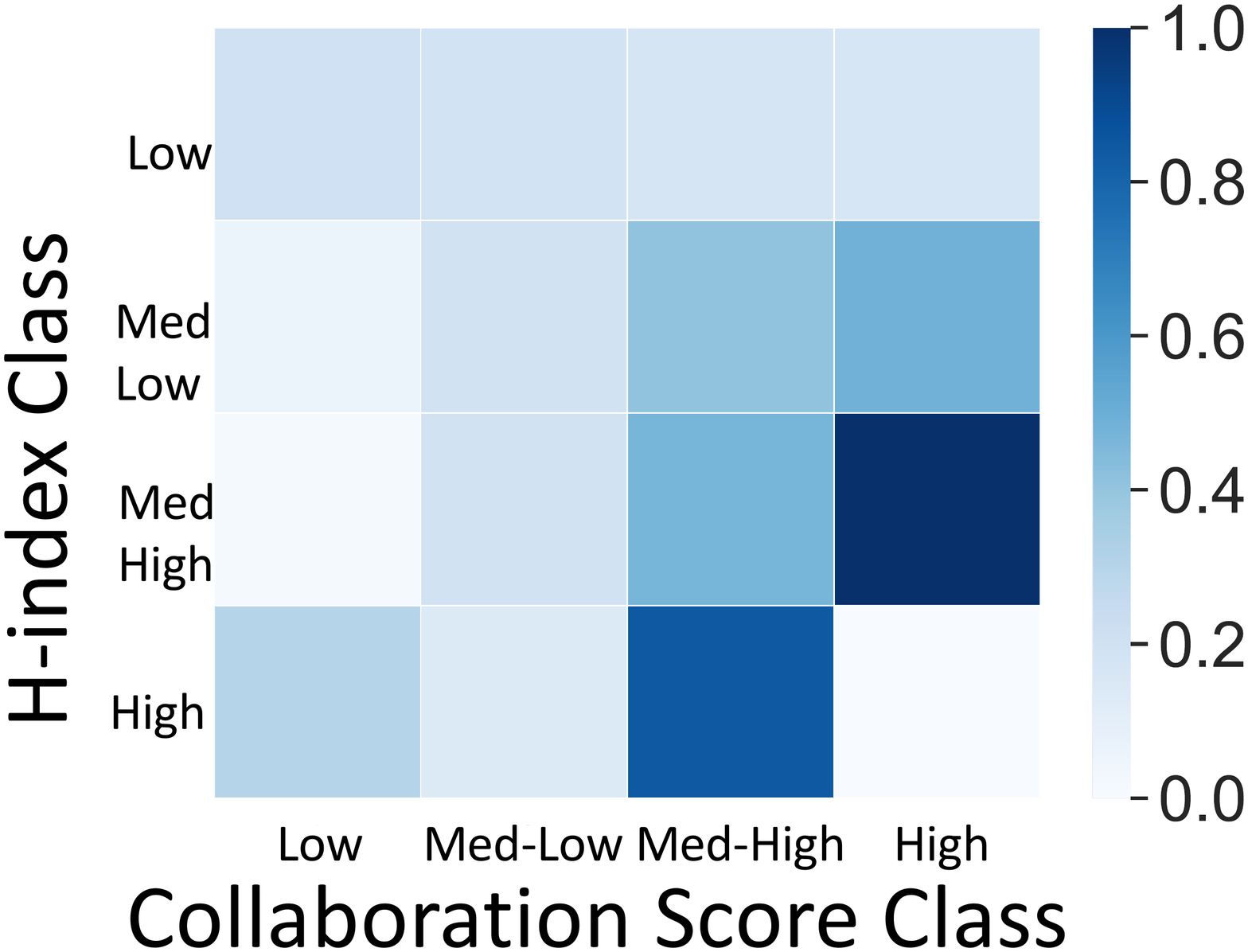}
  \caption{1986-1995}
  \end{subfigure}
    \begin{subfigure}[b]{0.19\textwidth}  \includegraphics[width=\textwidth]{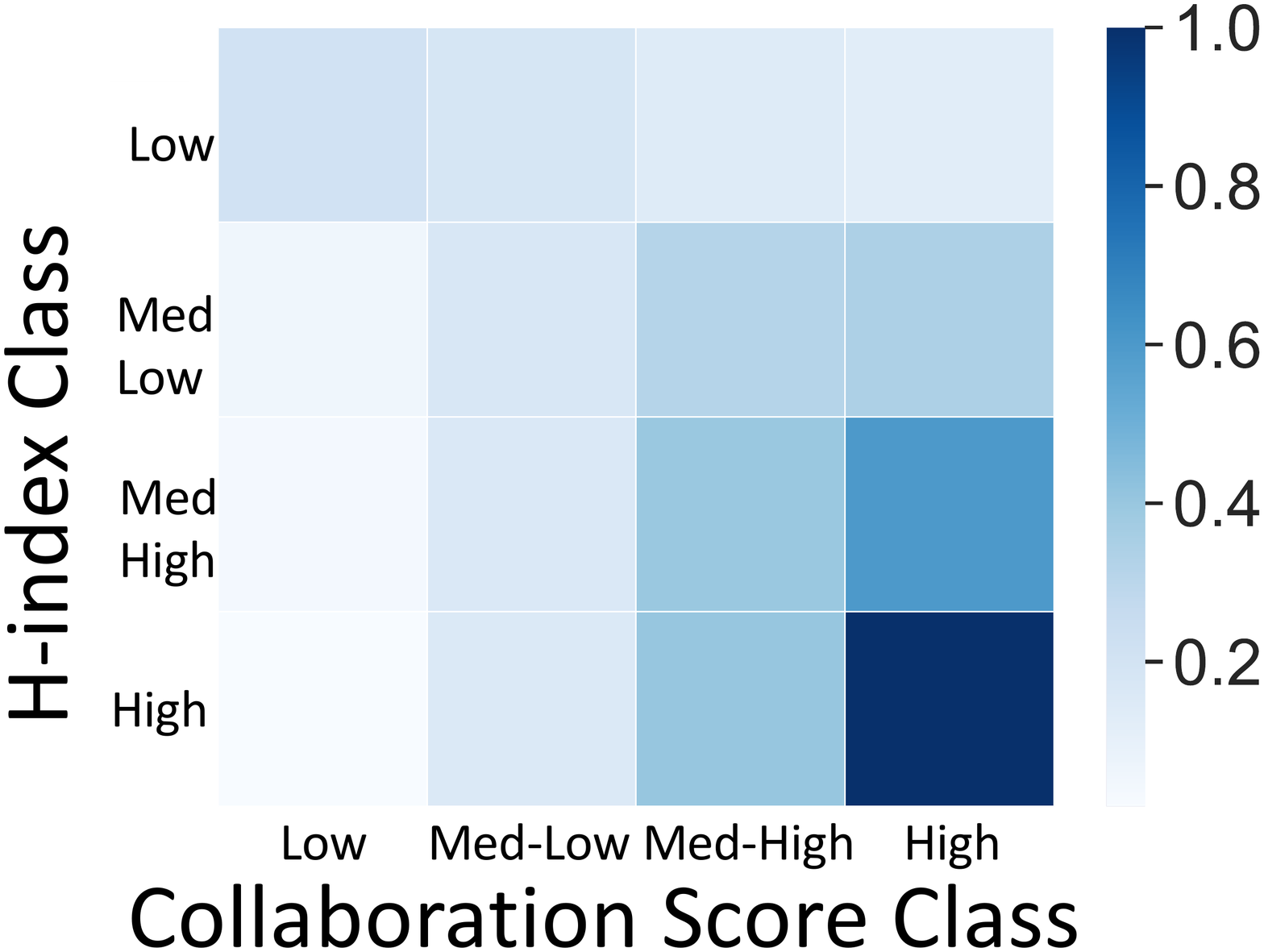}
  \caption{1996-2005}
  \end{subfigure}
    \begin{subfigure}[b]{0.196\textwidth}  \includegraphics[width=\textwidth]{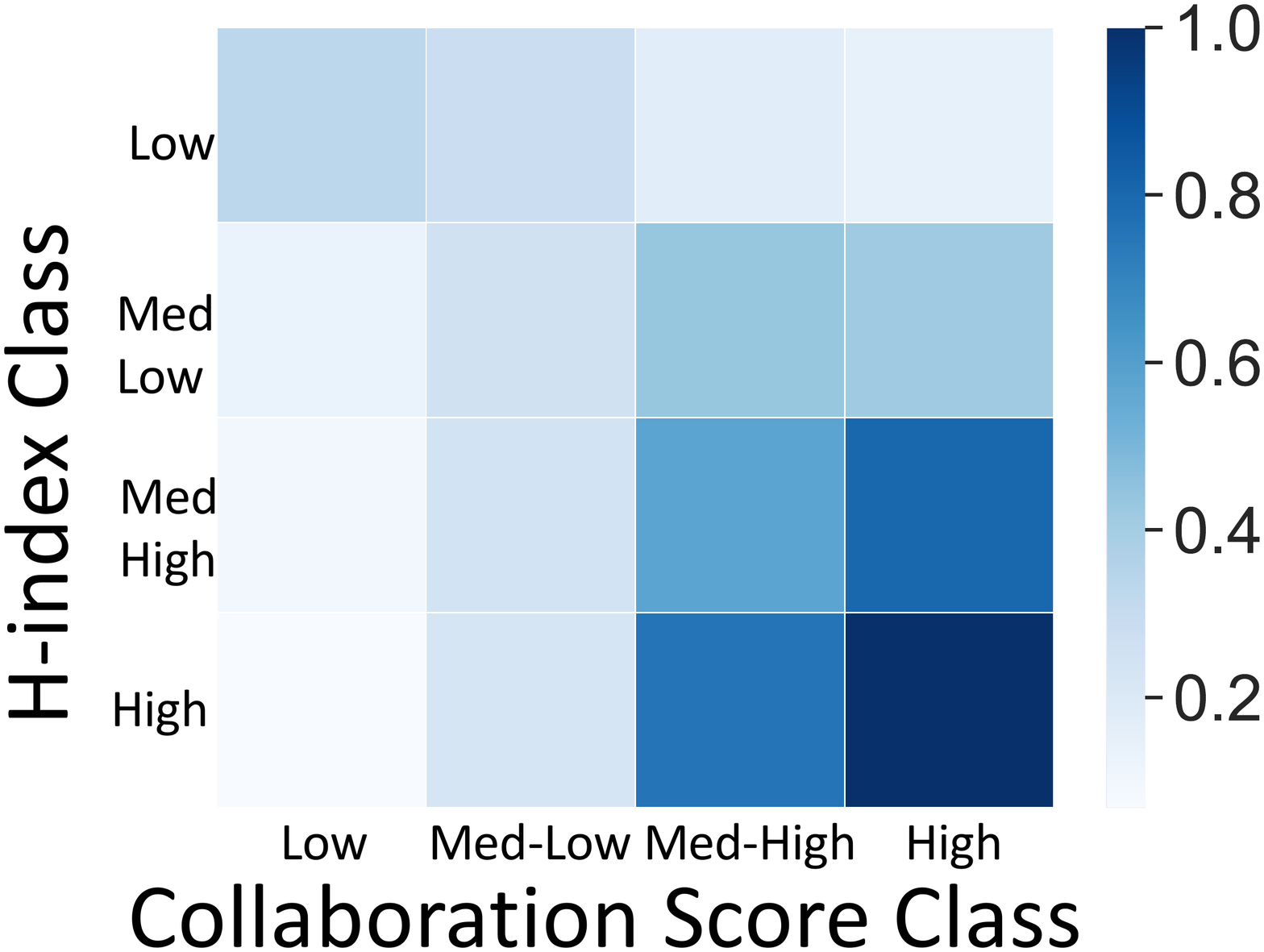}
  \caption{2006-2015}
  \end{subfigure}
  \caption{Relationship between entrance collaboration score of nodes and their h-index after 10 years for Biomedical Engineering}
  \label{fig:cor_be_se}
  \vspace{0.5cm}
\end{figure*}

\subsubsection{StA vs Other Assortativity Metrics}

Figure~\ref{fig:metrics} compares \texttt{StA} to \textit{DAC} and \textit{SAC} using 5 year snapshots. For \textit{DAC} and \texttt{StA}, we used the four fixed classes described in the previous section (for consistency of analysis: these boundaries are similar to those found by \texttt{ MaxStrat}, but the optimal boundaries vary slightly from year to year); scalar assortativity was computed using the $h$-index scores directly.  Unlike \texttt{StA}, \textit{DAC} and \textit{SAC} increase over time, and we examine each of these cases in turn below.

The primary driver behind the different behaviors of \textit{DAC} and $StA$ is that \textit{DAC} is based on over all inter- vs. intra- class connections, and so the sizes of classes (in terms of number of edges) weight the overall score, while \texttt{StA} is based on the average score of the classes and thus considers the impact of each class separately.

As an extreme case, consider a network in which there is only one class, and so all connections are intra-class connections.  In this case, \textit{DAC} is maximized ($DAC=1$). However, this network is not highly stratified because there are no medium- or high- score nodes in the network ($StA=0.25$).

We see a slightly less-extreme version of this phenomenon in the co-authorship networks.  When the fields are young, the the upper classes are very small compared to the lower classes.  Most of the edges are thus between nodes in the lower classes.  The classes are not segregated from one another; but the sheer size difference between the classes ensures that most edges are between nodes in the same (lower) class.  Thus, \textit{DAC} is high (because most edges are intra-class edges), but stratification is low, because the classes are not actually segregated from one another.  

As the fields age, the upper classes fill out, and segregation increases.  Because of this, \texttt{StA} increases.  \textit{DAC}, however, decreases, because as the class sizes become more balanced, the relative fraction of intra-class edges decreases more rapidly than the segregation between the classes increases.

Turning our attention to \textit{SAC}, the differences between \texttt{StA} and \textit{SAC} when fields are young is primarily due to class size.  At this point, most researchers have very low $h$-indices and a large fraction of edges are between nodes with similar scores, so \textit{SAC} is high.  However, because the upper classes and lower classes are not segregated from one another, \texttt{StA} is low.

As the fields age, \textit{SAC} decreases because the distribution and range of $h$-indices becomes wider, and so individuals may develop more diverse connections; for example, a node with an $h$-index of 30 might connect to a node with an $h$-index of 20.  However, because these nodes are in the upper class, and so are considered similar by \texttt{StA}.  Thus, while the \textit{absolute} difference between scores of neighbors increases (on the whole), and so \textit{SAC} decreases, the \textit{class} difference between scores of neighbors decreases (on the whole), and so \texttt{StA} increases.

\subsubsection{StA and Social Mobility}

A significant potential consequence of social stratification is a reduction of social mobility. We find that as networks age, the entrance point of new nodes has a larger effect on their trajectories through the field. For instance, Figure~\ref{fig:cor_be_se} shows the relationship between entrance collaboration scores and h-index of researchers after ten year in the Biomedical Engineering field (results in other fields are provided in the Appendix)\footnote{For authors in years 2012-2015 we examined their current $h$-index (2021)}. Cell $(c_i, c_j)$ is the normalized number of authors with starting collaboration score from class $c_i$ and $h$-index of class $c_j$ after 10 years, ($cell (c_i, c_j) = \frac{|(c_i, c_j)|}{|c_i|\cdot|c_j|}$, where $|c_i|$ is the number of authors in class $c_i$).
The $x$-axis shows the class of collaboration scores and the $y$-axis the class of $h$-indices (as before, each is divided into 4 tiers).

We observe that as the network ages, entrance point increasingly matters; and those who start their career by collaborating with high score nodes become much more likely to achieve a high $h$-index themselves.  A relationship between entry point and trajectory has been observed before~\cite{li2019early}); but we demonstrate that this tendency tends to increase over time.

More analysis of \texttt{StA} on  different numbers of classes/tiers is 
provided in the appendix.

\subsection{Discussion}
There are several major takeaways from our analysis:

\begin{enumerate}
\item Stratification in scientific fields tends to increase over time.  From Figure~\ref{fig:cor_nlp}, we see that this is due primarily to the increasing tendency of high $h$-index nodes to collaborate with other high $h$-index nodes.  

\item Even as stratification increases, other assortativity metrics decrease: this occurs because even as nodes connect more within their own class, the diversity of their intra-class connections increases.  This is true particularly for high $h$-index nodes.

\item As social stratification increases, social mobility decreases.  This is a key observation, and has major implications for the career trajectory of new researchers.  Simply put: when a field is new, a researcher's trajectory is not strongly determined by their first few collaborators; but as the field ages, these early connections become increasingly important.
\end{enumerate}

\subsubsection{Explaining Stratification}
We hypothesize that a major cause of stratification is related to access to collaborators.  It is known that network proximity is important to researchers when forming new collaborations (e.g., through triadic closure~\cite{kim2017over}).  When a field is young, and the collaboration network is less structured, a new researcher can find strong collaborators near them in the network: these collaborators can then help that new research boost her skills and profile.  However, as the field ages, researchers with high $h$-index begin to find one another; and as they congregate, new researchers in other areas in the network have a much greater network distance from these older, successful researchers.  This, in turn, makes it more difficult for those new researchers to join established projects, obtain mentoring from successful senior researchers, and so on.  In contrast, researchers who happen to join the network in proximity to these high $h$-index researchers (e.g., the PhD students of such researchers) get a substantial head start.  

The social science literature describes this phenomenon as \textit{social distance}, which has long been understood as a cause of stratification~\cite{bottero2003social}.
In an extreme case, a stratified network might be divided into separate connected components corresponding to the various score intervals (tiers).  This, naturally, has consequences for social mobility.  

Outside of co-authorship networks, social distance is defined very broadly, and can encompass characteristics like ethnicity, socio-economic status, occupation, etc.  A high social distance between classes suggests that individuals lack access- both directly and indirectly- to those in other classes.  In a network setting, the simplest way to determine `access' is by the existence of paths: if there is no path between two nodes, then by standard network evolution processes (e.g., triadic closure and the like), it is virtually impossible for them to connect in the future (this is not to say that they cannot connect; but if they do, it is likely because of processes external to the network topology).  

To further investigate this hypothesis, we examined the properties of connected components in the co-authorship network.  If connected components exhibit very different score-related characteristics, this is indicative of social distance driving the increase in stratification.

For each node $u$ in the network, we compute a \textit{collaboration score}, defined as the average $h$-index of the four highest-scoring collaborators of $u$ (we only consider the top neighbors because access to higher-class individuals is more important than access to lower-class individuals for upwards social mobility.)  Note, importantly, that we are not simply using the scores of the nodes themselves: we are examining whether nodes in the various components have \textit{access} to high $h$-index nodes.  

Next, for each connected component, we compute the average of the collaboration scores of all nodes in that component.  We refer to this as the \textit{component score}.  A higher value indicates that on average, nodes in the connected component have collaborations with high scoring nodes, while a lower average indicates that on average, nodes in the component lack connections to high scoring nodes.  Note that a component with many low $h$-index nodes can still have a high component score, as long as those low $h$-index nodes have collaborations with high $h$-index nodes.

\begin{figure*}
  \centering
  \begin{subfigure}[b]{0.3\textwidth}  \includegraphics[width=\textwidth]{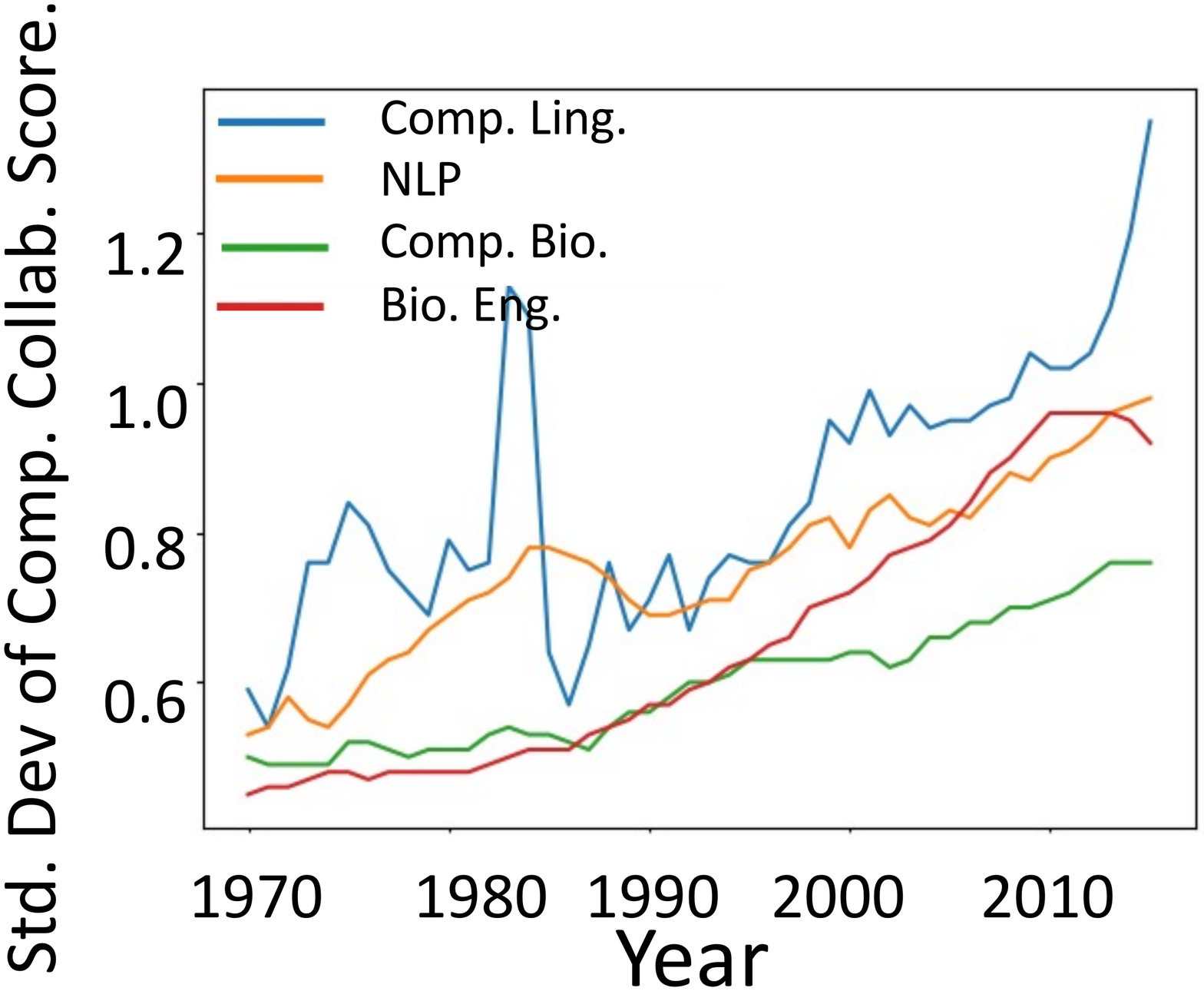}
  \caption{Standard Deviation}
  \end{subfigure}
  \begin{subfigure}[b]{0.31\textwidth}  \includegraphics[width=\textwidth]{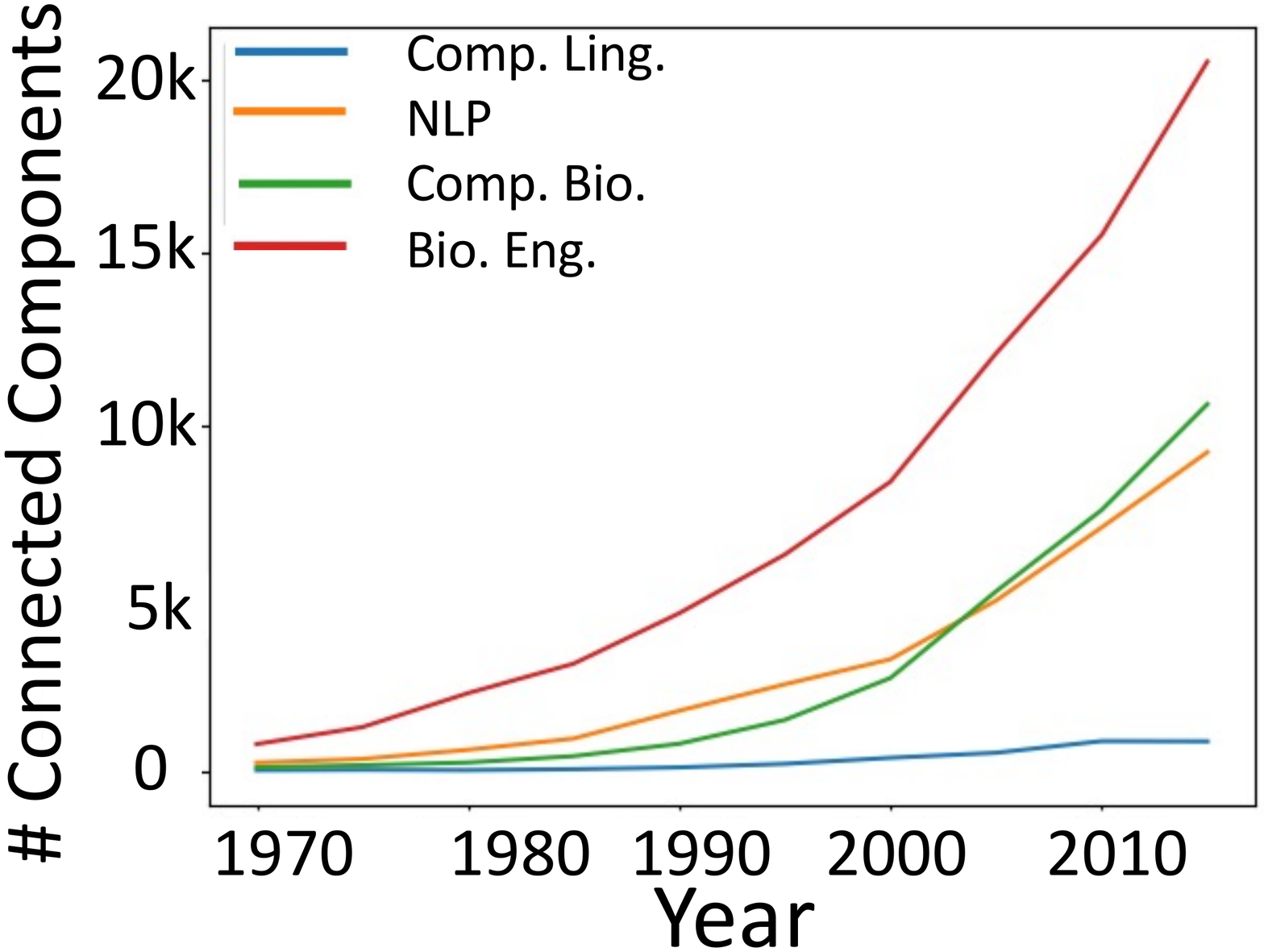}
  \caption{Number of components}
  \end{subfigure}

  \caption{Standard deviation of average component collaboration scores over all components, and number of connected components.  As the networks get older, standard deviation increases.}
  \label{fig:stdev}
  \vspace{0.5cm}
\end{figure*}

If social distance is indeed a driver of stratification, we expect these component scores to vary significantly across components: in some areas of the networks, nodes have access to high $h$-index nodes (high component scores), and in other areas, nodes do not (low component scores).  Accordingly, we compute the standard deviation across the set of component scores.  A low standard deviation indicates that nodes have similar collaboration patterns across different components of the network, and a high standard deviation indicates that the network has some components in which nodes tend to have higher collaboration scores and some components in which nodes tend to have lower collaboration scores. The former means that there are no major differences in social distance, suggesting that the network is not stratified; while the latter implies the converse.  Note that we are not measuring stratification directly: rather, we are exploring whether there is a major difference in access to high $h$-index nodes in different regions of the network.  

Figure~\ref{fig:stdev} shows these results.  In all cases, as the network gets older, the number of components increases and the standard deviation across components increases. In other words, over time, the components show increasingly different behaviors from one another: in some, nodes have access to high-scoring nodes, and in others, nodes do not.  This matches the earlier stratification results exactly.

This analysis suggests a mechanism by which stratification in co-authorship networks could occur: if researchers have a preference for connecting to high $h$-index researchers, and there is a practical limitation on the number of researchers that one can collaborate with, then those researchers with a high $h$-index will tend to congregate with one another, leaving regions of the network devoid of such researchers.  This then affects new researchers who enter the network in these areas, leading to decreased mobility.

{{Moreover, our study shows that the position of researchers in the network has a key role in their eventual success. In addition to our hypothesis, this can be due to different reasons such as privilege or discrimination}.
{To explain this, we draw an analogy to results from \textit{Stratification Economics}, a sub-field of economics that explains inter-group inequalities as caused, in part, by `uneven intergenerational transmission of resources and advantage'}}~\cite{darity2022position,darity2008stratification, burnazoglu2022editorial}.
{{{Stratification economics examines differences between social classes, nations and regions, racial and ethnic groups, etc., in a competitive, cooperative, or collaborative environment}}~\cite{darity2022position}. {{Studies in this area consider both the relative position of individuals within a social group as well as the absolute position of social groups, because both factors affect in the lives, rewards, and satisfaction of individuals.}}~\cite{darity2022position}.}

{{{Like the societies studied in stratification economics, scientific collaboration takes place in a cooperative and collaborative environment, and inequalities in access to resources and prestige can be `inherited' through one's collaborators. The citations that we use in this study are not directly inherited, but resources, prestige, and status related to citations are unevenly transmitted across generations.  Such factors can play a key role in facilitating the stratification process of collaboration networks, contributing to the development of hegemonical epistemologies that further reinforce stratification and,}} {{consequently, may}} {{lead to discrimination. For instance, certain research groups or universities have access to modern equipment, facilities and resources that gives their scholars an initial research advantage. These scholars may come to set the global research agenda, such that the work of researchers who lie beyond this network stratum becomes increasingly marginalized. For instance, Obeng-Odoom reviews research highlighting how the editorial boards of leading journals in economics are dominated by a small groups of (primarily white, Eurocentric) editors from mainstream economic departments}~\cite{obeng2019intellectual}. {As a result, mainstream journals tend to exclude research contributions that focus on topics that are of interest of African scholars.}} {{Although our study did not explore racial discrimination,}}
{{it is clear that from this example how discriminatory social consequences can arise from a network-based stratification process.}}}

\subsubsection{Application and Limitations of Measuring Stratification in Other Domains}

While our analysis here has focused on co-authorship networks, the \texttt{StA} metric can be used to measure stratification in any system that has a network describing social or other relevant interactions, in which nodes are annotated with an appropriate score (the meaning of this score can vary by domain).  It would be of particular interest to use it to study socio-economic stratification: in particular, to gain another perspective on the extent to which personal and professional networks may affect class mobility, and to compare across different societies.

There are some important limitations in applying the \texttt{StA} metric.  First, while the mathematics can be adapted to directed graphs, one should be careful in interpreting such results.  On Twitter, for instance, there are many low-status individuals following a high-status individual, but this has very little relevance toward stratification as a whole, as it is very easy to follow someone, and reciprocity/consent are not required.  Additionally, for similar reasons, it is important to consider the meaning of edge information.  When studying online social media data, for instance, many platforms allow an effectively unlimited number of friends, and so high-status individuals may be willing to connect to low-status individuals; but such connections may not be meaningful.  In such cases, the existence of communication or meaningful interaction may be more relevant than the existence of a `friendship'.

\section{Conclusion and Future Work}

In this paper,  we proposed \texttt{StA}, a novel algorithm that measures {network social stratification} by evaluating the tendency of the network to be divided into ordered classes.
Then, we performed a case study on several co-authorship networks and examined the evolution of these networks over time and showed that networks evolved into highly stratified states.
In future work, we plan to study social stratification in other types of social networks, explore reasons behind stratification in different network types, and see how stratification can be prevented.

\bibliographystyle{unsrt}  
\bibliography{references}  
\input{Appendix.tex}

\end{document}

%% file: Appendix.tex
\newpage
\section{Appendix}

\subsection{Assortativity/Modularity metrics}

\begin{itemize}
\item The \textit{network modularity} metric measures the quality of a partitioning of the nodes in a network.  A network partitioning has high modularity if there are many intra-class edges and few inter-class edges, compared to what one would expect in a random network with the same degree distribution~\cite{newman2006modularity}. An inter/intra-class measurement very similar to modularity is used to examine social stratification in communications networks~\cite{leo2016socioeconomic}.   

\item The \textit{discrete assortativity coefficient (DAC)} measures the similarity of connections in the network. The \textit{DAC} of graph $G(V, E)$ is a normalized version of modularity
~\cite{newman2003mixing, gera2018leading}.

\textit{Modularity} and the \textit{DAC} are based on an unordered categorical attribute such as race and gender.

 $$DAC(G) = \frac{S_{discrete}(G) - ES_{discrete}(G')}{Max(S_{discrete}(G)) - ES_{discrete}(G')},$$
 
where $S_{discrete}$ is the \textit{DAC} of graph $G$ with adjacent matrix $\mathbf{A}$ ($a_{uv}$ is the element uv ), $ES(G')$ is the expected \textit{DAC} of a random network $G'$ with the same degree distribution as $G$, and $\sigma(u,v) =1$ if nodes $u$ and $v$ are from the same class and 0 otherwise.

 $$S_{discrete}(G)  =\frac{ \Sigma_{u, v \in V} (a_{uv})\cdot \sigma(u,v)}{2m },$$
 
 $$ES_{discrete}(G')  =\frac{ \Sigma_{u, v \in V} (deg(u) \cdot deg(v))\cdot \sigma(u,v)}{4m^2 },$$
 
 $$Max(S_{discrete}(G)=1.$$ 

\item The \textit{scalar assortativity coefficient (SAC)}, measures the similarity of connections in the network with respect to a numeric attribute, but does not allow for specification of groups. 

$$SAC(G) = \frac{S_{scalar}(G) - ES_{scalar}(G')}{Max(S_{scalar}(G)) - ES_{scalar}(G')}.$$

where $S_{scalar}$ is the `scalar assortativity score' of graph $G$, $ES(G')$ is the expected `scalar assortativity score' of a random network $G'$ with the same degree distribution as $G$, and $Max(S_{scalar}(G)$ is the maximum  `scalar assortativity score' possible.

$$S_{discrete}(G)  =\frac{ \Sigma_{u, v \in V} (a_{uv})\cdot s(u) s(v)}{2m },$$
 
$$ES_{discrete}(G')  =\frac{ \Sigma_{u, v \in V} (deg(u) \cdot deg(v))\cdot s(u) s(v)}{4m^2 },$$
$$
Max(S_{scalar}(G)) =\frac{ \Sigma_{u, v \in V} (a_{uv})\cdot s(u)^2}{2m }.$$

\end{itemize}

\subsection{StA computation}
As we explained $StA$ is computed as follows:

$$StA(G) = \frac{S_{strat}(G) - E(S_{strat}(G'))}{Max(S_{strat}(G)) - E(S_{strat}(G'))}.$$  

Here we explain in details how $S_{strat}(G)$ and $E(S_{strat}(G')$ are computed:

 $$S_{strat}(G) = \sum_{c_i \in C} \sum_{(u,v) \in E} \frac{sim(u,v, c_i)}{sim(u,v,c_i) + dissim(u,v,c_i)}$$

{where $sim(u,v,c_i)= w(u,v) \cdot \alpha(u, v, c_i)$ is the similarity of nodes u and v and  $dissim(u,v,c_i) = (1-w(u,v)) \cdot (1-\alpha(u, v, c_i)))$ is  the dissimilarity of nodes u and v Here, $w(u,v)$ is the weight of edge $(u,v)$ as computed based on similarity and $\alpha(u, v, c_i)=1$ if $c(u)=c(v)= c_i$ and $0$ otherwise.
 
 $S_{strat}$ is computed as the sum of scores of each class. In computing the score of each class, the numerator represents the sum of similarity weights of intra-class connections and the denominator represents the sum of similarity weights of intra-class connections plus the sum of dis-similarity weights (1-similarity) of inter-class connections.  This score is based on both scalar characteristics (actual value) and specified classes (\textbf{Properties 1, 2}).

Using similarity weights of intra-class connections ensures that connections within the same class are rewarded (the higher the similarity, the higher the reward) and using dis-similarity weights of inter-class connections ensures that connections within different classes are penalized (the higher the dis-similarity, the higher the penalty) (\textbf{Property 3}).

Using the sum of scores of each class takes
mutual segregation into account
and prevents the impact of a class with more connections be more than other classes (\textbf{Property 4}).}

 $$ES_{strat}(G) = \sum_{c_i \in C} \sum_{(u,v) \in E} \frac{E(sim(u,v,c_i))}{E(sim(u,v,c_i)) + E(dissim(u,v,c_i))}.$$

{$E(S_{strat}(G))$ is computed similarly to $S_{strat}(G)$, except that it considers expected similarity and dissimilarity. $E(sim(u,v,c_i))= w'(u,v) \cdot \alpha(u, v, c_i)$ is the expected similarity of nodes u and v and  $E(dissim(u,v,c_i) = (1-w'(u,v))) \cdot (1-\alpha(u, v, c_i)))$ is the expected dissimilarity of nodes u and v and computed using $w'(u,v)$.  $w'(u,v)$ is the expected weight of an edge between nodes $u$ and $v$ in networks with the same weighted degree distribution as $G$. $w'(u,v) = \frac{sw_u \cdot sw_v}{ (\Sigma_{x \in V} sw_x)^2}$}

 Maximum social stratification (\texttt{StA}) happens when there are no inter-class connections and all edges are intra-class connections. In this case, for each $c_i \in C$ the score of each class is one  (numerator and denominator in $S_{strat}$ formula are the same per each class). Then, as there are $k$ classes in the network, $Max(S_{strat}(G))=k$.

\subsection{{Class Boundaries}\label{sec:unknown1}}

 Depending on the application under study, the social class hierarchy might be known or unknown.  In many real applications, these classes are known ahead of time. For example, it is conventional in economic analysis of Western societies to define a lower, middle, and upper class~\cite{cannadine1998beyond}.  However, in other applications like meetings or conferences between individuals~\cite{gupte2011finding}, it is possible that neither the class boundaries nor even the number of classes is known ahead of time.  

If classes are not known ahead of time, one must first find the social class boundaries and then use {\texttt{stratification assortativity}} metric for known classes.
As explained,  social class refers to hierarchical social categories arising from different relationships~\cite{krieger1997measuring}. 
Here, we are interested in social classes that result in the greatest \texttt{stratification assortativity}. As explained before, stratification has consequences, and if there is a setting for classes that has high levels of stratification, then that setting is of great importance. For instance, in a society that is divided into lower, middle and upper class, there might be many ways of partitioning the network into non-stratified classes, but the existence of a stratified partition is of great importance.

The problem is similar to the problem of community detection in the network that requires partition of the network into components of densely connected nodes~\cite{blondel2008fast}. Here, we are interested in finding components of ordered nodes  with more intra-component interaction and less inter-component interactions normalized by weight and class sizes.

In this section, we introduce  \texttt{MaxStrat}, a heuristic to find the class boundaries that result in the greatest \texttt{stratification assortativity}.  \texttt{MaxStrat} requires that the user specify $k$, the number of classes; but if this is not known, one can consider all possible values of $k$ within whatever range is desired.  

Finding classes that maximize \texttt{stratification assortativity} can also be used to find the hierarchy in a social network. The hierarchy underlying a social network can be used in different applications including link recommendation (friendship recommendation)~\cite{gupte2011finding}.

\noindent \textbf{MaxStrat: } 
 \texttt{MaxStrat} assumes as input an undirected, unweighted network $G$ with $n$ nodes, where each node $u$ has score $s(u)$.  The user also specifies a desired number of classes $k$ (if no $k$ is specified, one can run \texttt{MaxStrat} for different values of $k$ to find the one that results in the highest social stratification). 
 
 The \texttt{MaxStrat} heuristic is based on maximizing the non-normalized version of the $StA$  ($StA'= {S_{strat}(G) - E(S_{strat}(G'))}$ ) .

\texttt{MaxStrat} consists of two steps. First, finding ordered classes that maximize $StA'$ and second, in an iteration, scanning the classes and moving nodes with the upper-bound or lower-bound scores to adjacent classes if it improves $StA$. Pseudo-code for  \texttt{MaxStrat} is shown in Algorithm~\ref{alg:MaxStrat} .
 
\noindent \textbf{Step 1: } 
Let $h$ represent the number of distinct scores.  Without loss of generality, assume that the set of distinct scores is $\{1, 2, ..., h\}$. 
The algorithm computes matrices $T_b$, for $b \in \{0, 1, ..., k\}$ which will keep track of intermediate solutions.  
 $T_0[i, j]$  denotes the amount that this interval would contribute to the network's $StA'$ score if $i$ and $j$ were the boundaries of a class: 

\vspace{0.2cm}
\tiny
\noindent $$T_0[i, j] 
 = \Sigma_{(u,v) \in E} \frac{w(u,v) \cdot \alpha(u, v, C_{[i,j]})}{w(u,v) \cdot \alpha(u, v, C_i) + (1-w(u,v)) \cdot (1-\alpha(u, v, C_{[i,j]}))} -  \frac{w'(u,v) \cdot \alpha(u, v, C_{[i,j]})}{w(u,v) \cdot \alpha(u, v, C_{[i,j]}) + (1-w'(u,v)) \cdot (1-\alpha(u, v, C_{[i,j]}))}.$$
\normalsize

For $b>0$, define $T_b[i, j]$ to be the maximum $StA'$ contribution that can be obtained by splitting the interval $[i, j]$ into $m$ sub-intervals.  $T_b[i, j]$ is clearly undefined if $j - i + 1 < b$ and $T_1[i, j] = T_0[i, j]$.  For larger values of $m$, $T_b[i, j]$ can be found by considering all indices $r$ between $i$ and $j$, computing the sum $ T_0[i, r] + T_{b-1} [r+1, j]$, and taking the maximum of these values. These values are computed up to $b = k$, and finally, $T_k[i, j]$ is returned. 

As described above, the algorithm computes $T_0$ to $T_k$.  However, this can be sped up further by only computing those elements of matrices that are actually needed (See Algorithm~\ref{alg:Step1}  to compute only necessary elements, using recursion with memoization.
Note that $T_0$ is passed to the recursive implementation as input to avoid extra computation.

\noindent \textbf{Step2:} The output of step 1 is a set of ordered tiered intervals. Step 1 is based on $StA'$ and the final result maximizes  $StA'$. As $StA$ is the normalized version of $StA'$, there might be few cases that the classes found in step 1 are not the optimal solutions for $StA$. To address this problem, \texttt{MaxStrat} contains a scanning step to make sure that the final results are close to optimal. 

The process consists of several iterations. Suppose that the intervals are $\{[i_1,j_1],$ $..., [i_k, j_k]\}$. In each iteration, for each two consecutive intervals  $[i_a,j_a]$, $[i_{a+1},j_{a+1}]$ $(i_{a+1} = j_a +1)$,
check whether substituting intervals $a$ and $a+1$ with either $[i_a,j_a-1]$, $[i_{a+1}-1,j_{a+1}]$ or $[i_a,j_a+1]$, $[i_{a+1}+1,j_{a+1}]$ will give a higher $StA$. And if it does, substitute the current intervals with the one where intervals $a$ and $a+1$ are updated to the maximum one possible.
Continue the process until no substitution will give a better result in an iteration.  Pseudo-code for this step is shown in Algorithm~\ref{alg:Step2}.

\begin{algorithm*}[] 
\footnotesize
\caption{{{MaxStrat }}}
\label{alg:MaxStrat} 
\begin{algorithmic}[1] 
\REQUIRE ~~\\ 
 Network $\textbf{G}$ with $n$ nodes and attribute (Score), Desired number of classes k
\STATE i = min score of nodes
\STATE j = max score of nodes
\STATE $T_0 $: empty matrix
\FOR{ $l$ in range $(i,j)$}
\FOR{ $r$ in range $(l,j)$}
\STATE \tiny
 $T_0[i, j] = \Sigma_{(u,v) \in E} \frac{w(u,v) \cdot \alpha(u, v, C_{[i,j]})}{w(u,v) \cdot \alpha(u, v, C_i) + (1-w(u,v)) \cdot (1-\alpha(u, v, C_{[i,j]}))}-  \frac{w'(u,v) \cdot \alpha(u, v, C_{[i,j]})}{w(u,v) \cdot \alpha(u, v, C_{[i,j]}) + (1-w'(u,v)) \cdot (1-\alpha(u, v, C_{[i,j]}))}.$
\normalsize
\ENDFOR
\ENDFOR
\STATE intervals, sum = Step1$(G, k, [i,j], T_0)$
\STATE intervals = Step2$(G, intervals, k)$
\RETURN $intervals$ 
\end{algorithmic}
\end{algorithm*}

\begin{algorithm*}[] 
\footnotesize
\caption{{{Step1 }}}
\label{alg:Step1} 
\begin{algorithmic}[1] 
\REQUIRE ~~\\ 
 Network ${G}$ with $n$ nodes and attribute (Score), Desired number of classes k' at current step, interval [i,j], $T_0$
\STATE score = distinct scores of nodes
\IF{$k =1$ or $i=j$}
\STATE \textbf{return} $[(i,j)], T_0[i,j]$ 
\ELSE
\STATE $interval =[], maxSum=0$
\FOR{ $c$ in range $(j-i)$}
\STATE $leftInterval = [i, i+c]$, $leftSum = T_0[i, i+c]$
\STATE $rightIntervals, rightSum = Step1(G, k'-1, [i+c+1, j] ,T_0)$
\IF{$leftSum + rightSum > maxSum$}
\STATE $maxSum = leftSum + rightSum$
\STATE $Intervals = rightInterval.append(leftInterval) $
\ENDIF 
\ENDFOR 
\ENDIF

\RETURN $Intervals, sum$ 
\end{algorithmic}
\end{algorithm*}

\begin{algorithm*}[] 
\footnotesize
\caption{{{Step2 }}}
\label{alg:Step2} 
\begin{algorithmic}[1] 
\REQUIRE ~~\\ 
 Network $\textbf{G}$ with $n$ nodes and attribute (Score), intervals, k
 \STATE score = StA(G, intervals),  flag = True, checked = [intervals]
 \WHILE{flag}
  \STATE flag = False
 \FOR{ $a$ in range $k-1$}
    \STATE $[i_a, j_a] = intervals[a], [i_{a+1}, j_{a+1}] = intervals[a+1]$
    \STATE $intervals_1, intervals_2 = intervals.copy()$
      \STATE $intervals_1[a] = [i_a, j_a+1] , intervals_1[a+1] = [i_{a+1}+1, j_{a+1}] $
    \IF{$intervals_1$ not in $checked$ \& $i_a \leq j_a +1$ \& $i_{a+1}+1 \leq j_{a+1}$}
        \STATE $checked.append(intervals_1)$,  $Score_1 = StA(G, intervals_1)$
         \IF{$Score_1>Score$}
        \STATE $intervals = intervals_1, flag = True$
        \ENDIF
    \ENDIF
  
    \STATE $intervals_2[a] = [i_a, j_a-1] , intervals_2[a+1] = [i_{a+1}-1, j_{a+1}] $
          \IF{$intervals_2$ not in $checked$ \& $i_a \leq j_a -1$ \& $i_{a+1}-1 \leq j_{a+1}$}
        \STATE $checked.append(intervals_2)$,  $Score_2 = StA(G, intervals_2)$
         \IF{$Score_2>Score$}
        \STATE $intervals = intervals_2, flag = True$
        \ENDIF
    \ENDIF
 \ENDFOR 
\ENDWHILE
\RETURN $Intervals$ 
\end{algorithmic}
\vspace{0.5cm}
\end{algorithm*}

\subsubsection{Time Complexity}
    Let $n$ denote the number of nodes in the graph, $h$ denote the number of distinct scores (without loss of generality, assume the scores are $\{1, ..., h\}$, and $k$ denote the desired number of classes.

	From the equation, it is clear to see that measuring \texttt{stratification assortativity} takes $O(n^2)$ time, where $n$ is the number of nodes in the graph.  For Algorithm~\ref{alg:MaxStrat}, the first step is to compute $T_0[i, j]$ for each interval $[i, j]$ in $\{1, ..., h\}$.  This takes $O(n^2h^2)$ time.  Then, the algorithm computes matrices $T_b$ for $b \in \{1,2,.. ,k\}$. Computing each element $T_b[i, j]$ is done by iterating over all indices $r$ between $i$ and $j$ and taking the sum of other values of $T$, so this is done in $O(j - i) = O(n)$ time.  There are $O(h^2k)$ elements in $T$, so this takes $O(nh^2k)$.  Adding on the time required to compute the $Cont$ values, this becomes $O(nh^2k + n^2h^2) = O(n^2h^2)$ time. 
	The scanning process is a heuristic that scans over the $k$ existing interval and it usually terminates very quickly.

\subsection{Sensitivity  of Stratification}
\begin{figure*}
  \centering
   \centering
   \begin{subfigure}[b]{0.6\textwidth}  \includegraphics[width=\textwidth]{figs/leg1.pdf}
  \end{subfigure}
  
  \vspace{-0.04cm}
  \begin{subfigure}[b]{0.24\textwidth}  \includegraphics[width=\textwidth]{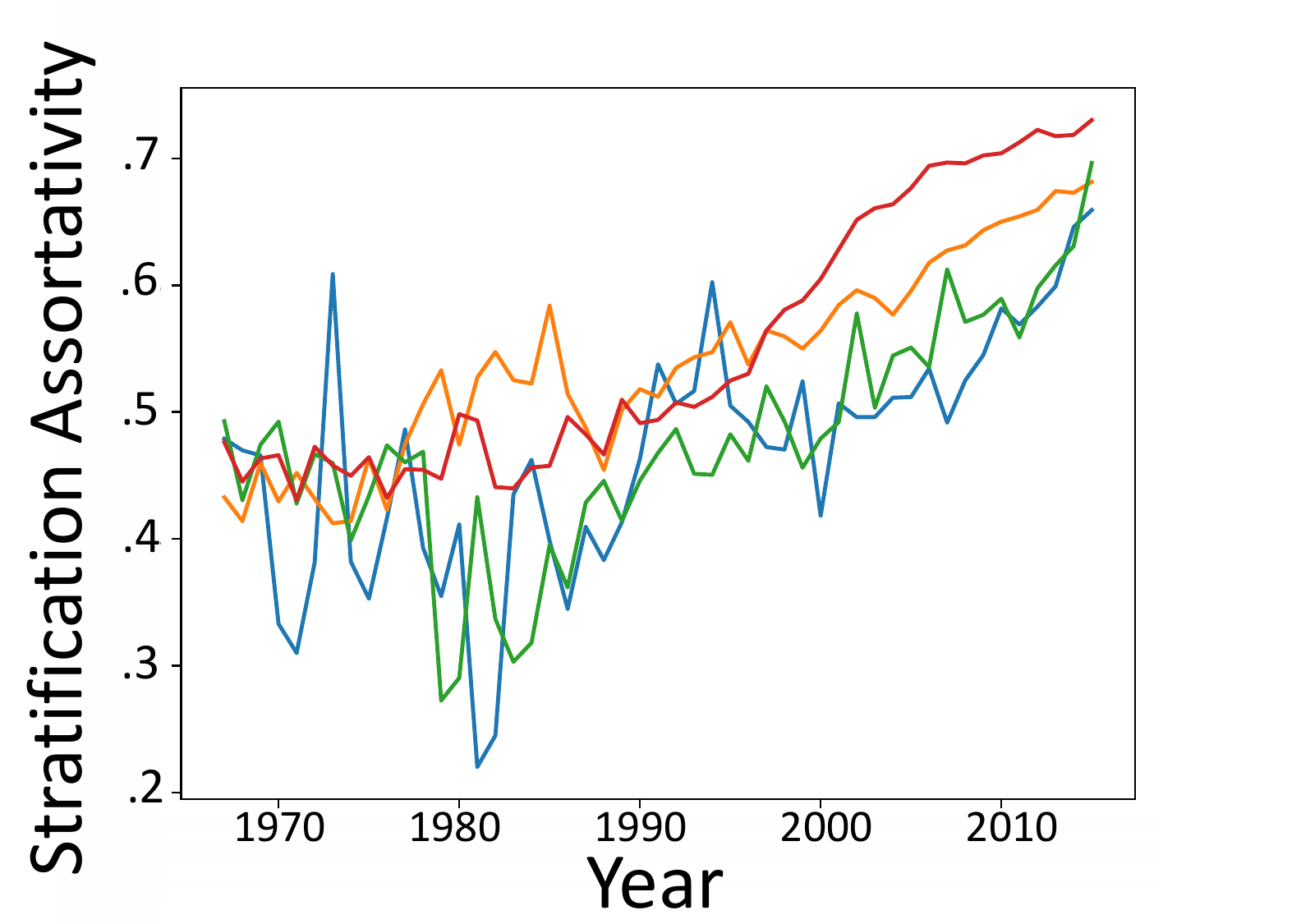}
  \caption{2-year intervals}
  \end{subfigure}
  \begin{subfigure}[b]{0.24\textwidth}  \includegraphics[width=\textwidth]{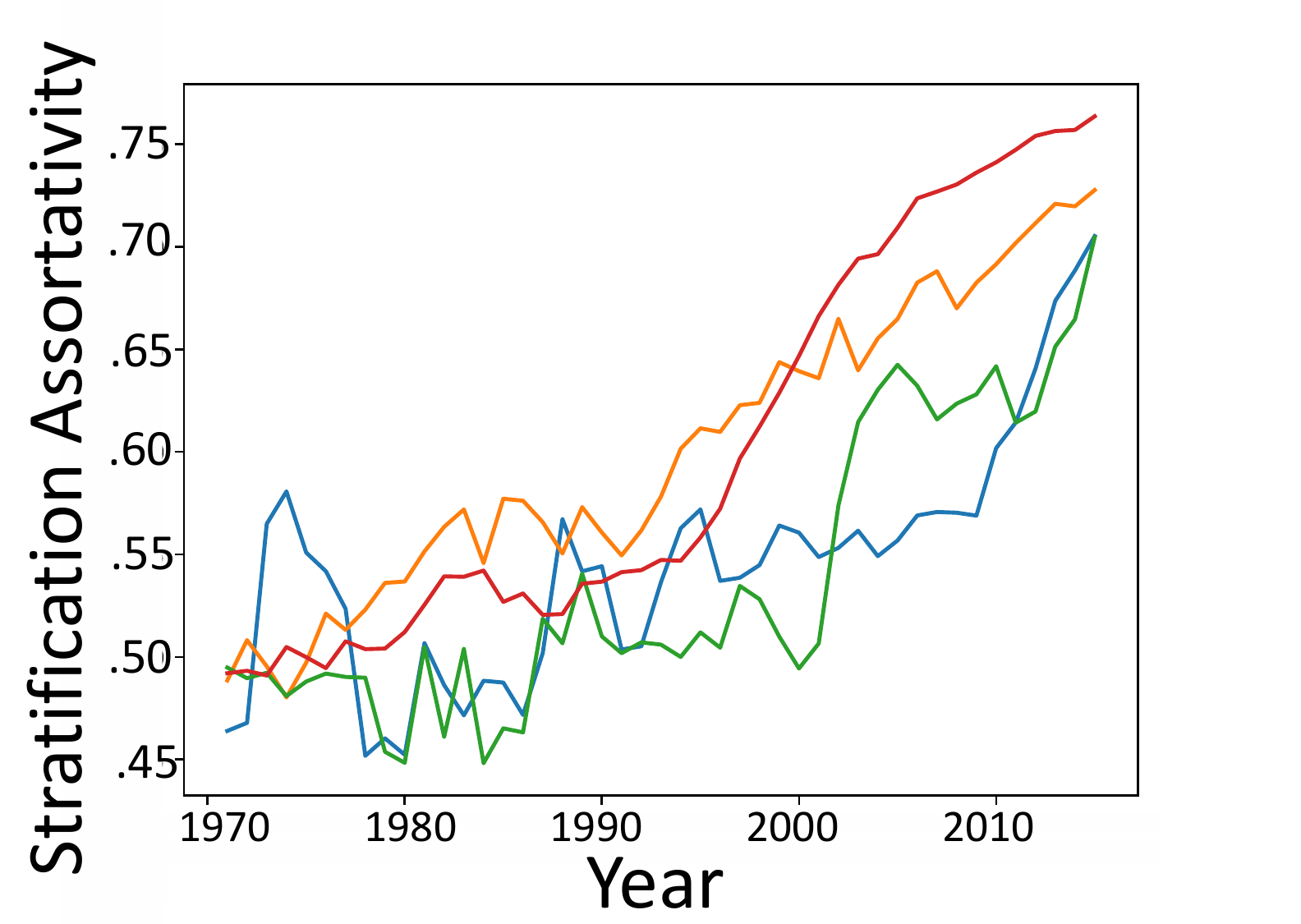}
  \caption{6-year intervals}
  \end{subfigure}
    \begin{subfigure}[b]{0.24\textwidth}  \includegraphics[width=\textwidth]{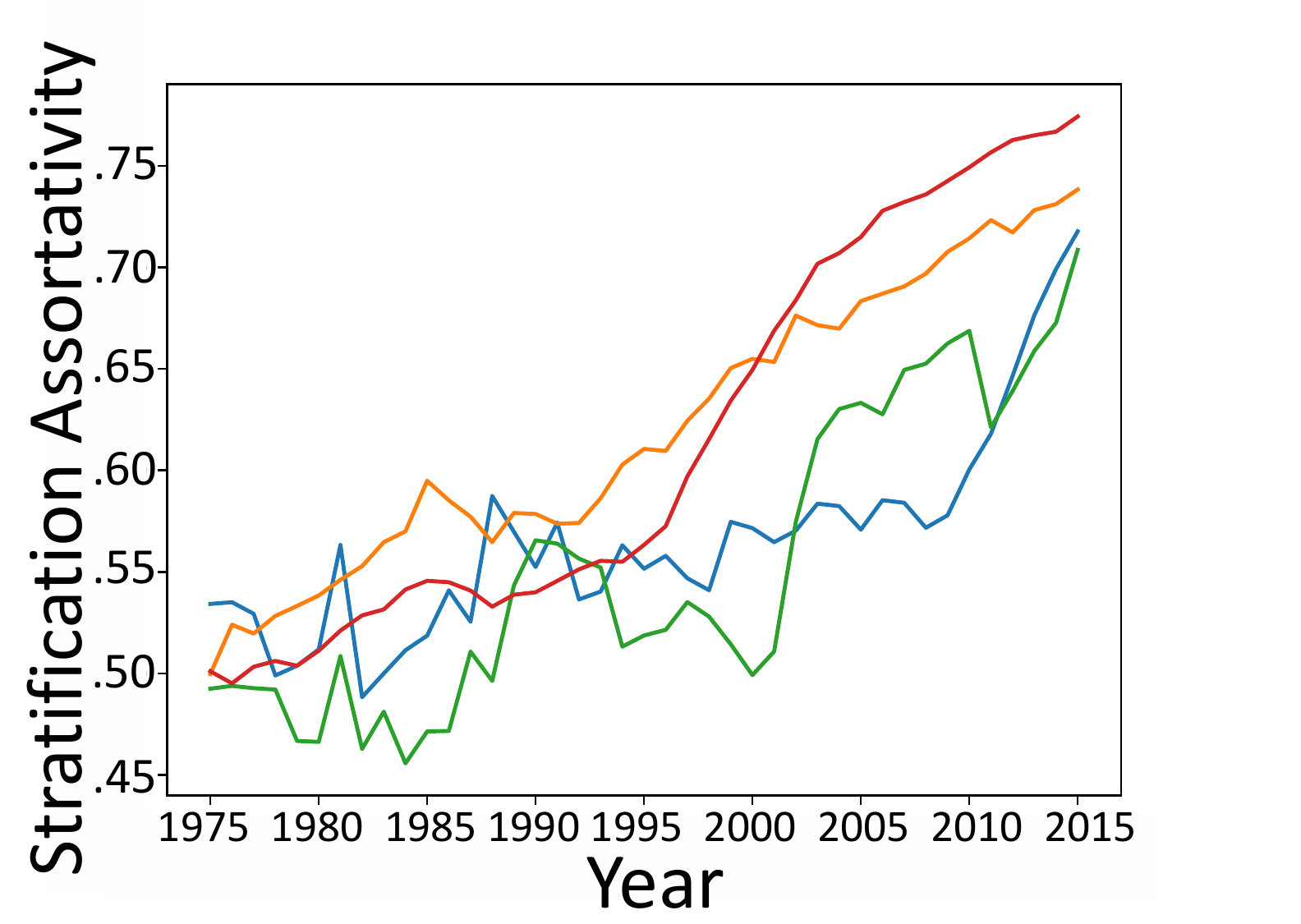}
  \caption{10-year intervals}
  \end{subfigure}
  \caption{Stratification assortativity over time for 2-year, 6-year and 10-year intervals.}
  \label{fig:SA_intervals}
  \vspace{0.5cm}
\end{figure*}

In this section, we examine  network social stratification with different numbers of classes/tiers.

to explore whether results are sensitive to the interval snapshot (5 years) used earlier, we generate networks showing patterns on 2-year, 6-year and 10-year interval snapshots. For these intervals, the general trends are identical, though values are lower for the 2-year intervals than other intervals.  This may be because 2 years is insufficiently long to capture many active collaborations, and so patterns are simply weaker overall in these snapshots. Figure~\ref{fig:SA_intervals} shows the \texttt{stratification assortativity} of these networks.

\subsection{Results}
\begin{figure*}
\vspace{-0.5cm}
  \centering
  \begin{subfigure}[b]{0.19\textwidth}  \includegraphics[width=\textwidth]{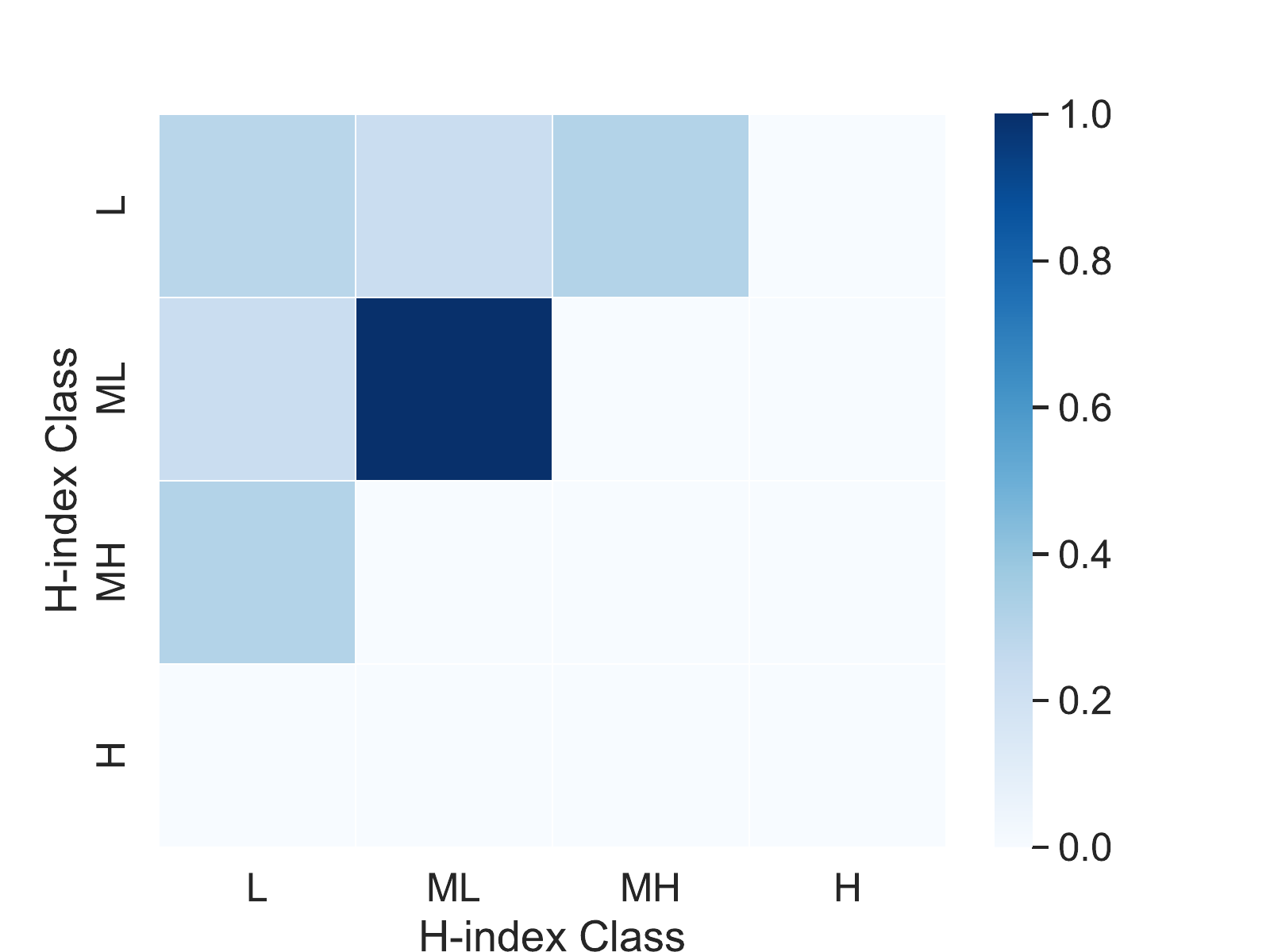}
  \caption{1966-1975}
  \end{subfigure}
  \begin{subfigure}[b]{0.19\textwidth}  \includegraphics[width=\textwidth]{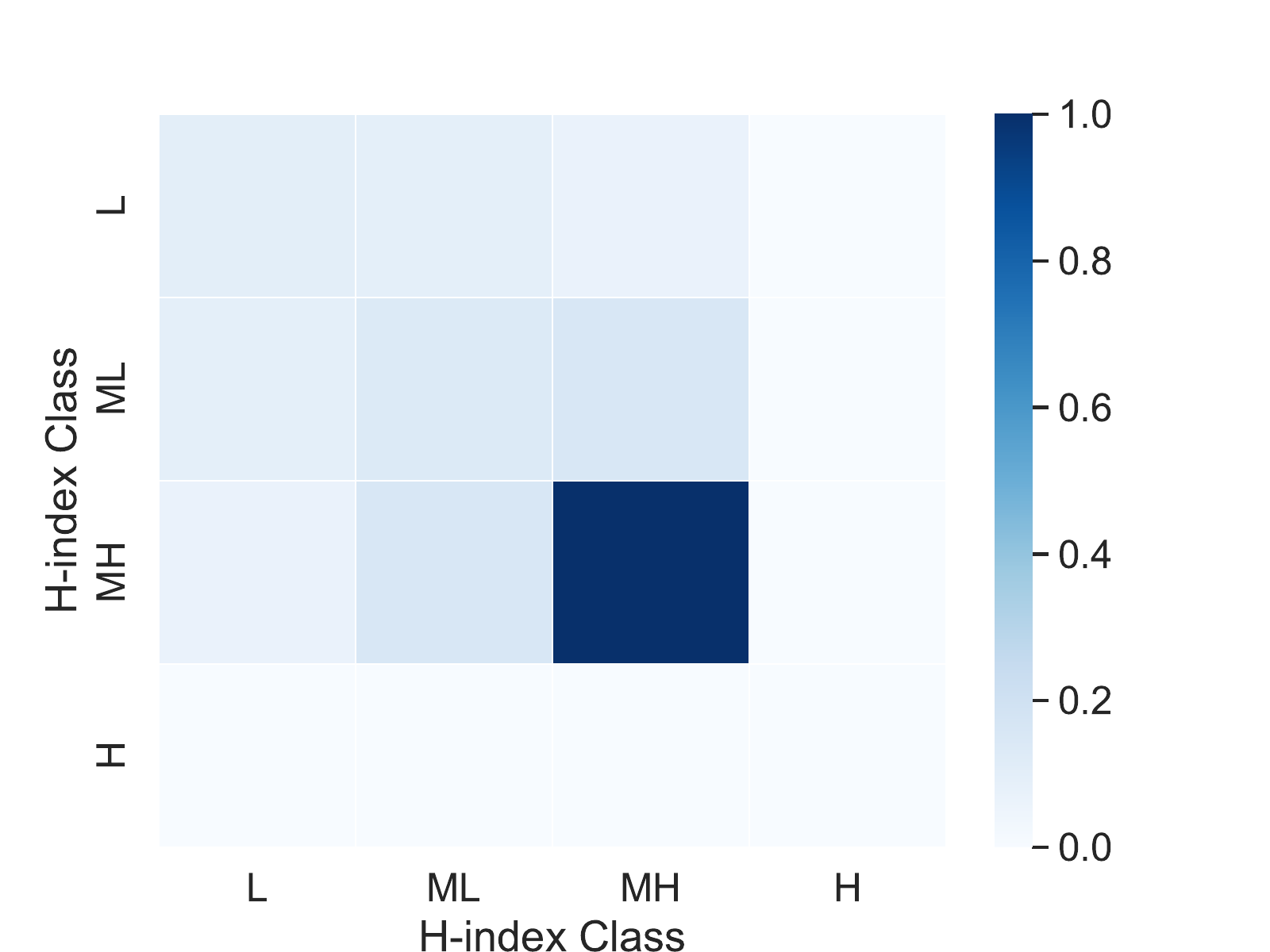}
  \caption{1976-1985}
  \end{subfigure}
    \begin{subfigure}[b]{0.19\textwidth}  \includegraphics[width=\textwidth]{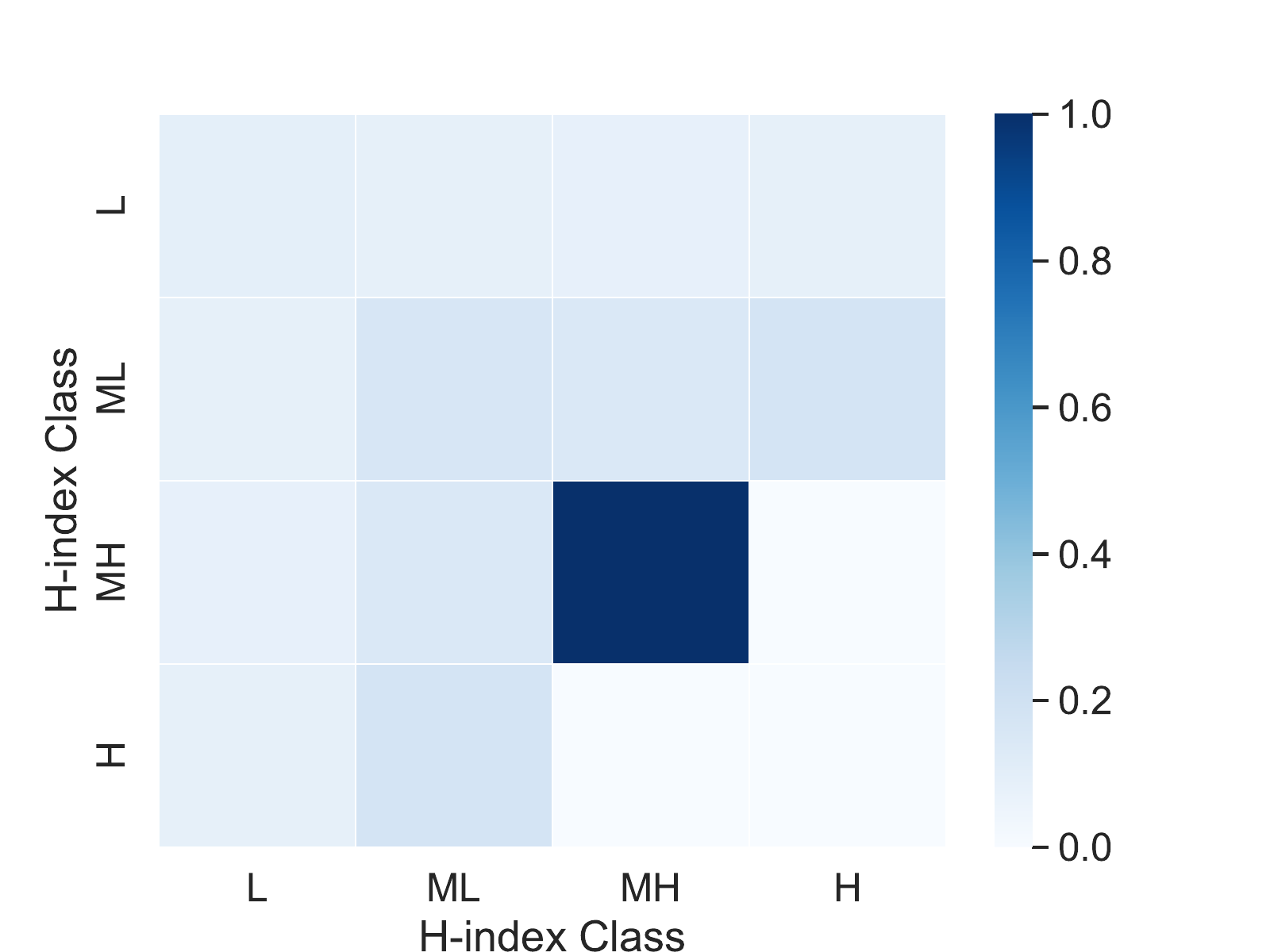}
  \caption{1986-1995}
  \end{subfigure}
    \begin{subfigure}[b]{0.19\textwidth}  \includegraphics[width=\textwidth]{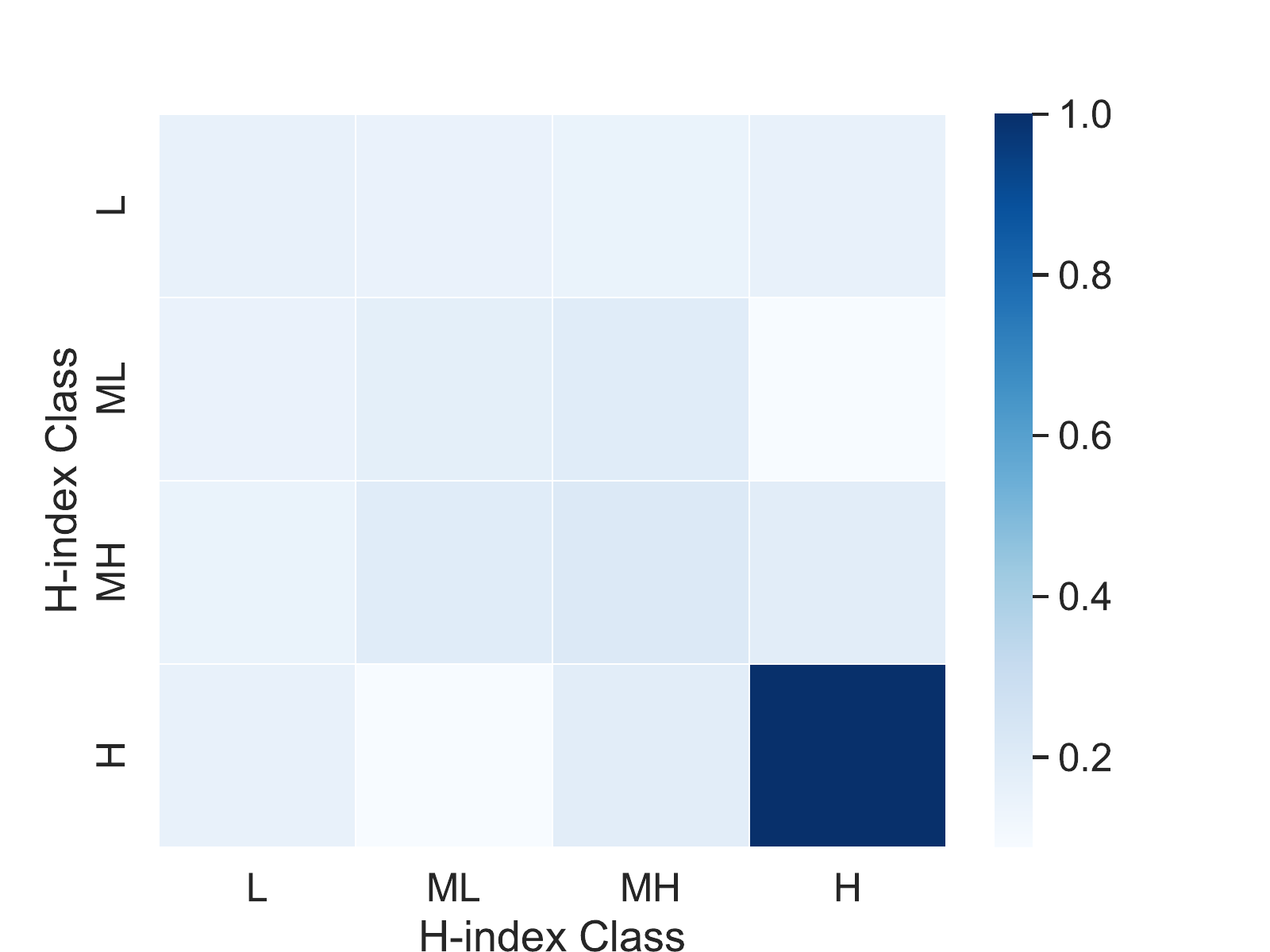}
  \caption{1996-2005}
  \end{subfigure}
    \begin{subfigure}[b]{0.19\textwidth}  \includegraphics[width=\textwidth]{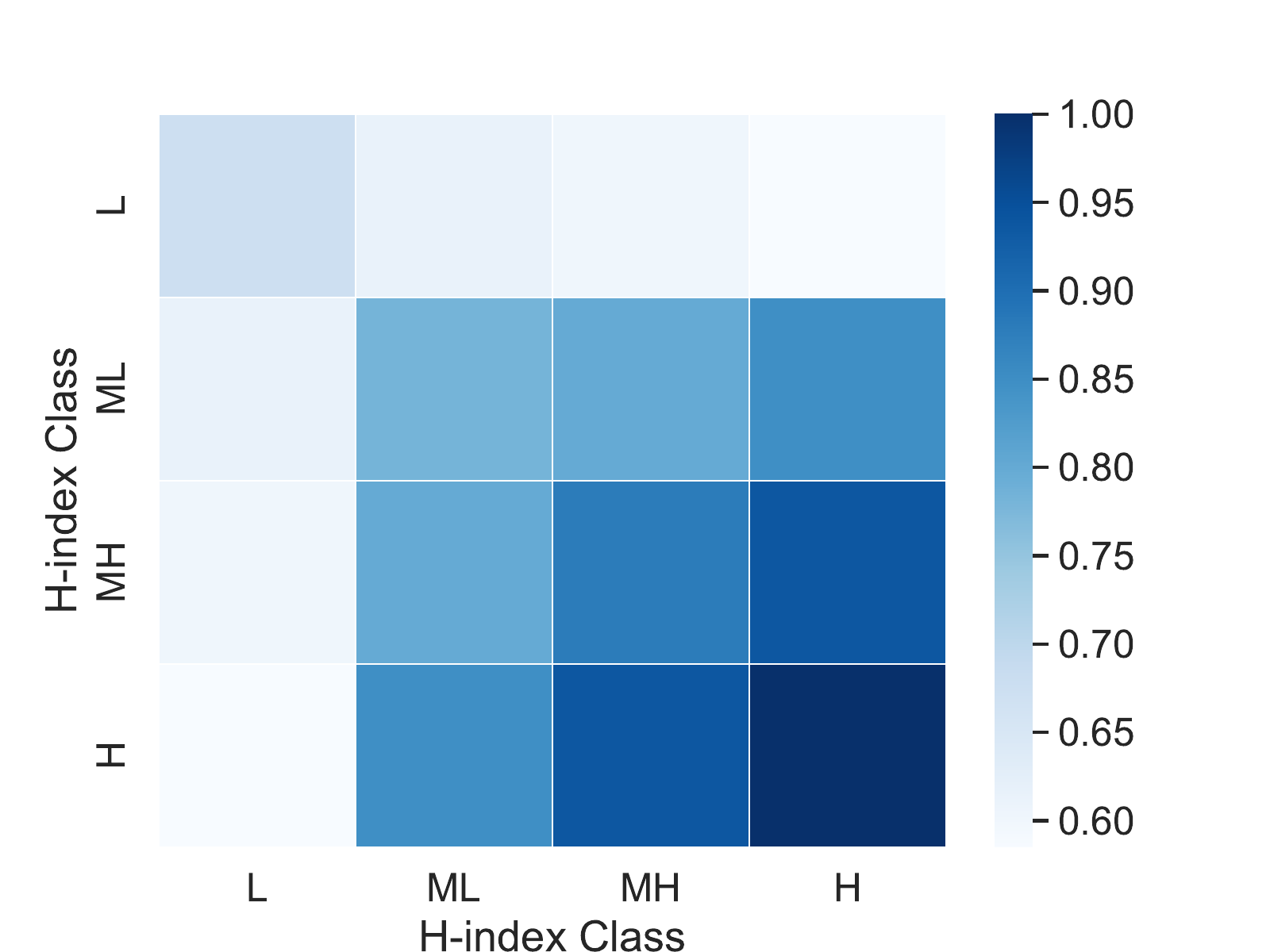}
  \caption{2006-2015}
  \end{subfigure}
  \caption{Collaborations from different h-index classes for Computational Linguistics, normalized by degree.}
  \label{fig:cor_cl}
  \vspace{0.5cm}
\end{figure*}

\begin{figure*}
\vspace{-0.5cm}
  \centering
  \begin{subfigure}[b]{0.19\textwidth}  \includegraphics[width=\textwidth]{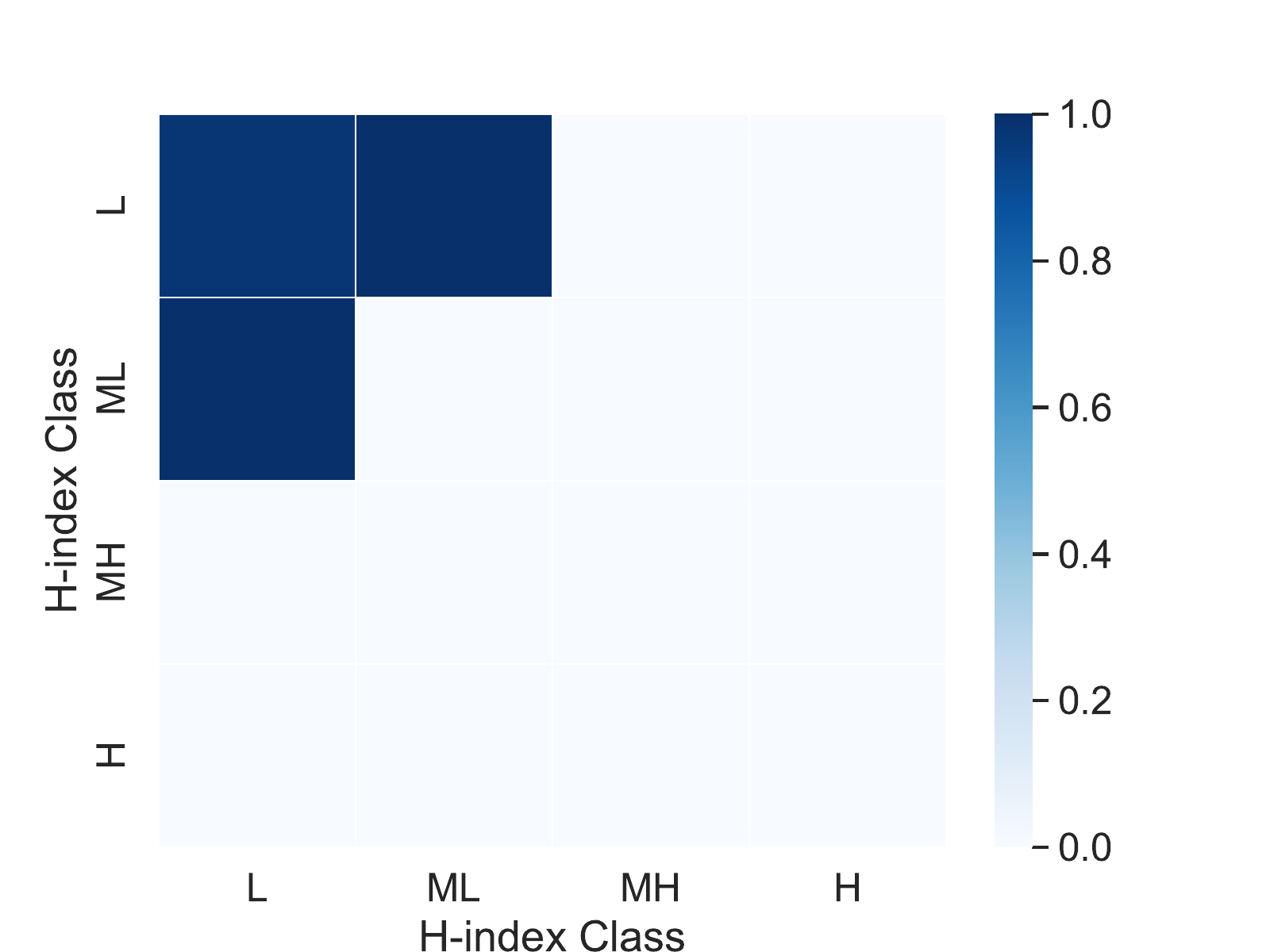}
  \caption{1966-1975}
  \end{subfigure}
  \begin{subfigure}[b]{0.19\textwidth}  \includegraphics[width=\textwidth]{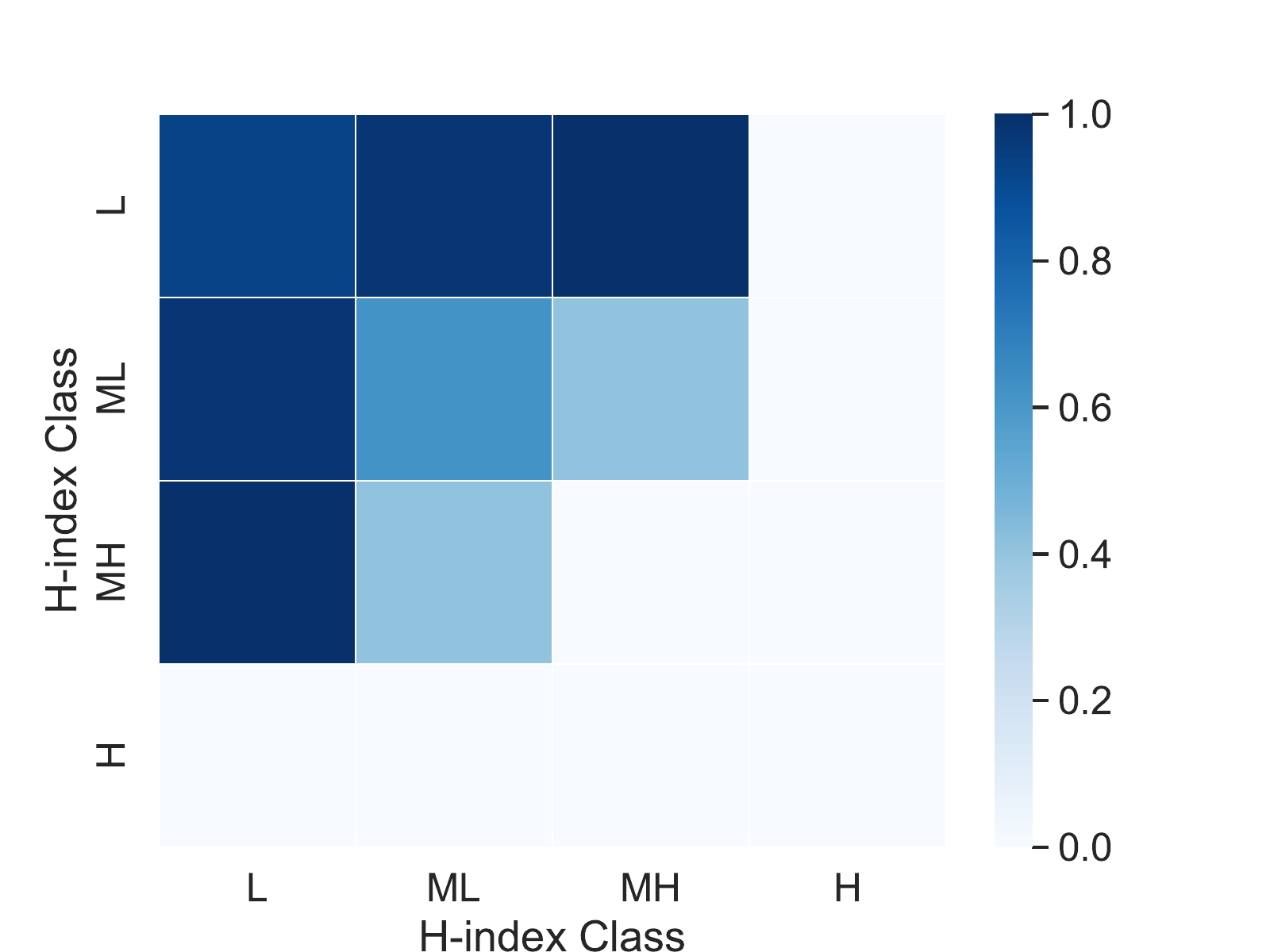}
  \caption{1976-1985}
  \end{subfigure}
    \begin{subfigure}[b]{0.19\textwidth}  \includegraphics[width=\textwidth]{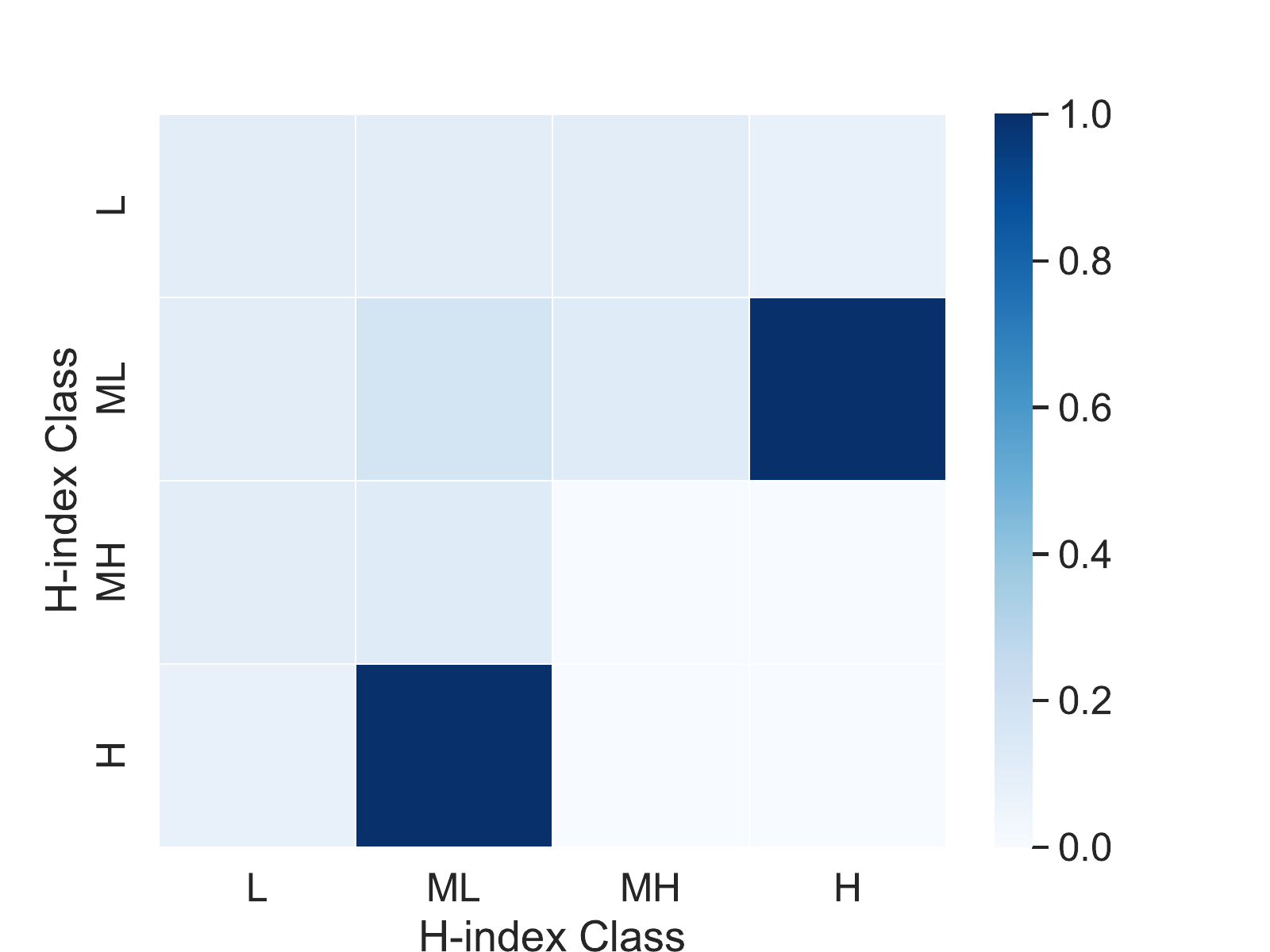}
  \caption{1986-1995}
  \end{subfigure}
    \begin{subfigure}[b]{0.19\textwidth}  \includegraphics[width=\textwidth]{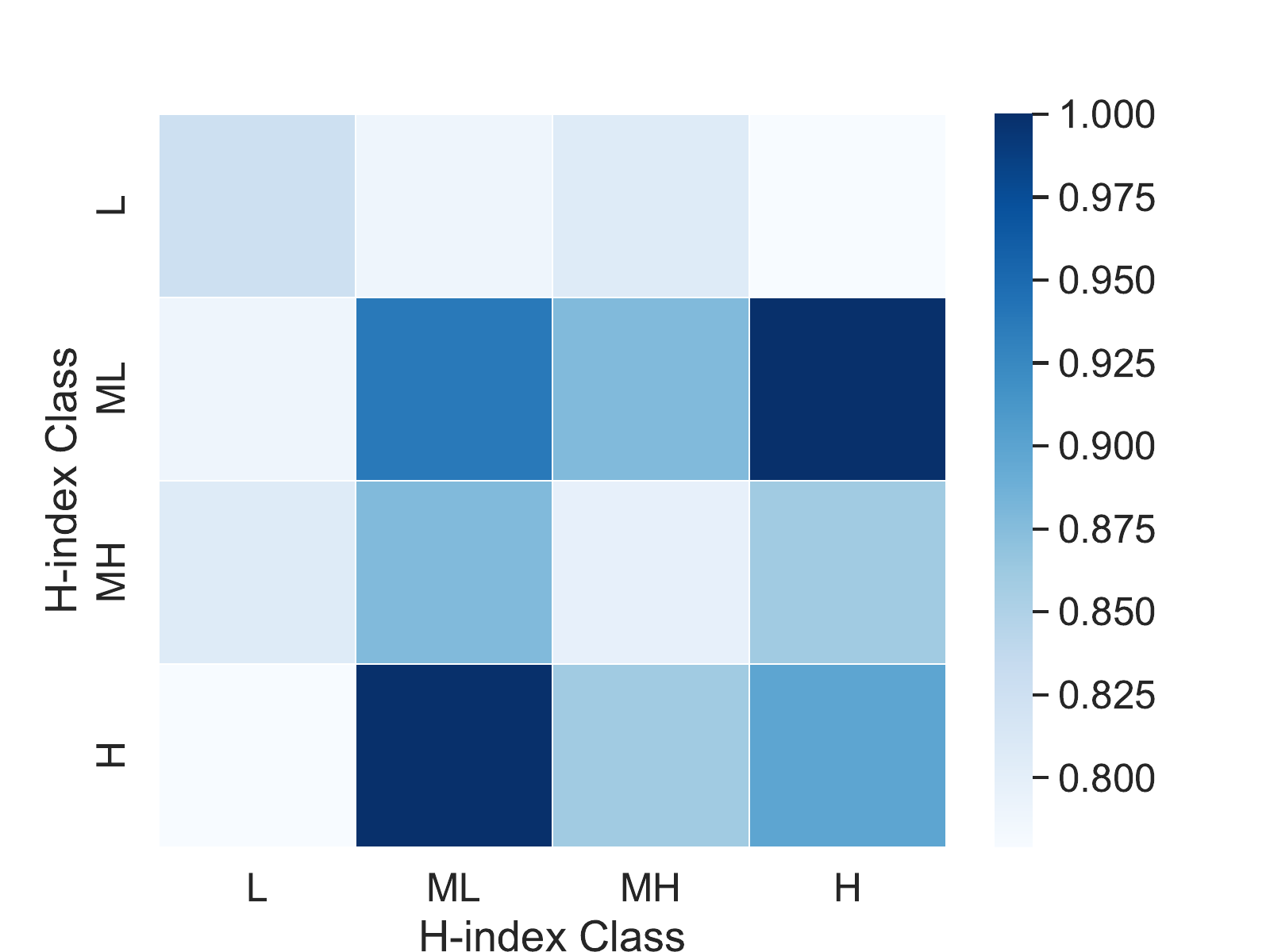}
  \caption{1996-2005}
  \end{subfigure}
    \begin{subfigure}[b]{0.19\textwidth}  \includegraphics[width=\textwidth]{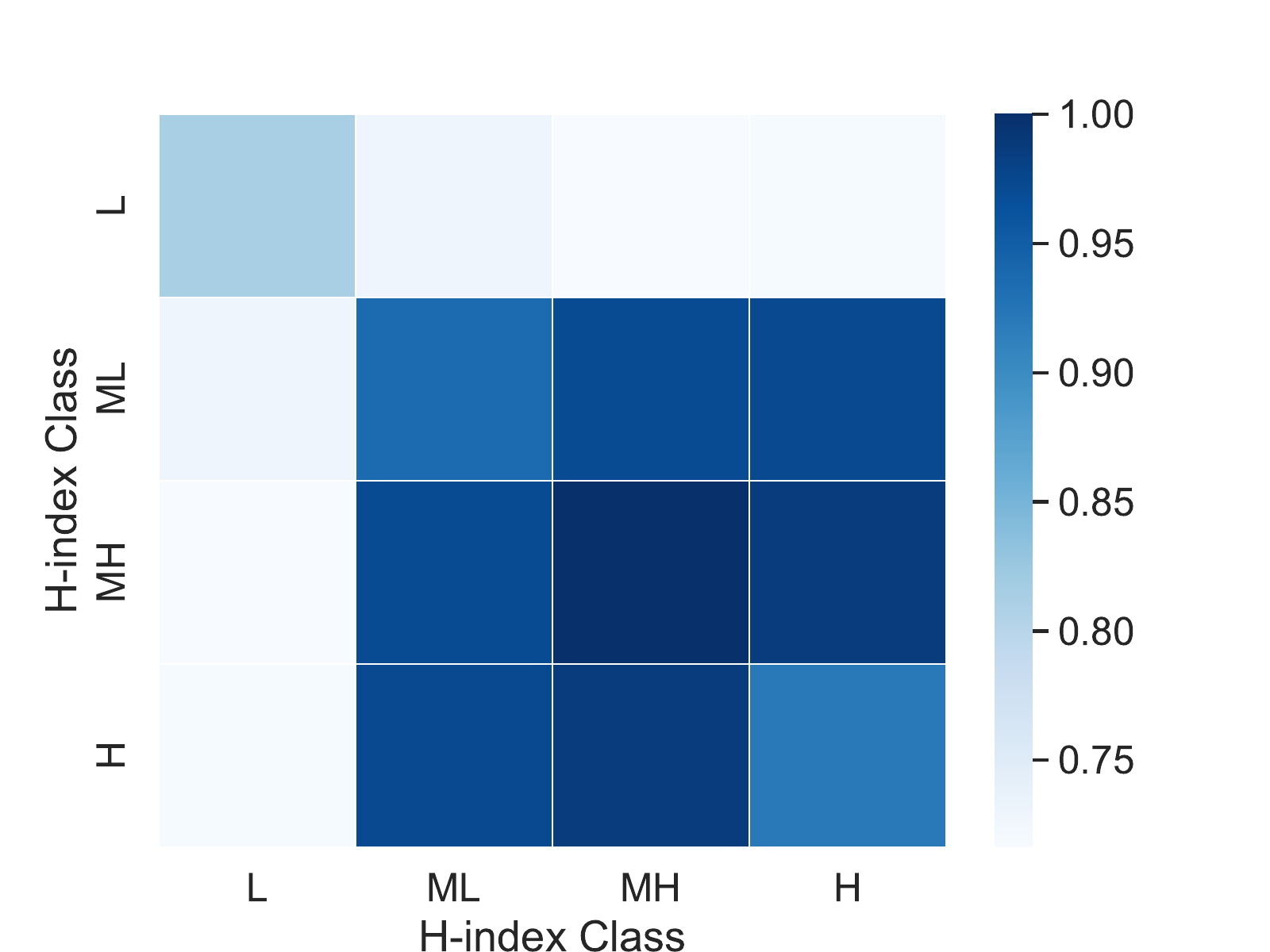}
  \caption{2006-2015}
  \end{subfigure} 
  \caption{Collaborations from different h-index classes for Computational Biology, normalized by degree.}
  \label{fig:cor_cb}
  \vspace{0.5cm}
\end{figure*}

\begin{figure*}
\vspace{-0.5cm}
  \centering
  \begin{subfigure}[b]{0.19\textwidth}  \includegraphics[width=\textwidth]{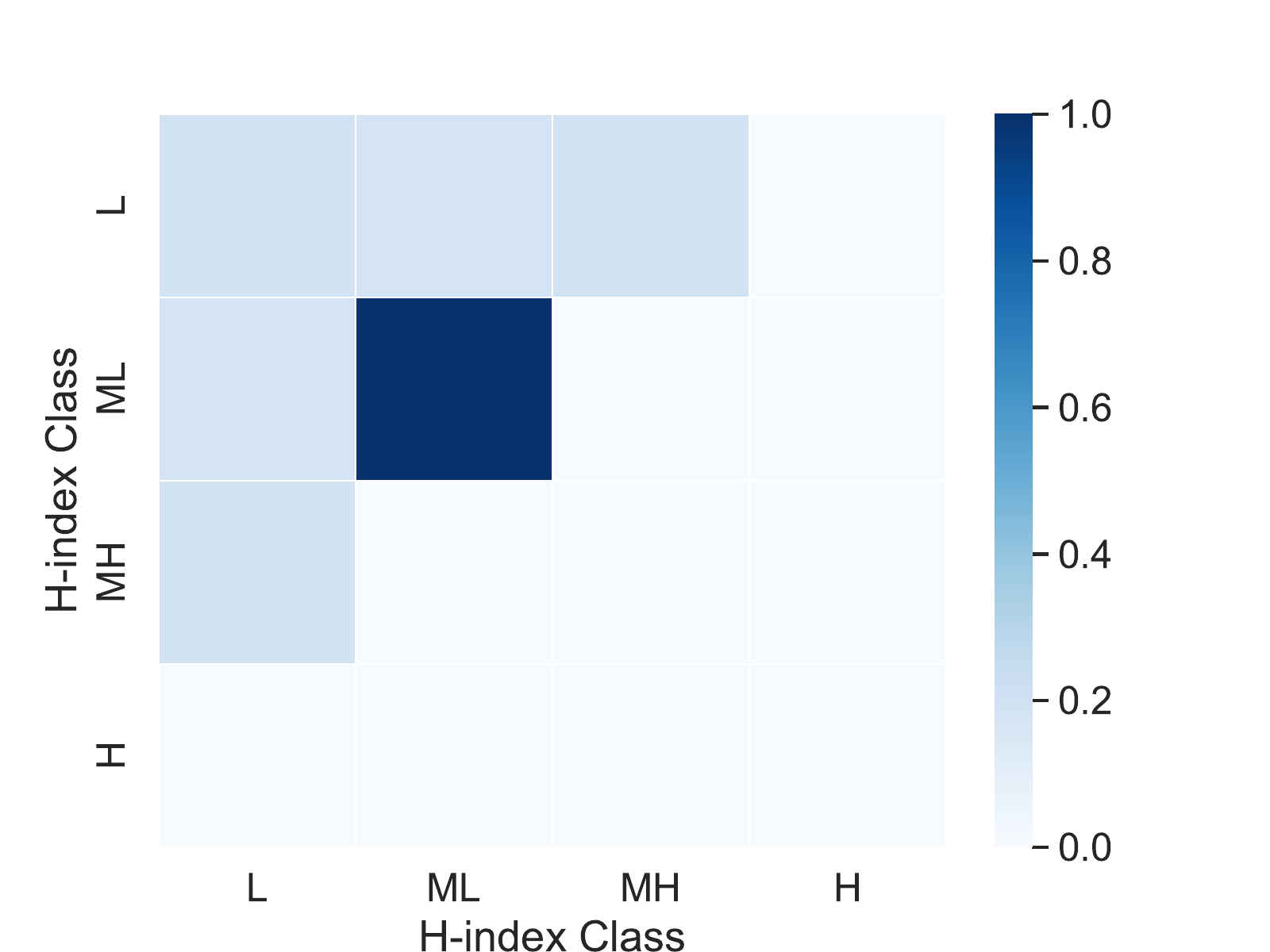}
  \caption{1966-1975}
  \end{subfigure}
  \begin{subfigure}[b]{0.19\textwidth}  \includegraphics[width=\textwidth]{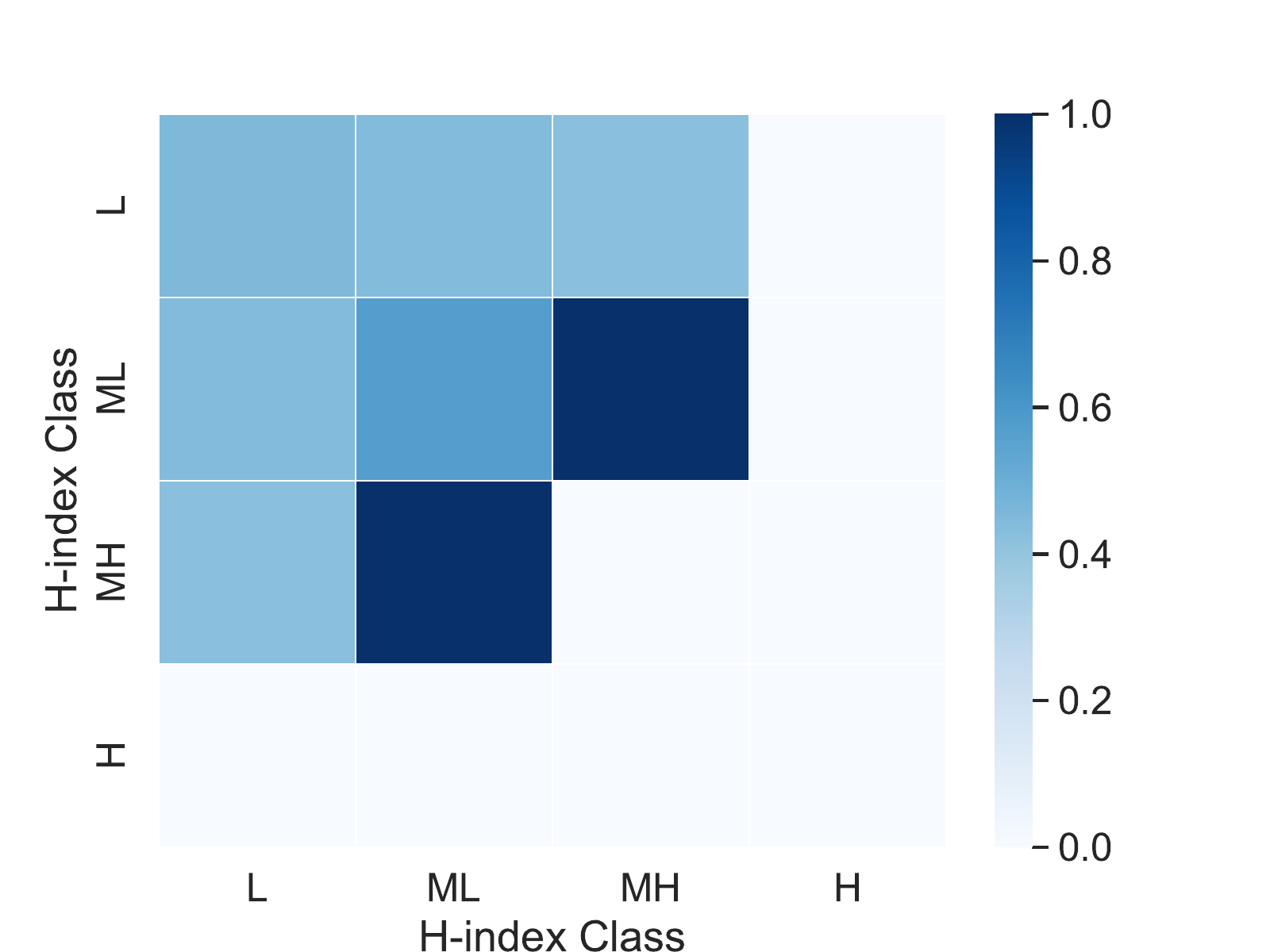}
  \caption{1976-1985}
  \end{subfigure}
    \begin{subfigure}[b]{0.19\textwidth}  \includegraphics[width=\textwidth]{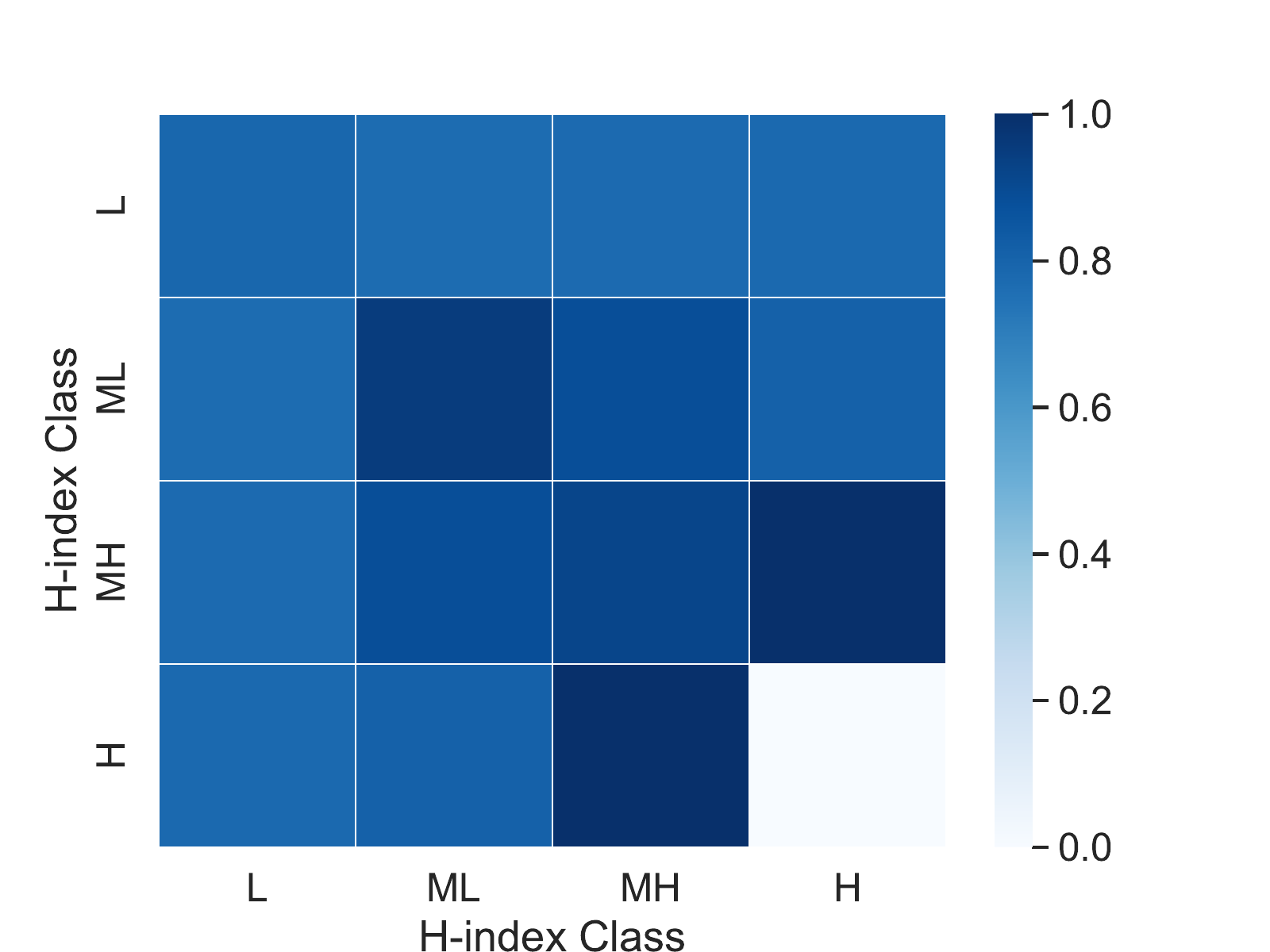}
  \caption{1986-1995}
  \end{subfigure}
    \begin{subfigure}[b]{0.19\textwidth}  \includegraphics[width=\textwidth]{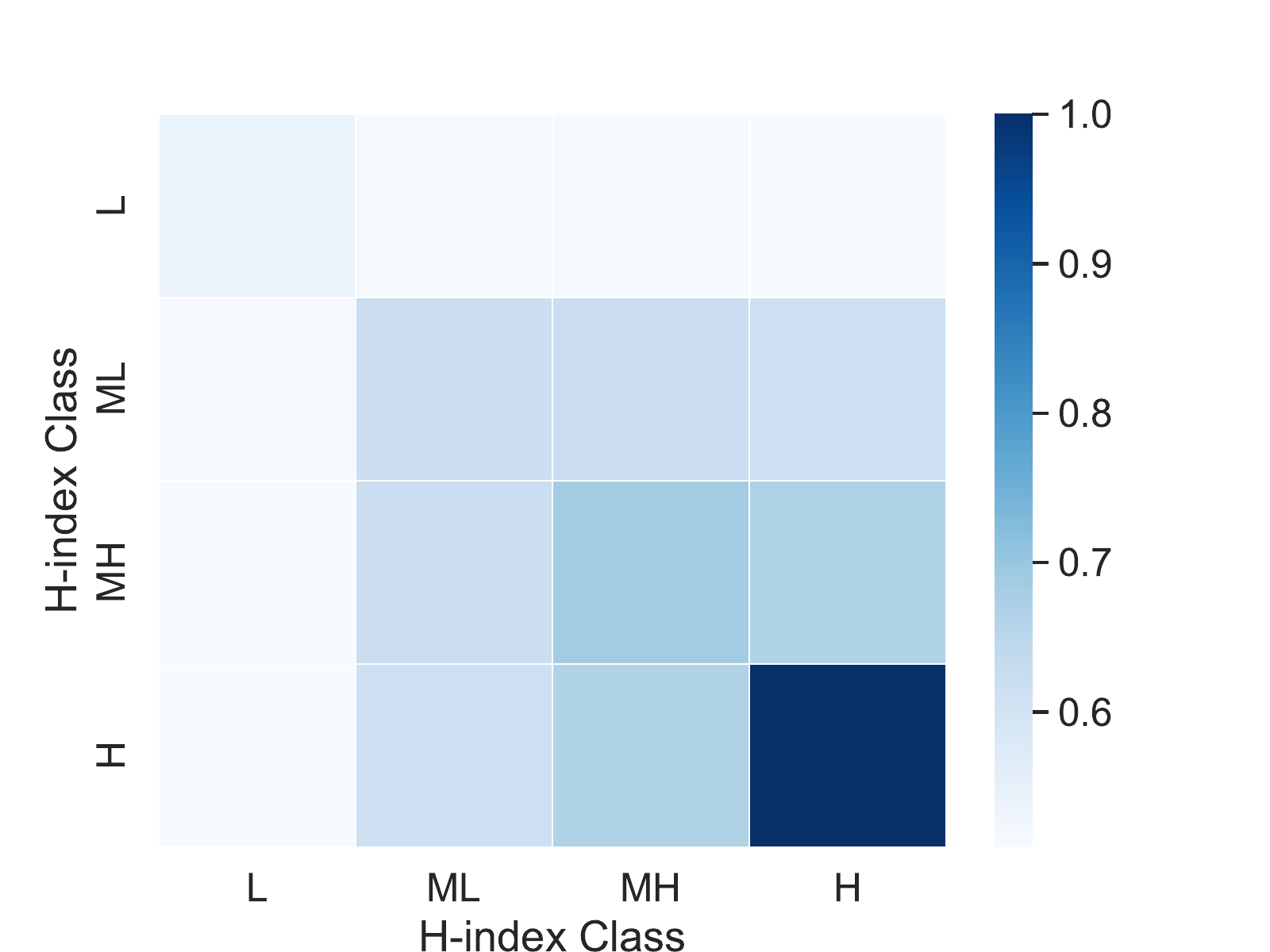}
  \caption{1996-2005}
  \end{subfigure}
    \begin{subfigure}[b]{0.19\textwidth}  \includegraphics[width=\textwidth]{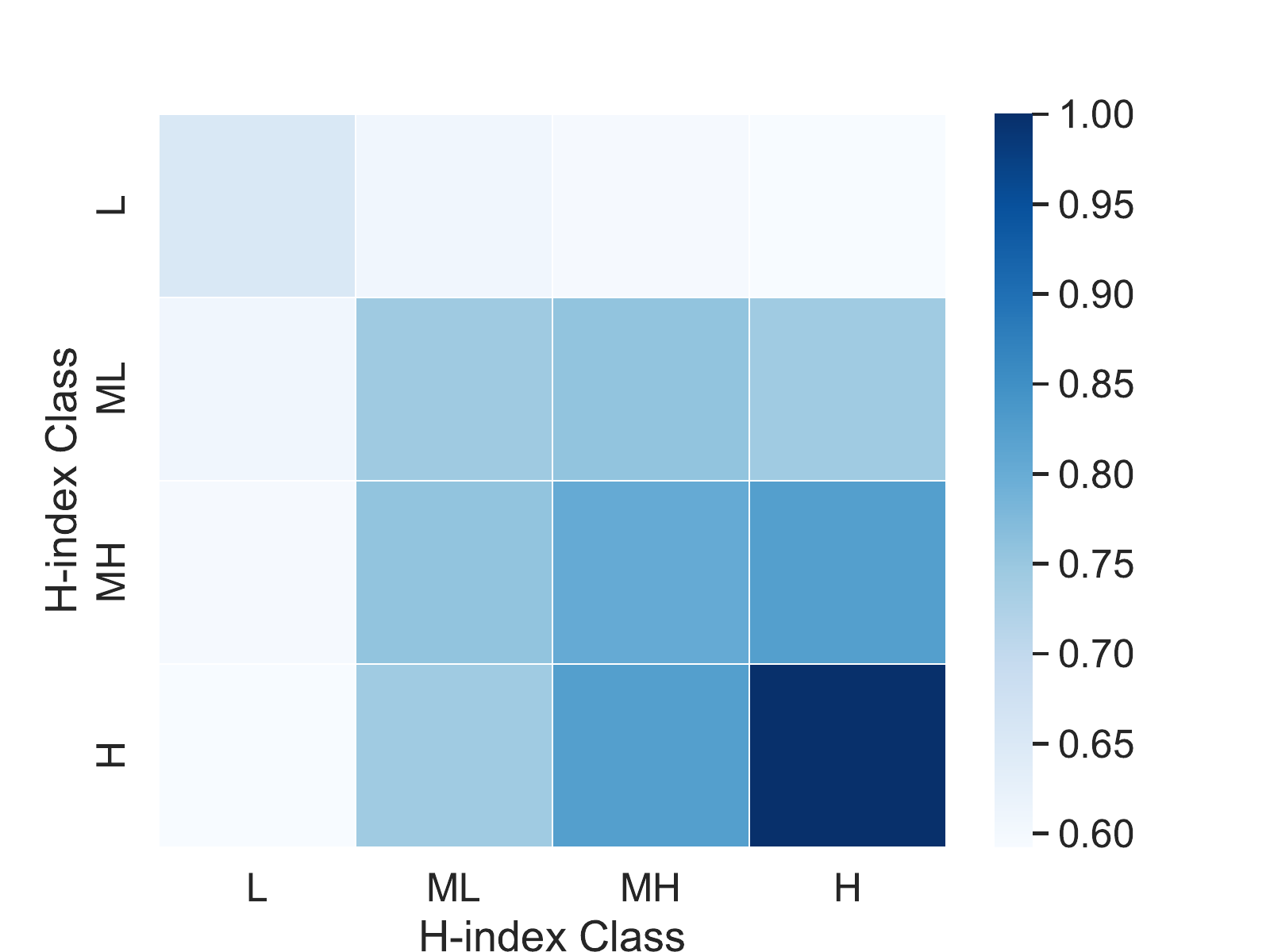}
  \caption{2006-2015}
  \end{subfigure}
  \caption{Collaborations from different h-index classes for Biomedical Engineering, normalized by degree.}
  \label{fig:cor_be}
  \vspace{0.5cm}
\end{figure*}

Figures~\ref{fig:cor_cl} to \ref{fig:cor_be} shows a heatmap describing results in the Computational Linguistics, Computational Biology, and Biomedical Engineering fields respectively. field. These plot shows the frequency of collaborations between classes of authors. Here, each cell $(c_i, c_j)$ is the number of connections from class $c_i$ to class $c_j$, ($|(c_i,c_j)|$), normalized by the number of connections of the two classes ($cell (c_i, c_j) = \frac{|(c_i, c_j)|}{|c_i|\cdot|c_j|}$, where $|c_i|$ is the number of connections where at least one side is in class $c_i$). The $x$-axis and $y$-axis show the class of h-indices (like before, we divided the h-index scores into 4 classes).

\begin{figure*}
  \centering
  \begin{subfigure}[b]{0.19\textwidth}  \includegraphics[width=\textwidth]{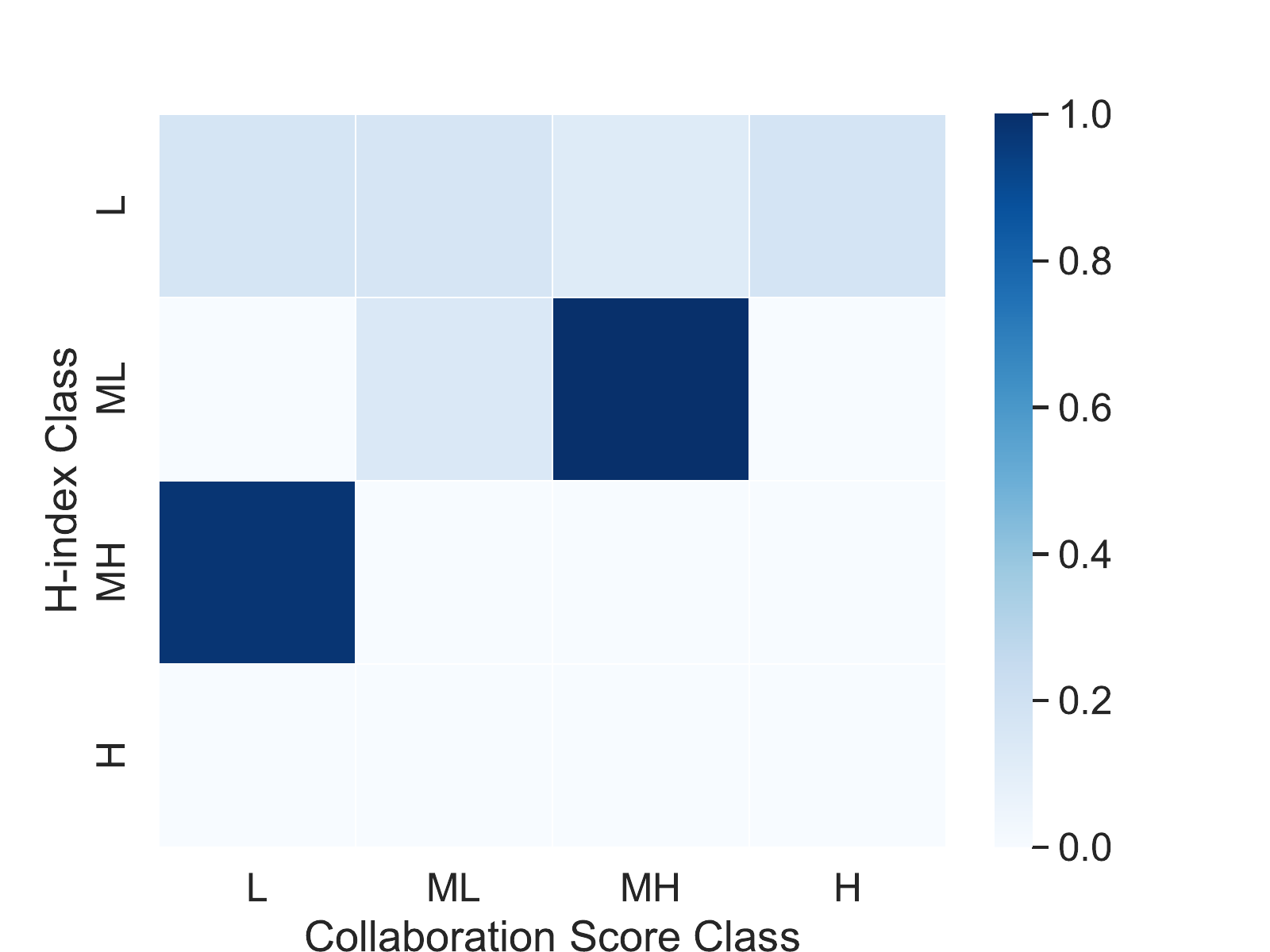}
  \caption{1966-1975}
  \end{subfigure}
  \begin{subfigure}[b]{0.19\textwidth}  \includegraphics[width=\textwidth]{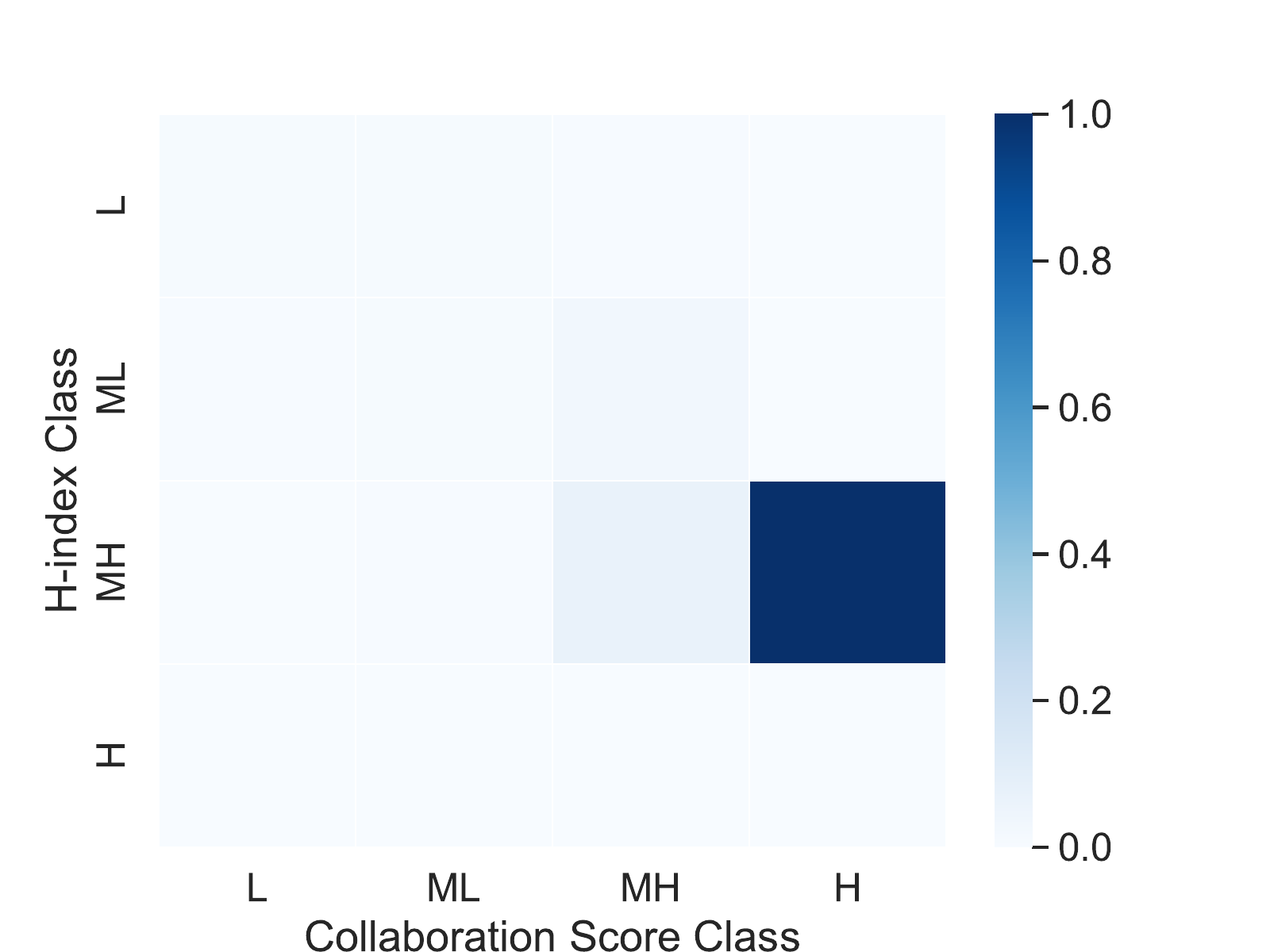}
  \caption{1976-1985}
  \end{subfigure}
    \begin{subfigure}[b]{0.19\textwidth}  \includegraphics[width=\textwidth]{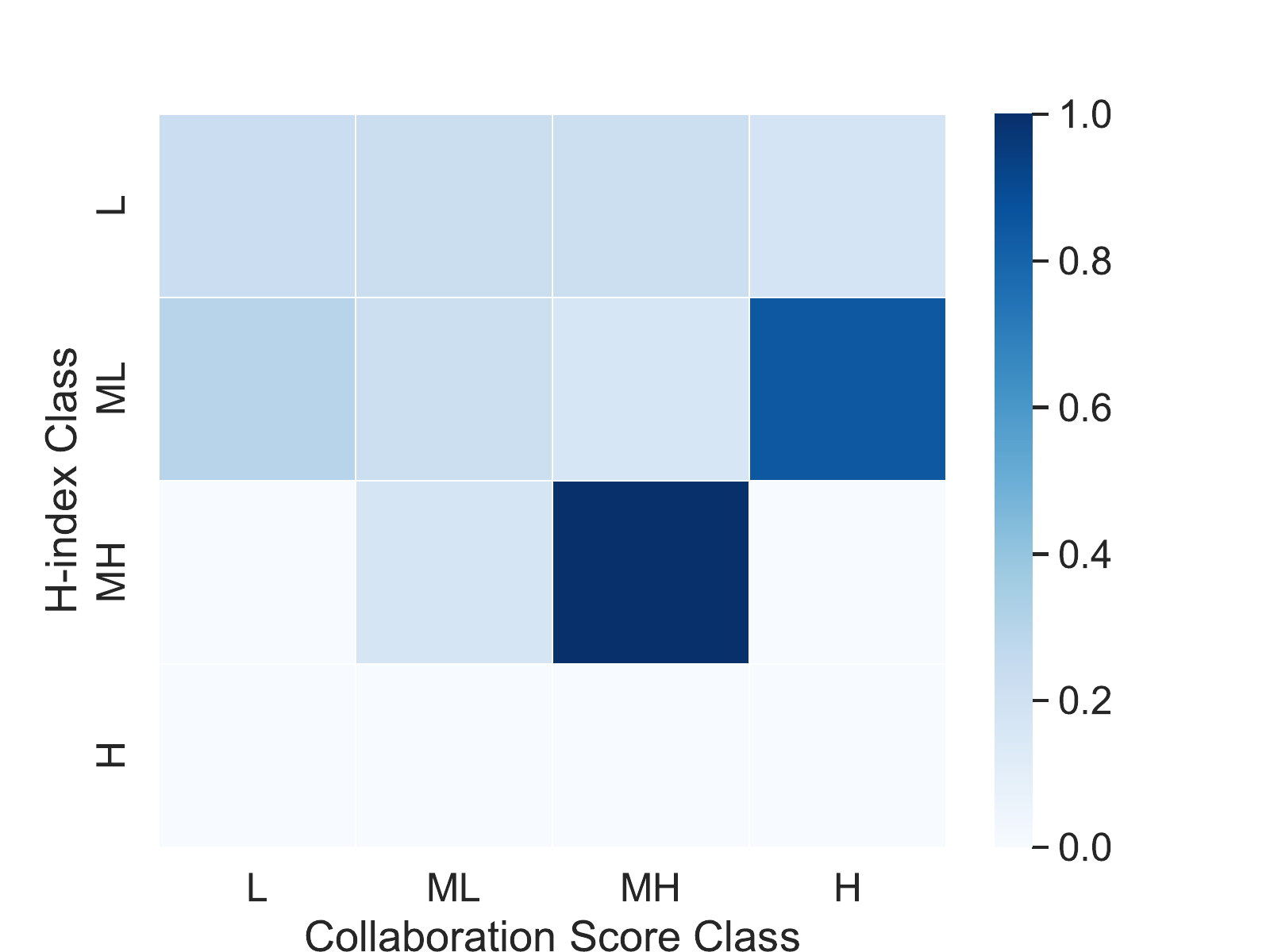}
  \caption{1986-1995}
  \end{subfigure}
    \begin{subfigure}[b]{0.19\textwidth}  \includegraphics[width=\textwidth]{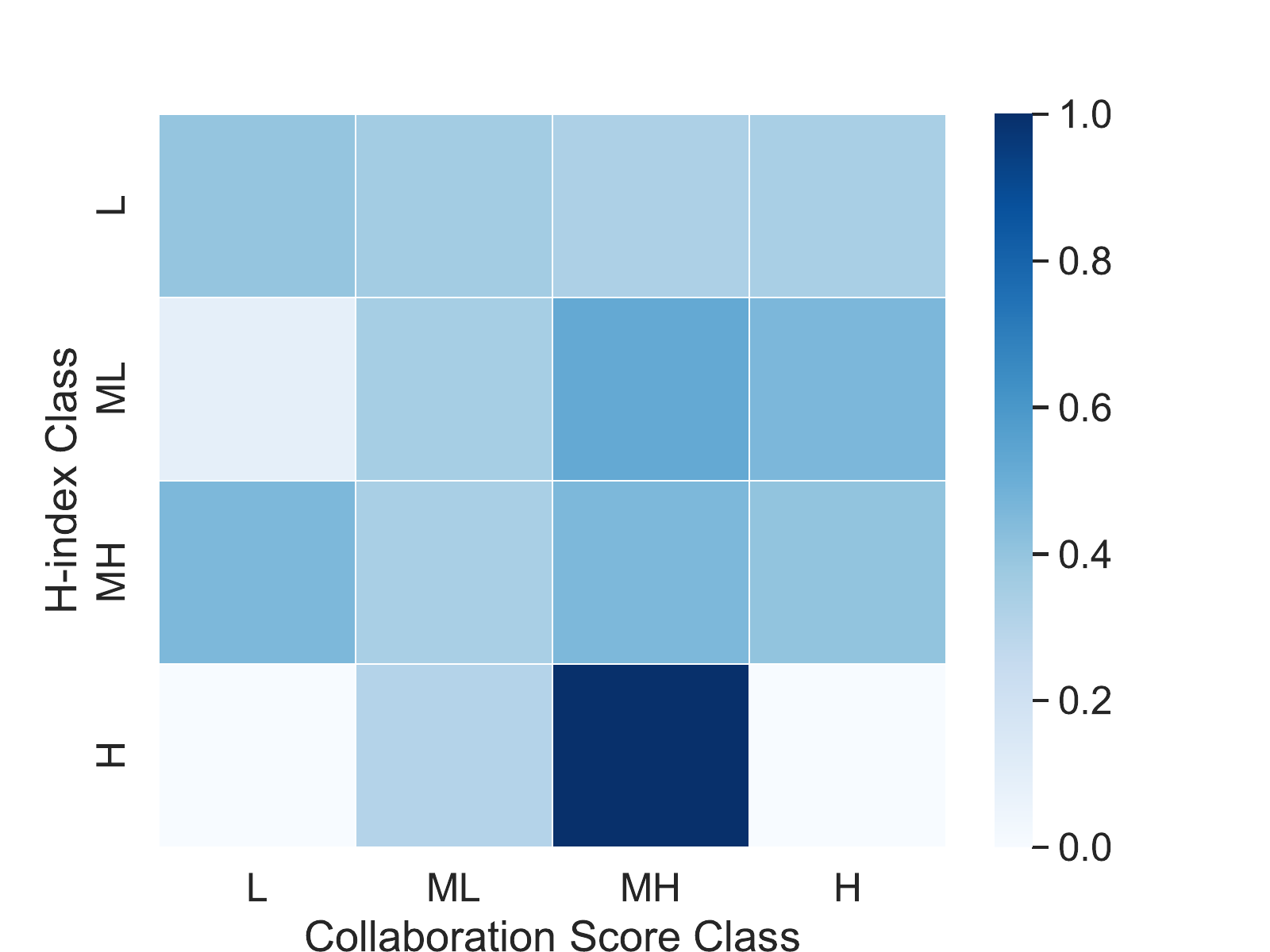}
  \caption{1996-2005}
  \end{subfigure}
    \begin{subfigure}[b]{0.196\textwidth}  \includegraphics[width=\textwidth]{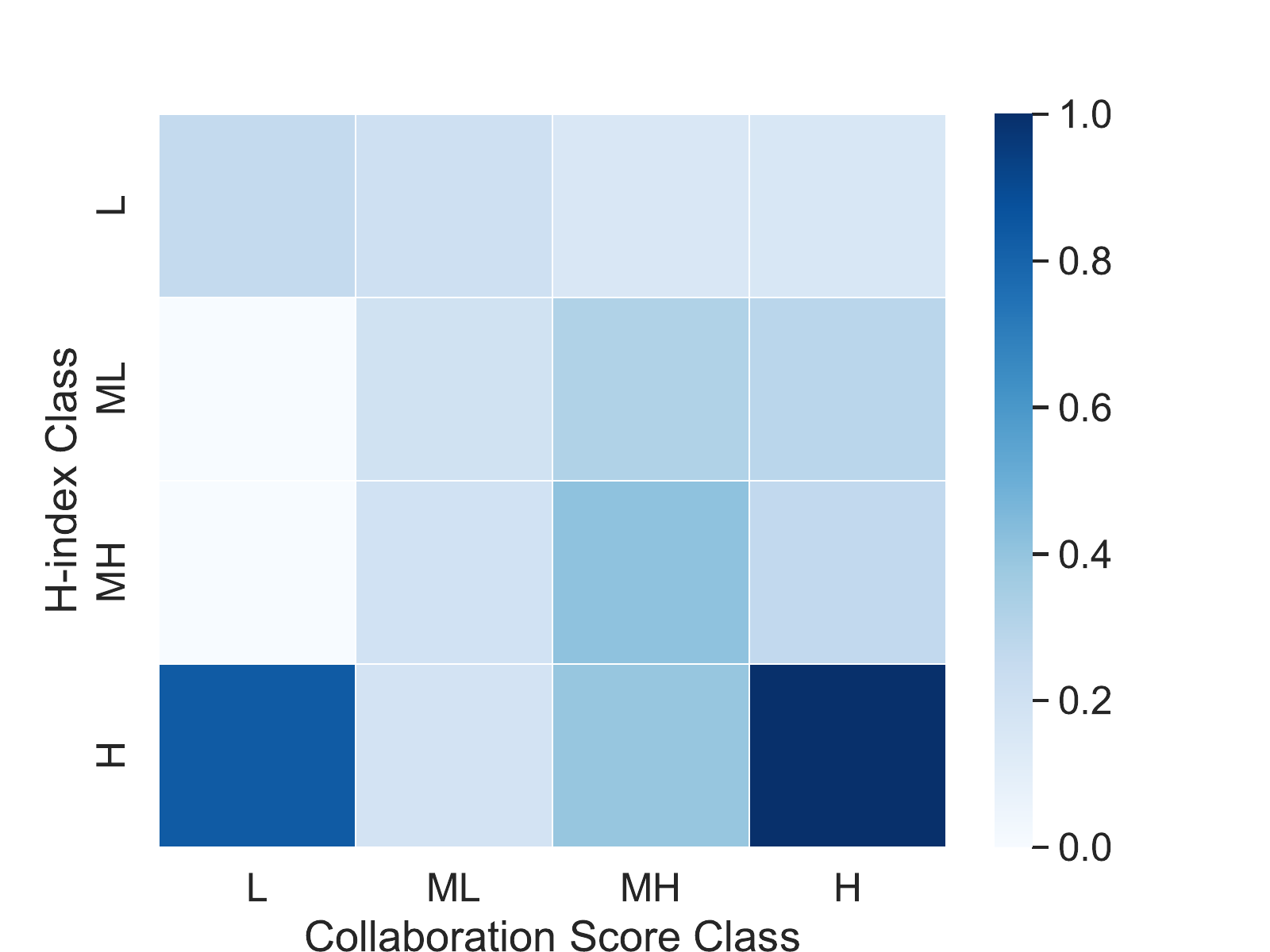}
  \caption{2006-2015}
  \end{subfigure}
  \caption{Relationship between entrance collaboration score of nodes and their h-index after 10 years for Computational Linguistics}
  \label{fig:cor_cl_se}
  \vspace{0.5cm}
\end{figure*}

\begin{figure*}
  \centering
  \begin{subfigure}[b]{0.19\textwidth}  \includegraphics[width=\textwidth]{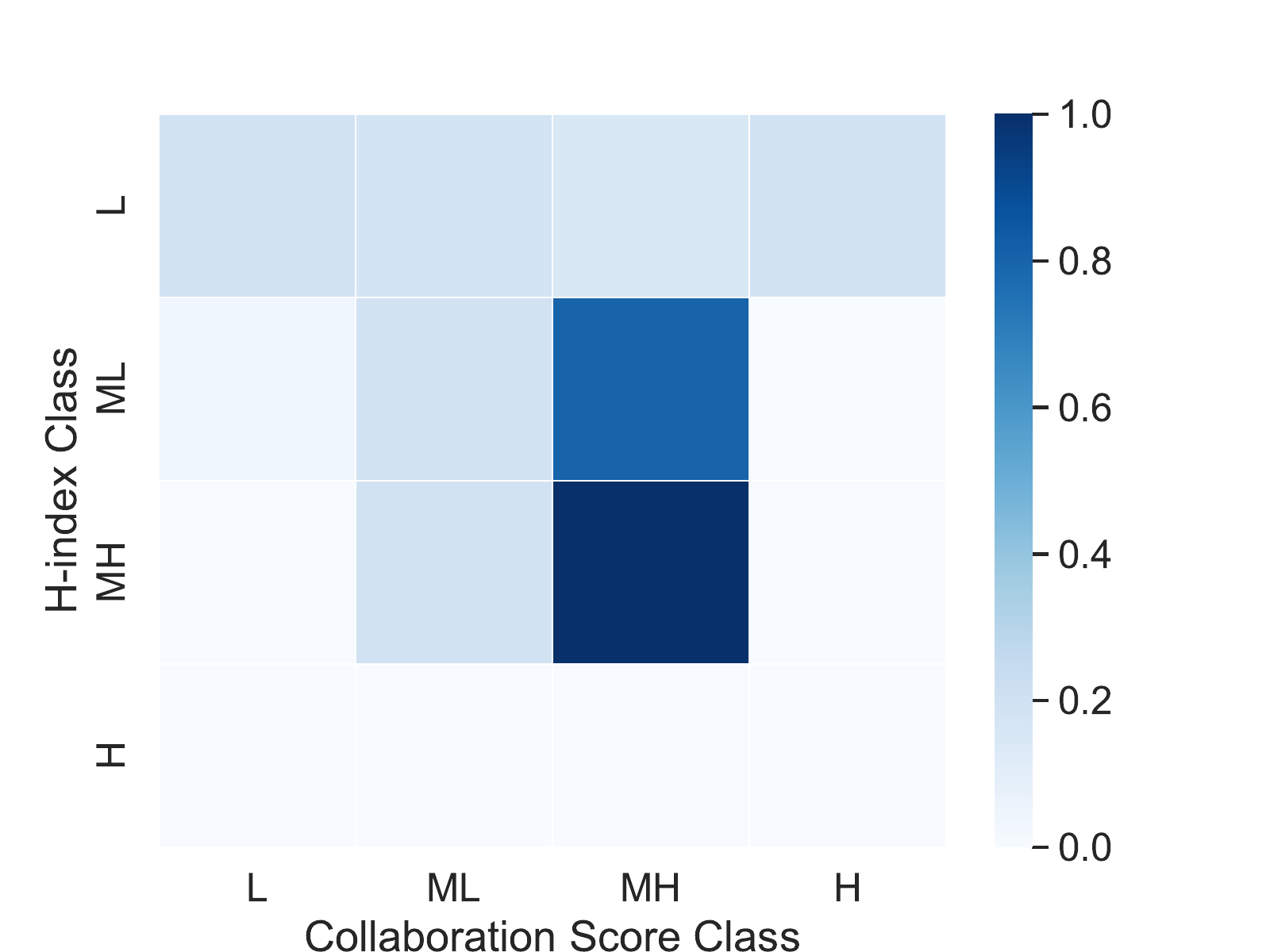}
  \caption{1966-1975}
  \end{subfigure}
  \begin{subfigure}[b]{0.19\textwidth}  \includegraphics[width=\textwidth]{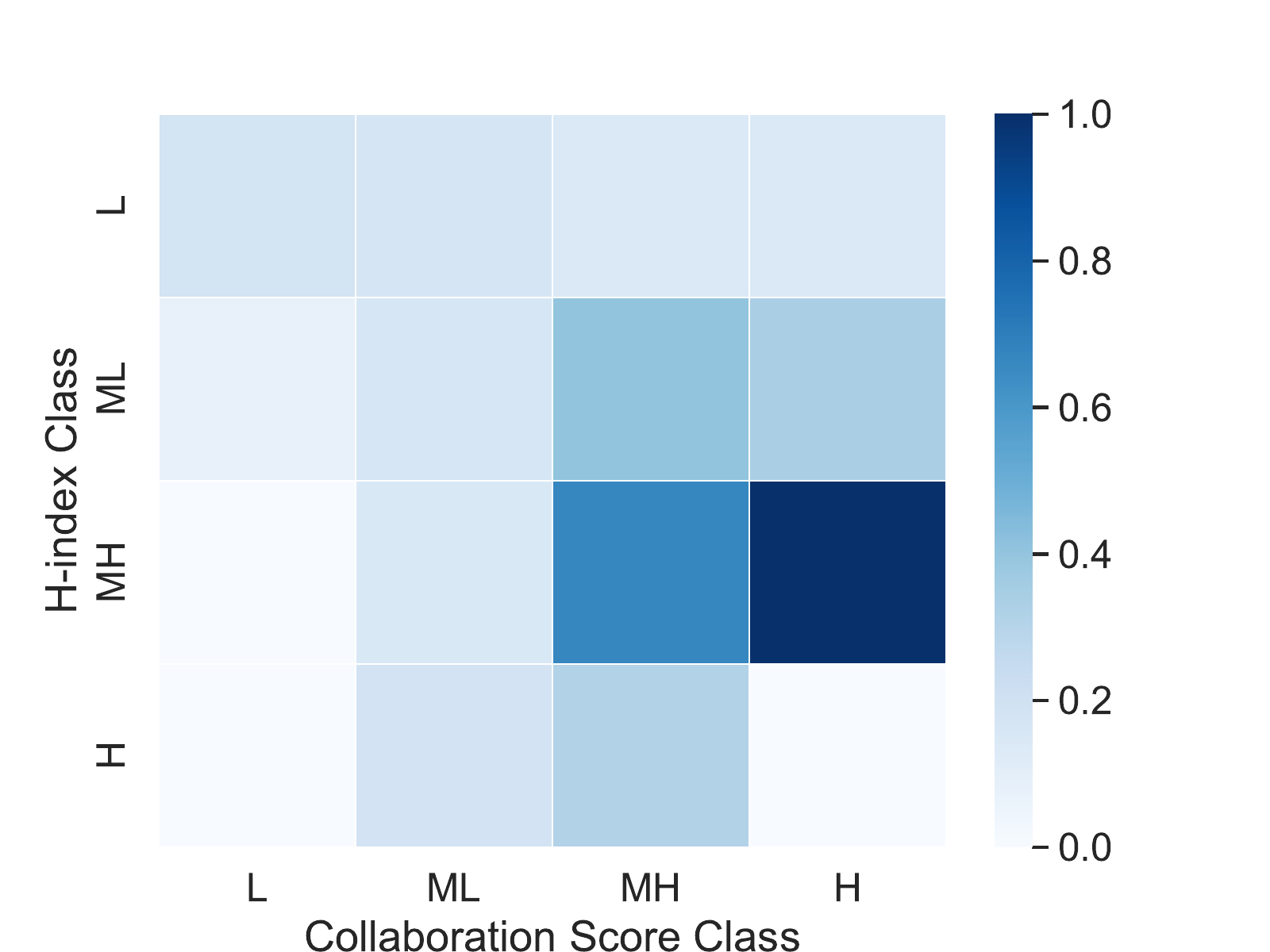}
  \caption{1976-1985}
  \end{subfigure}
    \begin{subfigure}[b]{0.19\textwidth}  \includegraphics[width=\textwidth]{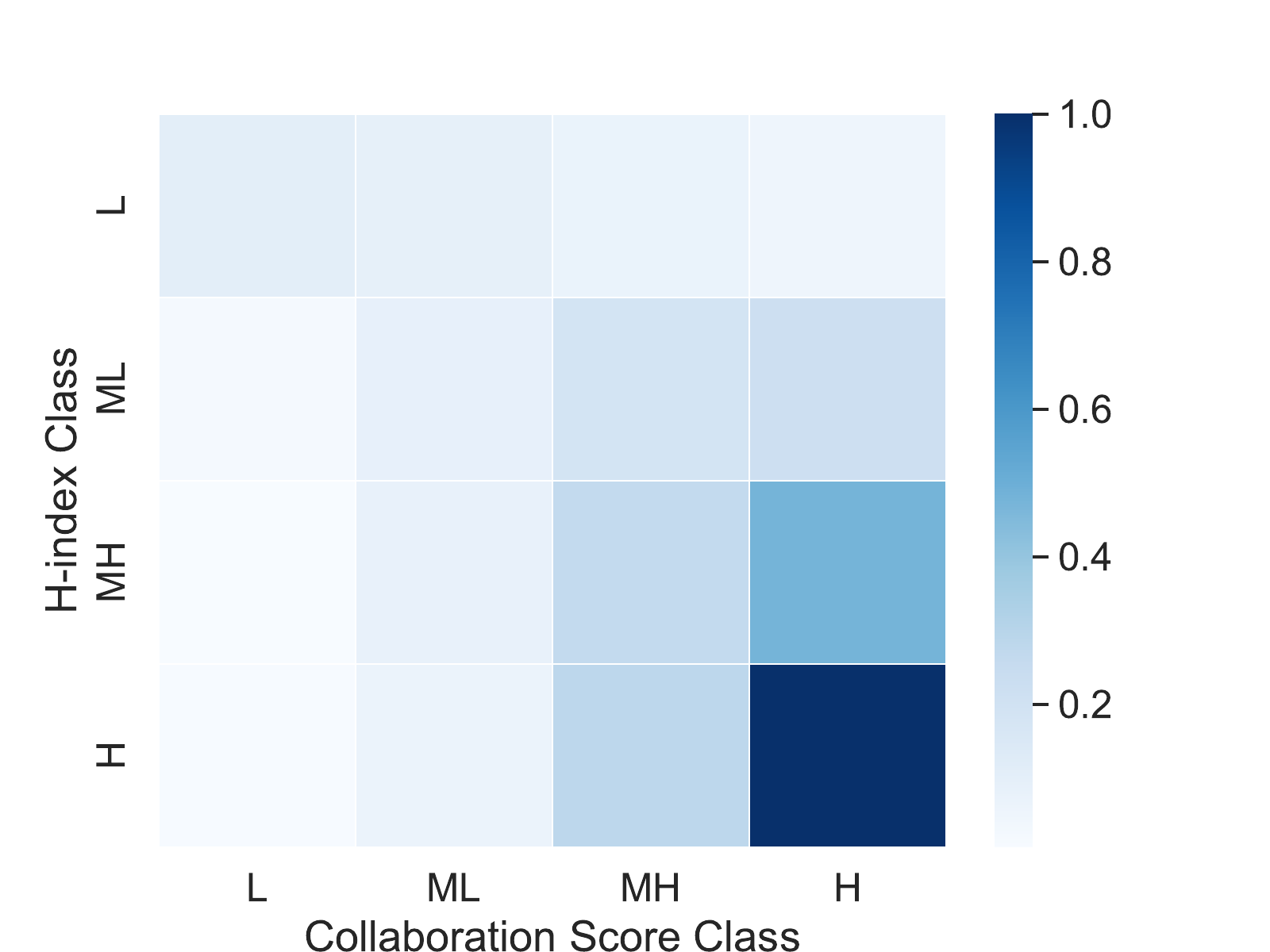}
  \caption{1986-1995}
  \end{subfigure}
    \begin{subfigure}[b]{0.19\textwidth}  \includegraphics[width=\textwidth]{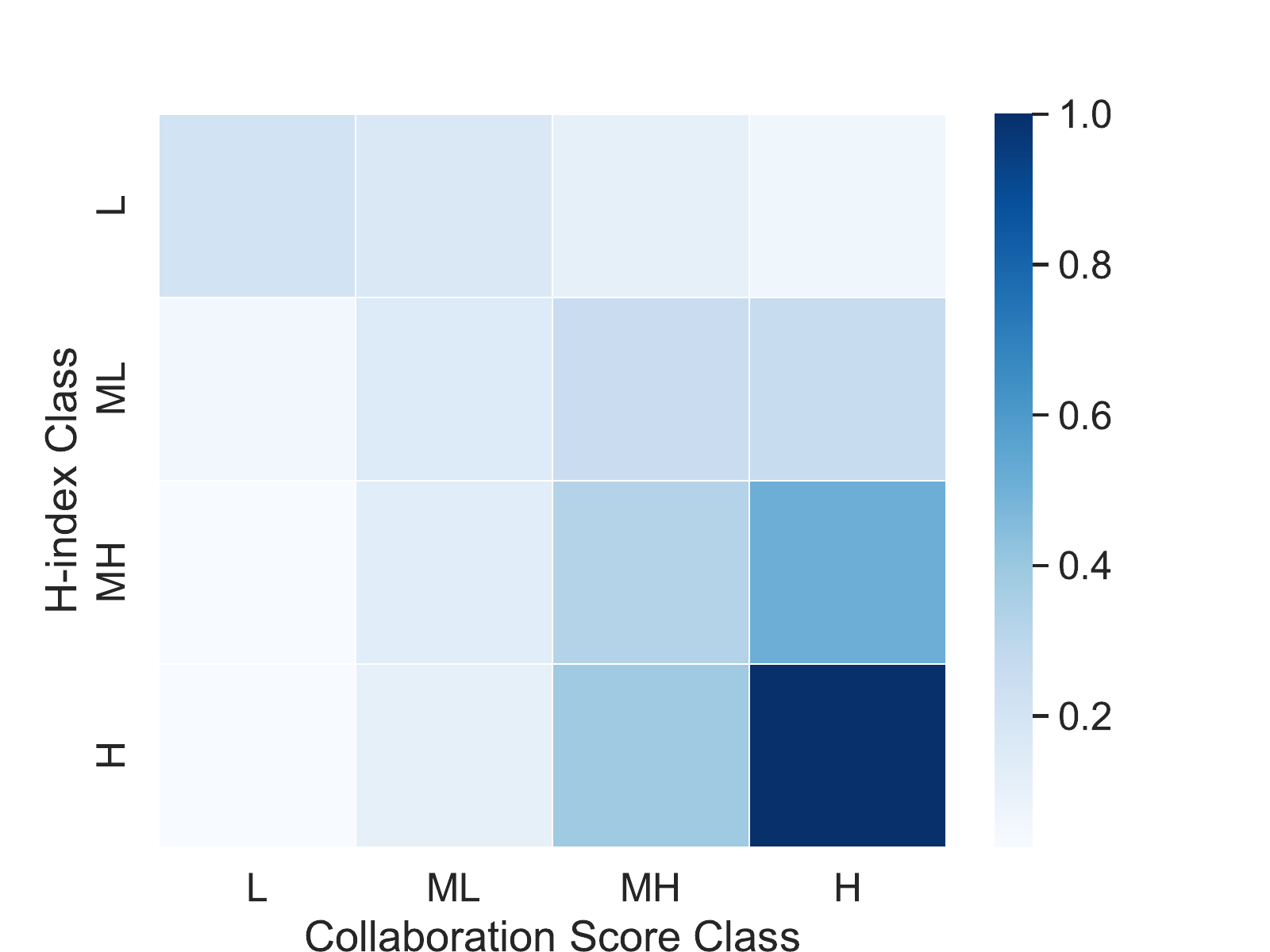}
  \caption{1996-2005}
  \end{subfigure}
    \begin{subfigure}[b]{0.196\textwidth}  \includegraphics[width=\textwidth]{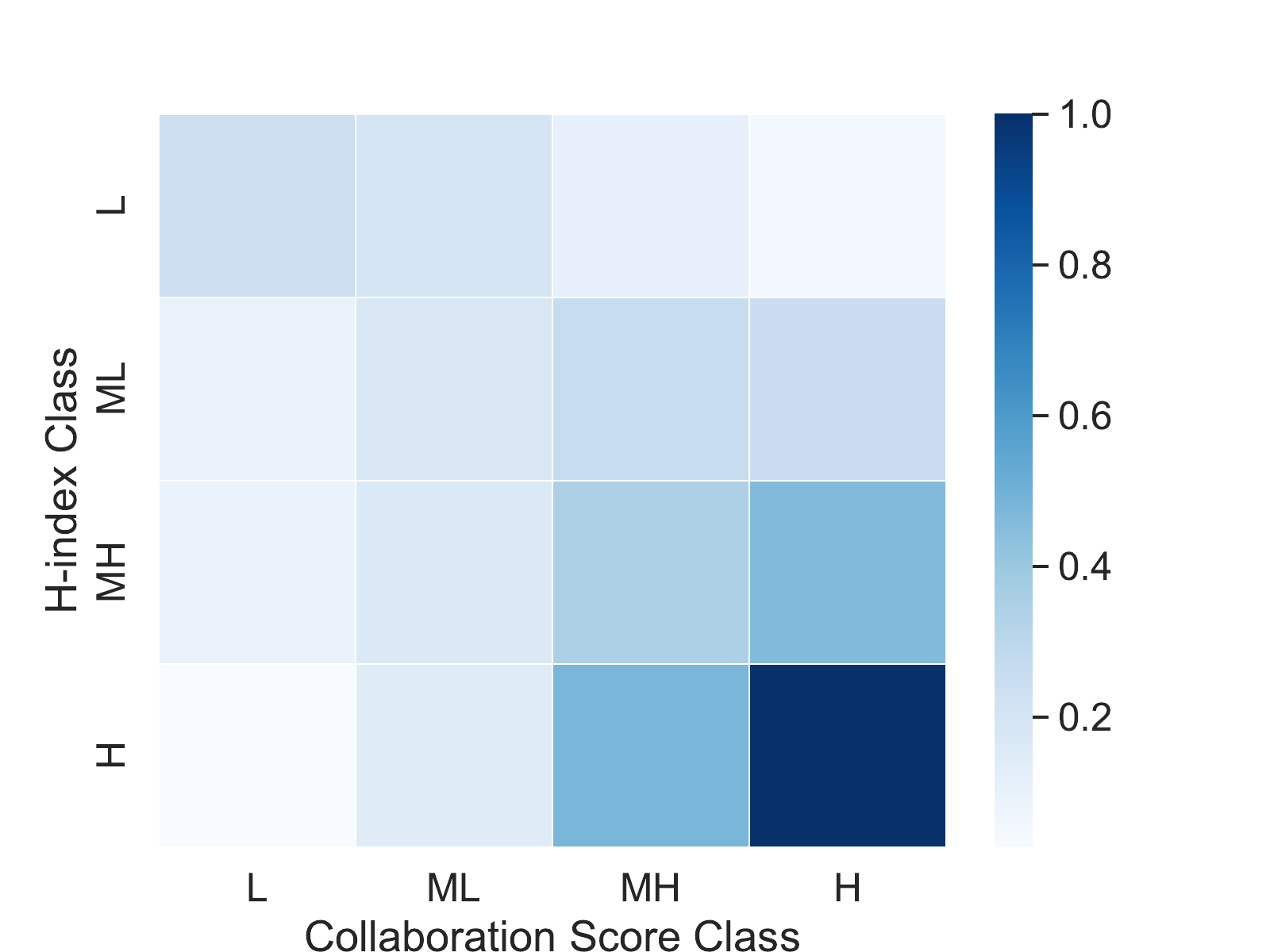}
  \caption{2006-2015}
  \end{subfigure}
  \caption{Relationship between entrance collaboration score of nodes and their h-index after 10 years for NLP }
  \label{fig:cor_nlp_se}
  \vspace{0.5cm}
\end{figure*}

\begin{figure*}
  \centering
  \begin{subfigure}[b]{0.19\textwidth}  \includegraphics[width=\textwidth]{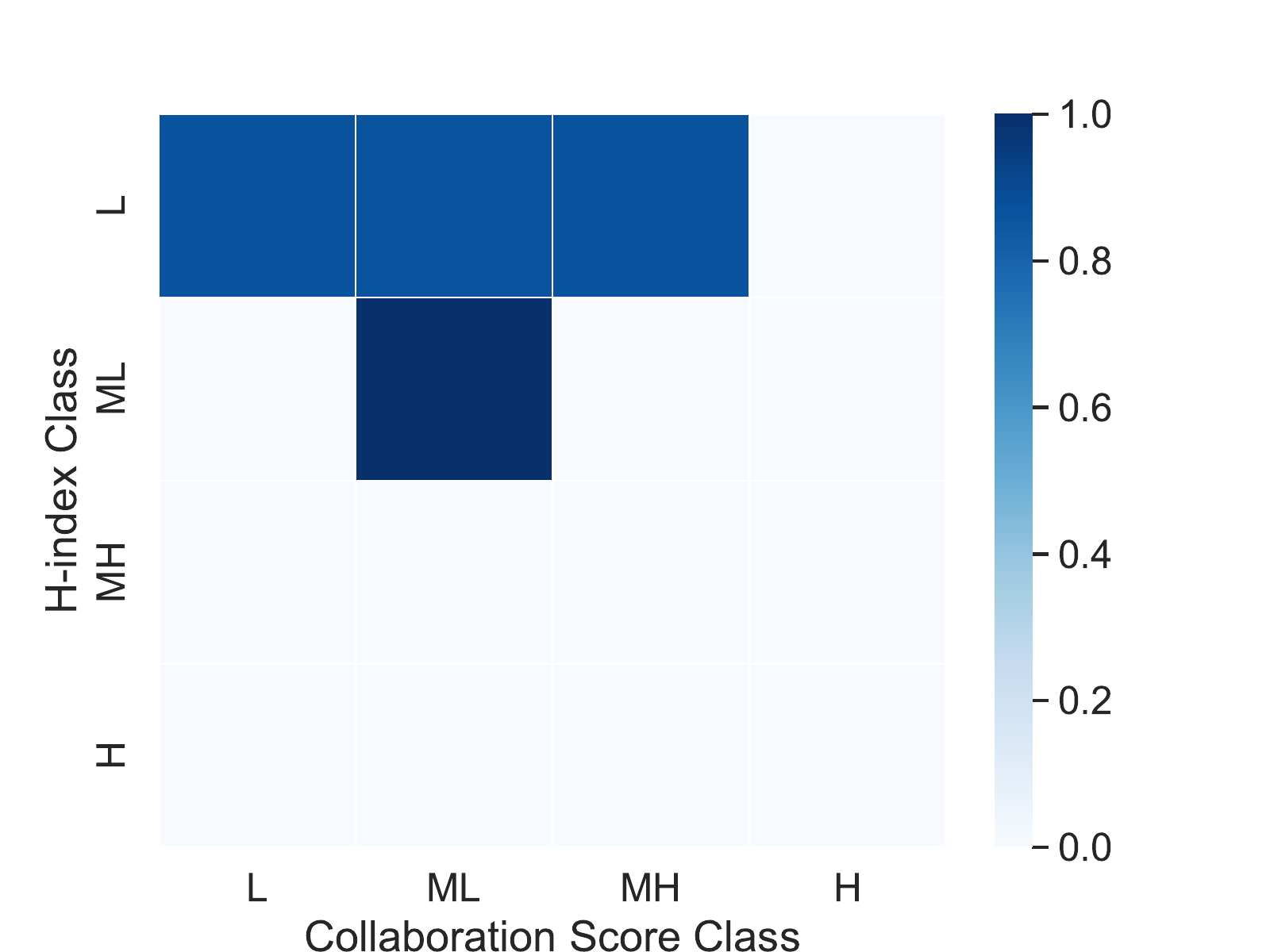}
  \caption{1966-1975}
  \end{subfigure}
  \begin{subfigure}[b]{0.19\textwidth}  \includegraphics[width=\textwidth]{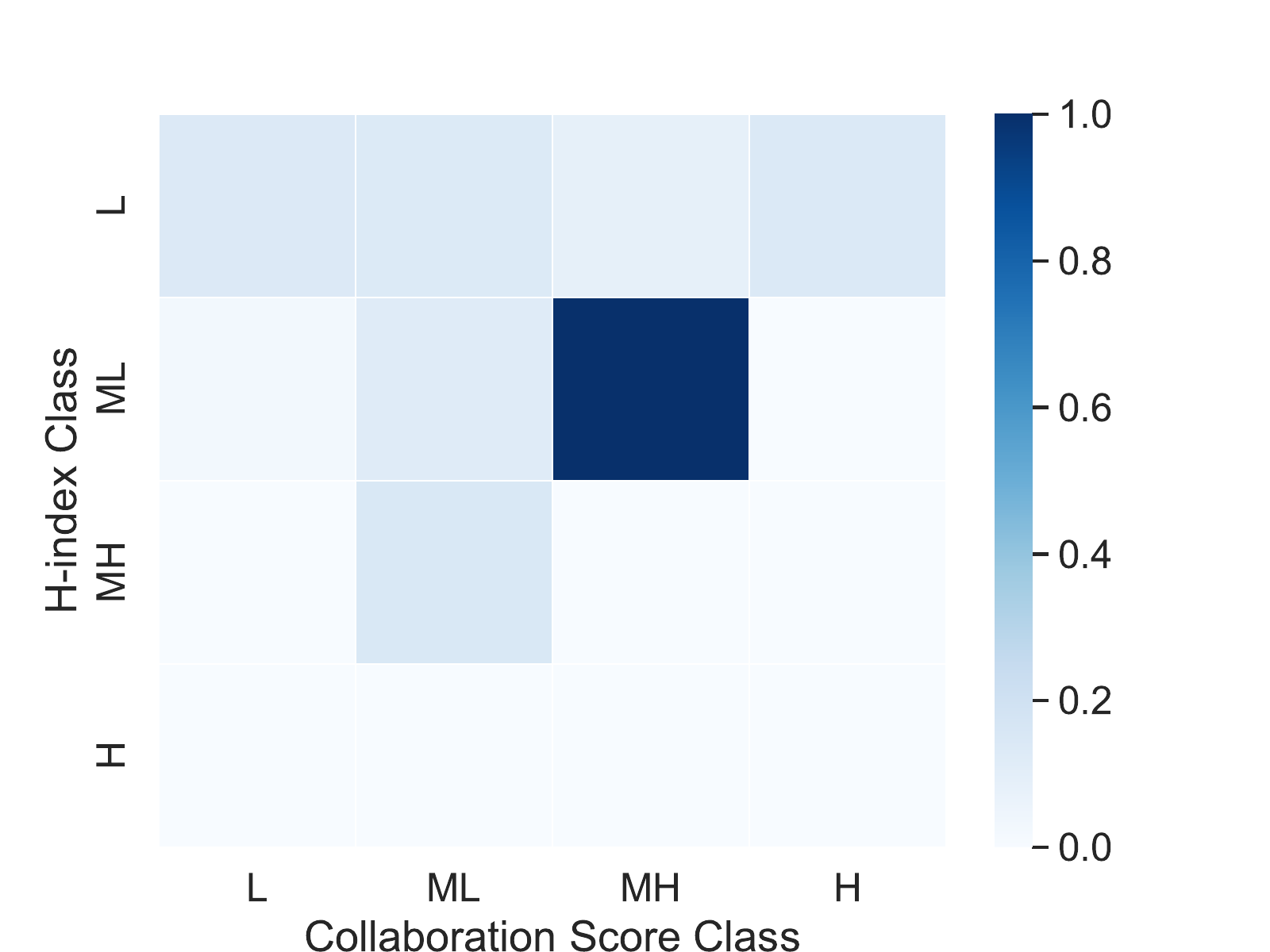}
  \caption{1976-1985}
  \end{subfigure}
    \begin{subfigure}[b]{0.19\textwidth}  \includegraphics[width=\textwidth]{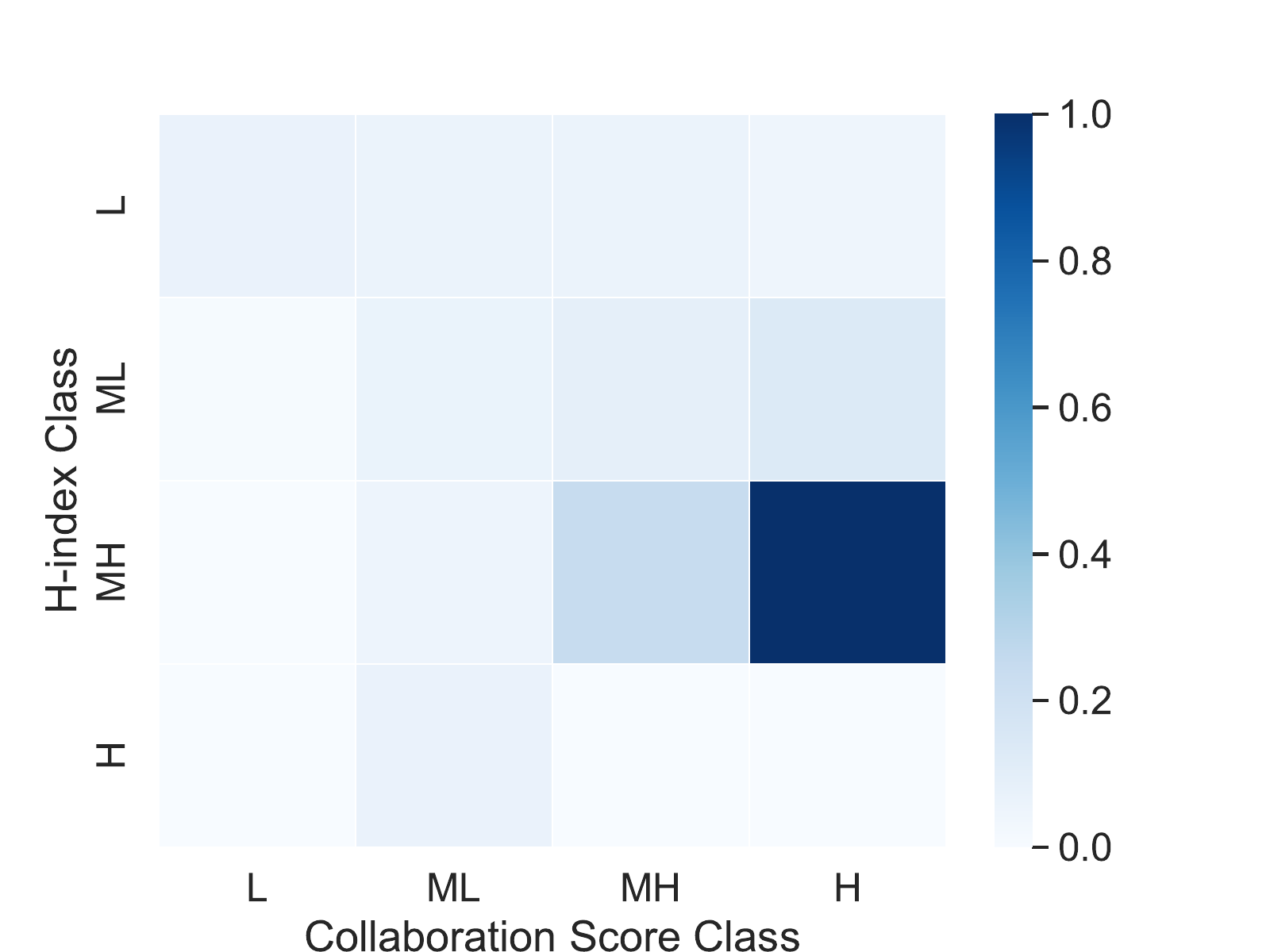}
  \caption{1986-1995}
  \end{subfigure}
    \begin{subfigure}[b]{0.19\textwidth}  \includegraphics[width=\textwidth]{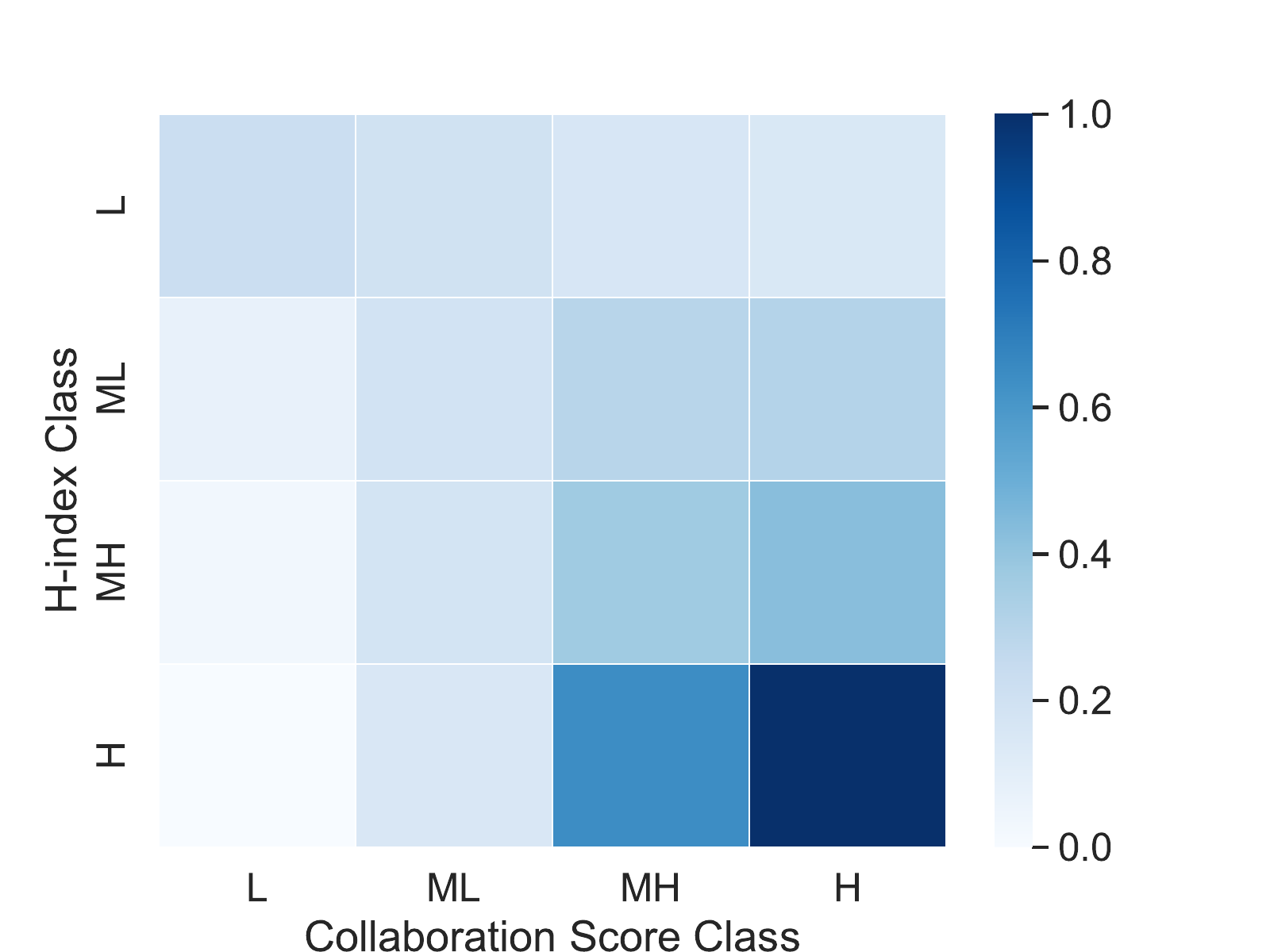}
  \caption{1996-2005}
  \end{subfigure}
    \begin{subfigure}[b]{0.196\textwidth}  \includegraphics[width=\textwidth]{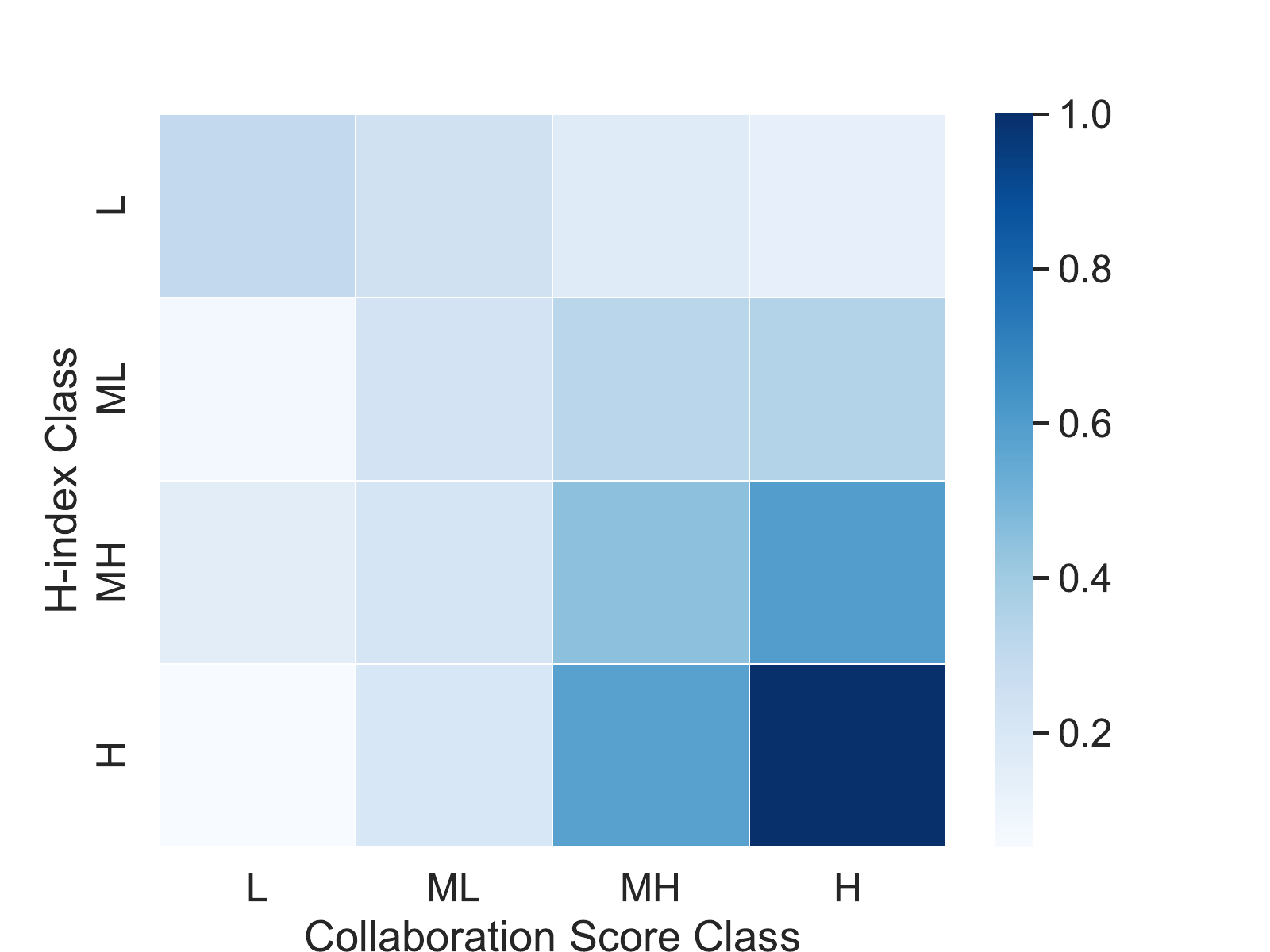}
  \caption{2006-2015}
  \end{subfigure}
  \caption{Relationship between entrance collaboration score of nodes and their h-index after 10 years for Computational Biology}
  \label{fig:cor_cb_se}
  \vspace{0.5cm}
\end{figure*}

Figures~\ref{fig:cor_cl_se} to \ref{fig:cor_cb_se} show the relationship between entrance collaboration scores and h-index of researchers after ten year in the Computational Linguistics, NLP, and  Computational Biology fields respectively.
 Cell $(c_i, c_j)$ is the normalized number of authors with starting collaboration score from class $c_i$ and $h$-index of class $c_j$ after 10 years, ($cell (c_i, c_j) = \frac{|(c_i, c_j)|}{|c_i|\cdot|c_j|}$, where $|c_i|$ is the number of authors in class $c_i$).
The $x$-axis shows the class of collaboration scores and the $y$-axis the class of $h$-indices (as before, each is divided into 4 tiers).

\newpage

